\documentclass[reqno,a4paper,11pt]{article}
\pdfoutput=1
\usepackage{xcolor}

\usepackage{graphicx}
\usepackage[textwidth = 430 pt, textheight = 630 pt]{geometry}

\definecolor{MyDarkBlue}{rgb}{0.15,0.25,0.45}
\usepackage{epsfig,rotating}
\usepackage{amsmath,amssymb}
\usepackage{amsfonts}
\usepackage{mathrsfs}
\usepackage{bbm}
\usepackage[normalem]{ulem}
\usepackage{booktabs}
\usepackage{enumerate}

\usepackage{booktabs}

\usepackage{latexsym}
\usepackage{amsthm}
\usepackage[all,knot]{xy}
\xyoption{arc}

\usepackage[utf8x]{inputenc}

\usepackage{hyperref}
\hypersetup{
hypertexnames=false,
colorlinks=true,
citecolor=MyDarkBlue,
linkcolor=MyDarkBlue,
urlcolor=MyDarkBlue,
pdfauthor={Christian S\"amann and Lennart Schmidt},
pdftitle={Towards an M5-Brane Model II: Metric String Structures},
pdfsubject={hep-th math-ph},
breaklinks=true
}

\usepackage{tikz}
\usetikzlibrary{cd} 
\usepackage{mathtools}
\usepackage[aligntableaux=center]{ytableau}
\usepackage[all,knot]{xy}
\xyoption{arc}

\let\fn\footnote
\renewcommand{\footnote}[1]{\linespread{1.1}\fn{#1}\linespread{1.29}}

\makeatletter\renewcommand{\section}{\@startsection
{section}{1}{\z@}{-3.5ex plus -1ex minus
    -.2ex}{2.3ex plus .2ex}{\bf }}
\makeatletter\renewcommand{\subsection}{\@startsection{subsection}{2}{\z@}{-3.25ex
plus -1ex minus
   -.2ex}{1.5ex plus .2ex}{\bf }}
\makeatletter\renewcommand{\subsubsection}{\@startsection{subsubsection}{3}{-2.45ex}{-3.25ex
plus -1ex minus -.2ex}{1.5ex plus .2ex}{\it }}
\renewcommand{\thesection}{\arabic{section}}
\renewcommand{\thesubsection}{\arabic{section}.\arabic{subsection}}
\renewcommand{\@seccntformat}[1]{\@nameuse{the#1}.~~}

\renewcommand{\theequation}{\thesection.\arabic{equation}}
\providecommand*{\xhookrightfill@}{%
  \arrowfill@{\lhook\joinrel\relbar}\relbar\rightarrow
}
\providecommand*{\xhookrightarrow}[2][]{%
  \ext@arrow 0395\xhookrightfill@{#1}{#2}%
}
\renewcommand{\twoheadrightarrow}{\mathrel{\mathrlap{\rightarrow}\mkern1mu\rightarrow}}
\makeatletter \@addtoreset{equation}{section}
\def\Ddots{\mathinner{\mkern1mu\raise\p@
\vbox{\kern7\p@\hbox{.}}\mkern2mu
\raise4\p@\hbox{.}\mkern2mu\raise7\p@\hbox{.}\mkern1mu}}
\usepackage[toc,page]{appendix}

\newtheorem{thm}{Theorem}[section]
\renewcommand{\thethm}{\thesection.\arabic{thm}}
\newtheorem{lemma}[thm]{Lemma}
\newtheorem{definition}[thm]{Definition}
\newtheorem{theorem}[thm]{Theorem}
\newtheorem{proposition}[thm]{Proposition}
\newtheorem{corollary}[thm]{Corollary}

\renewcommand{\appendices}{
\section*{Appendix}\label{appendices}\setcounter{subsection}{0}
\addcontentsline{toc}{section}{Appendix}
\setcounter{equation}{0}
\makeatletter
\renewcommand{\theequation}{\Alph{subsection}.\arabic{equation}}
\renewcommand{\thesubsection}{\Alph{subsection}}
\renewcommand{\thethm}{\Alph{subsection}.\arabic{thm}}
\@addtoreset{equation}{subsection}
\@addtoreset{thm}{subsection}
\makeatother
}

\usepackage{macros} 

\begin{document}
\begin{titlepage}
\begin{flushright}
 EMPG--19--20
\end{flushright}
\vskip2.0cm
\begin{center}
{\LARGE \bf Towards an M5-Brane Model II:\\[0.3cm] Metric String Structures}
\vskip1.5cm
{\Large Christian S\"amann$^a$ and Lennart Schmidt$^b$}
\setcounter{footnote}{0}
\renewcommand{\thefootnote}{\arabic{thefootnote}}
\vskip1cm
{\em ${}^a$ Maxwell Institute for Mathematical Sciences\\
Department of Mathematics, Heriot--Watt University\\
Colin Maclaurin Building, Riccarton, Edinburgh EH14 4AS,
U.K.}\\[0.5cm]
{\em ${}^b$ Department of Physics\\
National University of Singapore\\
2 Science Drive 3, Singapore 117551}\\[0.5cm]
{Email: {\ttfamily c.saemann@hw.ac.uk~,~phylen@nus.edu.sg}}
\end{center}
\vskip1.0cm
\begin{center}
{\bf Abstract}
\end{center}
\begin{quote}
In this paper, we develop the mathematical formulation of metric string structures. These play a crucial role in the formulation of certain six-dimensional superconformal field theories and we believe that they also underlie potential future formulations of the $(2,0)$-theory. We show that the connections on non-abelian gerbes usually introduced in the literature are problematic in that they are locally gauge equivalent to connections on abelian gerbes. Connections on string structures form an exception and we introduce the general concept of an adjusted Weil algebra leading to potentially interacting connections on higher principal bundles. Considering a special case, we derive the metric extension of string structures and the corresponding adjusted Weil algebra. The latter lead to connections that were previously constructed by hand in the context of gauged supergravities. We also explain how the Leibniz algebras induced by an embedding tensor in gauged supergravities fit into our picture.
\end{quote}
\end{titlepage}

\tableofcontents

\section{Introduction and results}

\subsection{Overview}

Our understanding of M-theory would be vastly improved by a clean picture of the effective dynamics of stacks of multiple M5-branes. These dynamics are governed by the so-called $(2,0)$-theory, a six-dimensional superconformal field theory, whose existence was postulated over 20~years ago~\cite{Witten:1995zh}. Attempts at constructing a classical Lagrangian of this theory have so far failed, and it is believed that such a Lagrangian does not exist, see e.g.~\cite{Witten:2007ct}. On closer inspection, however, many of the arguments against its existence are not conclusive~\cite{Saemann:2017zpd} and there may still be hope if we can identify the correct mathematical framework.

The $(2,0)$-theory involves a 2-form potential and deforming the free abelian theory to an interacting one is already a challenge. As proved in~\cite{Bekaert:9909094,Bekaert:2000qx}, there is no continuous such deformation. But this may be too much to ask; the Lagrangian may be of Chern--Simons type and therefore demand for a discrete coupling constant. This is the case in the M2-brane models and higher Chern--Simons terms indeed are present in the $\CN=(1,0)$-supersymmetric model presented in~\cite{Saemann:2017zpd}.

Mathematically, the 2-form potential arising in the description of a single M5-brane is a connection on a gerbe, a higher or categorified notion of an abelian principal bundle. It is therefore reasonable to turn towards connections on the non-abelian generalizations of gerbes introduced in the literature~\cite{Breen:math0106083,Aschieri:2003mw}. These are given in terms of local 1- and 2-forms, where the additional 1-forms are required to circumvent the usual Eckmann--Hilton type argument that higher-dimensional parallel transport has to be abelian, cf.~\cite{Baez:2010ya,Saemann:2016sis}.

At an abstract level, such connections allow for an elegant construction of 6d superconformal field equations via a higher-dimensional Penrose--Ward transform~\cite{Saemann:2012uq,Saemann:2013pca,Jurco:2014mva,Jurco:2016qwv}. Looking at concrete examples, however, suggests that the solutions of these equations are not particularly interesting. Similarly, direct constructions of a Lagrangian involving connections on non-abelian gerbes led to negative results, see e.g.~\cite{Ho:2012nt}.

As we show in section~\ref{ssec:fake_flat_trivial}, the connections defined in~\cite{Breen:math0106083,Aschieri:2003mw} are locally gauge-equivalent to connections on abelian gerbes. While they are suitable for higher versions of Chern--Simons theory, they necessarily fail in the description of non-abelian field theories that may contain locally non-vanishing 2-form curvatures. This is, in fact, a rather general feature of connections on higher principal bundles. Higher gauge algebras are modeled by $L_\infty$-algebras, and each $L_\infty$-algebra comes with its own homotopy Maurer--Cartan theory, a generalization of Chern--Simons theory. For every $L_\infty$-algebra, we thus obtain gauge potentials, curvatures, gauge transformations and Bianchi identities; that is, a full set of kinematical data for a (higher) gauge theory. This straightforward categorification of connections leads precisely to kinematical data which is suitable for higher Chern--Simons theories, but fails for the purposes of non-topological higher gauge theories.

For certain $L_\infty$-algebras, however, there is a choice that one can make in the definition of the kinematical data, which allows for connections on non-abelian gerbes which are {\em not} gauge equivalent to connections on abelian gerbes. One class of such $L_\infty$-algebras are the string Lie 2-algebras, higher analogues of the Lie algebras $\aspin(n)$. These are particularly interesting since their appearance in the description of the $(2,0)$-theory is expected for a number of reasons~\cite{Saemann:2019leg}. Furthermore, the string group $\sString(3)$ is a categorified version of $\sSpin(3)\cong \sSU(2)$~\cite{Saemann:2017rjm}, the simplest, interesting non-abelian Lie group. Just as $\sSU(2)$ is the total space of the Hopf fibration and intimately linked to monopoles, $\sString(3)$ underlies a categorified Hopf fibration linked to the categorified monopoles known as self-dual strings, cf.~\cite{Saemann:2017rjm}.

In this paper, we derive in detail the local connection data, the appropriate notion of curvature, the gauge transformations as well as the Bianchi identities for two models of the string Lie 2-algebra, allowing for an interpolation to general string Lie 2-algebra models. We also develop the metric extensions which are required for an action principle, and point out the relation of the resulting local connection data with the higher form curvatures obtained in the tensor hierarchy of gauged supergravities.

\subsection{The problem with non-abelian connections}

A definition of connections that allows for a generalization to $L_\infty$-algebras~\cite{Sati:2008eg} was given long ago by Henri Cartan~\cite{Cartan:1949aaa,Cartan:1949aab}. In this approach, the dichotomy of Lie algebras and differential forms, the two basic ingredients in the local definition of connections, is overcome by moving from a Lie or $L_\infty$-algebra $\frg$ to its dual differential graded algebra (dga), known as the {\em Chevalley--Eilenberg algebra} $\sCE(\frg)$. Morphisms between $\sCE(\frg)$ and the dga of differential forms $(\Omega^\bullet(U),\dd)$ on a contractible patch $U$ of some manifold encode flat connections on $U$. Non-flat connections are obtained if one replaces $\frg$ with the corresponding $L_\infty$-algebra of inner derivations, $\inn(\frg)$, whose Chevalley--Eilenberg algebra is known as the {\em Weil algebra} $\sW(\frg)$ of $\frg$. To define global $\frg$-connection objects, one imposes constraints on the morphisms from $\sW(\frg)$ to $\Omega^\bullet(U)$. In particular, the morphism has to map a particular differential graded subalgebra of $\sW(\frg)$, the invariant polynomials $\inv(\frg)$ of $\frg$, to global objects and the images of a specific subset, that is, the reduced invariant polynomials, will form the topological invariants.

The invariant polynomials now sit in a complex,
\begin{equation}\label{eq:ses_1}
 \begin{tikzcd}
  0 & \arrow[l] \sCE(\frg) & \sW(\frg)\arrow[l] & \inv(\frg) \arrow[l] & 0 \arrow[l]~~,
 \end{tikzcd} 
\end{equation}
which already exhibits the problems arising in the straightforward categorification of connections based on $L_\infty$-algebras. Recall that the appropriate notion of isomorphism for $L_\infty$-algebra is that of a quasi-isomorphism and for the definitions of the Weil algebra and the invariant polynomials to be meaningful, they have to be compatible with these. That is, a quasi-isomorphism $\phi:\frg\rightarrow \tilde \frg$ has to induce a chain of (dual) quasi-isomorphisms,
\begin{equation}\label{eq:ses_2}
 \begin{tikzcd}
  0 & \arrow[l] \sCE(\frg) \arrow[d,leftrightarrow]{}{\approxeq} & \sW(\frg)\arrow[l] \arrow[d,leftrightarrow]{}{\approxeq}& \inv(\frg) \arrow[l]\arrow[d,leftrightarrow]{}{\approxeq} & 0 \arrow[l]\\
  0 & \arrow[l] \sCE(\tilde \frg) & \sW(\tilde \frg)\arrow[l] & \inv(\tilde \frg) \arrow[l] & 0 \arrow[l]
 \end{tikzcd}
\end{equation}
While this is always true for Lie algebras, it fails to hold for general $L_\infty$-algebras, where the quasi-isomorphism between the invariant polynomials fails to exist in general. The conclusion of~\cite{Sati:2008eg} was that the Weil algebra should be deformed by a topological invariant.

This, however, is not the only issue with higher connections: gauge transformations, in general, close only up to an equation of motion, which is known as the fake flatness condition, a particular restriction on the curvature of higher connections.

Physicists would say the BRST complex is open, and we have to lift to the BV complex, where the required equations of motion are imposed, cf.~the discussion in~\cite{Jurco:2018sby}. The problem with fake flatness, however, is that this equation effectively renders the higher connections abelian, which we prove in section~\ref{ssec:fake_flat_trivial}. As a consequence, it is virtually impossible to write down gauge-invariant equations of motion for interacting field theories.

The requirement that all fake curvature forms need to vanish has also been observed in the finite description of connections in terms of parallel transport functors~\cite{Baez:2004in,Baez:0511710}. Here it was found that a consistent parallel transport of strings is only invariant under surface reparametrizations if the fake curvature condition is met.

\subsection{String structures as an exception}

For particular $L_\infty$-algebras, such as the string Lie 2-algebras, however, the definition of the Weil algebra can be modified~\cite{Sati:2008eg}, guaranteeing the expected compatibility of the complex~\eqref{eq:ses_1} with (dual) quasi-isomorphisms. The dga-morphisms to differential forms then yield the connections on higher generalizations of spin structures known as {\em string structures}~\cite{Killingback:1986rd,Witten:1987cg}, see section~\ref{ssec:local_connections_ordinary_string_structures}. These were first discovered when trying to couple gauge potential 1-forms to the Kalb--Ramond $B$-field in supergravity~\cite{Bergshoeff:1981um,Chapline:1982ww}.

The discussion of the string Lie 2-algebra in~\cite{Sati:2008eg}, however, is incomplete for our purposes. First, string Lie 2-algebras can be modeled by various representatives in their quasi-isomorphism classes. The explicit formulas for string connections in~\cite{Sati:2008eg} as well as the classical formulas of~\cite{Bergshoeff:1981um,Chapline:1982ww} are given only for {\em minimal} or {\em skeletal models}. For these, the underlying graded vector space is minimal but the Jacobiator, which encodes the failure of the binary product to preserve the Jacobi identity, is non-trivial. As a consequence, these models are hard to integrate\footnote{While there is an abstract prescription for integrating $L_\infty$-algebras~\cite{Henriques:2006aa}, it is fair to say that concrete computations are hard, even for the simple example of the String Lie 2-group. For example, the string group model presented in~\cite{Schommer-Pries:0911.2483}, which integrates the skeletal model, contains essential structures which are only proven to exist, but not given explicitly.}, and any concepts involving finite transformations, such as the transition functions of an underlying principal 2-bundles, become difficult to work with. A consistency requirement for eventual higher gauge theories is that they are agnostic about which representative of the quasi-isomorphism class of an $L_\infty$-algebra is used in their definition. We thus need to extend the discussion to at least the other extreme, that is, the case of {\em strict models} of the string Lie-algebras, in which the Jacobiator is trivial.

Second, the graded vector spaces underlying string Lie 2-algebra models do not admit a symplectic structure, the necessary ingredient in defining a {\em cyclic structure}, the correct notion of an inner product for $L_\infty$-algebras. This is a severe gap if one wants to define natural action functionals for higher gauge theories with gauge algebra the string Lie 2-algebra.\footnote{See however~\cite{Fiorenza:2012tb} for an alternative way of defining an action functional for higher connections `by hand', without such a cyclic structure.} A solution to this problem is to use a procedure similar to introducing antifields in the BV formalism and to essentially double the relevant graded vector space by applying a degree-shifted cotangent functor~\cite{Saemann:2017zpd,Saemann:2017rjm}. This leads to metric string structures which arise naturally in 6d $\CN=(1,0)$ supersymmetric field theories~\cite{Saemann:2017zpd,Saemann:2017rjm}, and we believe these to be relevant to a potential future construction of a Lagrangian for the (2,0)-theory. The full development of these metric string structures, i.e.~the explicit derivation of the form of the connections and their curvature, their gauge transformations and the Bianchi identities, for general string Lie 2-algebra models is the main goal of this paper.

\subsection{Results}

In this paper, we try to be mostly self-contained; we intend to establish notation and conventions, carefully presenting our motivation, and providing relevant examples. The manuscript is rather long, and we therefore summarize our results in the following.
\begin{itemize}
    \item[$\circ$] Theorem~\ref{thm:equivalence_of_field_theories} states that two gauge theories over the same contractible manifold are equivalent if and only if their gauge $L_\infty$-algebras are quasi-isomorphic.
    \item[$\circ$] Definition~\ref{def:adjusted_Weil_algbra} formulates the concept of {\em adjusted Weil algebras}, which lead to higher connections whose gauge transformations close without imposing any further conditions such as fake flatness. That is, their BRST complex is closed. 
    \item[$\circ$] The relevance of this definition is underlined by Theorem~\ref{thm:fake_flat_trivial}, which states that higher fake-flat connections (i.e.~the connections usually found in the literature) are always gauge equivalent to abelian ones. This important statement is not particularly deep, but to the best of our knowledge, it has not been stated anywhere else so far in the literature.
    \item[$\circ$] Our main result are formulas~\eqref{eq:metric_string_structures_sk} and~\eqref{eq:metric_string_structures_lp}, which give the curvatures, gauge transformations and Bianchi identities for the skeletal and the loop model of metric string structures; formulas for other models follow from combining these. As stated above, we believe that these formulas are crucial in the potential future construction of an action for the (2,0)-theory.
    \item[$\circ$] In section~\ref{ssec:sd3forms}, we compare our formulas for curvatures to those derived from tensor hierarchies. Our formulas fill in some gaps in the definition of the latter curvatures and show that some modifications should be made.
    \item[$\circ$] An important result in this context is presented in section~\ref{ssec:tensor_hierarchies}, where we show that the structure encoded in the embedding tensor of supergravity is that of a weak Lie 2-algebra, from which recently given interpretations in terms of Leibniz algebras directly follow.
\end{itemize}

Minor results, which we did not find in the literature, but which we found to be relevant to the general discussion:
\begin{itemize}
    \item[$\circ$] Theorem~\ref{thm:quasi-iso_inv_inv} states that quasi-isomorphisms of $L_\infty$-algebras are compatible with both definitions of invariant polynomials of $L_\infty$-algebras.
    \item[$\circ$] In section~\ref{ssec:BRST_complex}, we explain how the BRST complex of a higher gauge theory is rapidly derived from a generalisation of the AKSZ-formalism. This is very helpful in the computation of adjusted higher connections.
    \item[$\circ$] The formulas for compositions of $L_\infty$-algebra 2-morphisms and quasi-isomorphisms in the dga-picture are worked out in appendix~\ref{app:A:concatenation}.
\end{itemize}

\section{\texorpdfstring{$L_\infty$}{L-infinity}-algebras and associated differential graded algebras}\label{sec:L_infty_algebras}

In this section, we review basic algebraic structures underlying the construction of (higher) gauge theories. We introduce $L_\infty$-algebras and their Chevalley--Eilenberg description in terms of a differential graded commutative\footnote{All our differential graded algebras will be commutative, and we shall mostly drop this adjective from here on.} algebra (dga). We also introduce the associated Weil and free algebras. Finally, we review how quasi-isomorphisms, which form the relevant type of isomorphisms for $L_\infty$-algebras, are described in this picture.\footnote{Let us stress from the beginning that the nomenclature, the conventions and the notation in this paper differ to some degree from our previous papers~\cite{Saemann:2017rjm,Saemann:2017zpd,Saemann:2019leg}. We hope that our new choice is more consistent, easier to work with and more future proof.}

\subsection{Chevalley--Eilenberg algebra of an \texorpdfstring{$L_\infty$}{L-infinity}-algebra}\label{ssec:CE-algebras}

A natural and convenient categorification of the notion of a Lie algebra is given by what are called strong homotopy Lie algebras, or $L_\infty$-algebras for short. 

\begin{definition}
 An \underline{$L_\infty$-algebra} $\frg$ consists of a $\RZ$-graded vector space $\frg = \bigoplus_{k\in \RZ} \frg_k$ together with a set of totally antisymmetric, multilinear maps or \underline{higher products} $\mu_i :  \wedge^i \frg \to \frg,$ $i\in \NN^+$, of degree~$2-i$, which satisfy the \underline{higher} or \underline{homotopy Jacobi identities}
\begin{equation}\label{eq:hom_rel}
 \sum_{i+j=n}\sum_{\sigma\in S_{i|j}}\chi(\sigma;a_1,\ldots,a_{n})(-1)^{j}\mu_{j+1}(\mu_i(a_{\sigma(1)},\ldots,a_{\sigma(i)}),a_{\sigma(i+1)},\ldots,a_{\sigma(n)})=0
\end{equation}
for all $n\in\NN^+$ and $a_1,\dots,a_n\in \frg$, where the second sum runs over all $(i,j)$-unshuffles $\sigma\in S_{i|j}$. An \underline{$n$-term $L_\infty$-algebra}, or \underline{Lie $n$-algebra}\footnote{Strictly speaking, a Lie $n$-algebra is a $(n-1)$-fold categorification of a Lie algebra, but a restriction of the general categorification is categorically equivalent to $n$-term $L_\infty$-algebras. For the details in the case $n=2$, see~\cite{Baez:2003aa}. We therefore use the terms interchangeably. A general categorification will be called a {\em weak Lie $n$-algebra}.}, is an $L_\infty$-algebra that is concentrated (i.e.~non-trivial only) in degrees $-n+1,\dots,0$. The \underline{trivial $L_\infty$-algebra} is the $L_\infty$-algebra $\frg=\bigoplus_{k\in\RZ}\frg_k$ with $\frg_k=\{0\}$.
\end{definition}
Here, an {\em unshuffle} $\sigma\in S_{i|j}$ is a permutation whose image consists of ordered tuples $\big(\sigma(1),\dots,\sigma(i)\big)$ and $\big(\sigma(i+1),\dots,\sigma(n)\big)$. Moreover, $\chi(\sigma; a_1,\dots,a_n)$ denotes the {\em graded antisymmetric Koszul sign} defined by the graded antisymmetrized products
\begin{equation}
a_1 \dots   a_n = \chi(\sigma;a_1,\dots,a_n)a_{\sigma(1)} \dots  a_{\sigma(n)}~,
\end{equation}
where any transposition involving an even element acquires a minus sign.

In the categorification of a Lie algebra to a Lie 2-algebra, we relax the Jacobi identity to hold up to a natural transformation. This is evident in the lowest few homotopy Jacobi relations, i.e.
\begin{equation}\label{eq:hJacobi_Lie_2}
\begin{aligned}
0&=\mu_1\left(\mu_1\left(a_1\right)\right)~,\\
0&= \mu_1\left(\mu_2\left(a_1,a_2\right)\right)-
\mu_2\left(\mu_1\left(a_1\right),a_2\right)+
(-1)^{\left| a_1\right|  \left| a_2\right| } \mu_2\left(\mu_1\left(a_2\right),a_1\right)~,\\
0 &=\mu_1\left(\mu_3\left(a_1,a_2,a_3\right)\right)-\mu_2\left(\mu_2\left(a_1,a_2\right),a_3\right)+(-1)^{|a_2| |a_3|} \mu_2(\mu_2(a_1,a_3),a_2)-\\
&~~~~-(-1)^{|a_1|(|a_2|+ |a_3|)} \mu_2(\mu_2(a_2,a_3),a_1)-
(-1)^{|a_1| |a_2|} \mu_3(\mu_1(a_2),a_1,a_3)+\\
&~~~~+
\mu_3(\mu_1(a_1),a_2,a_3)+
(-1)^{(|a_1|+|a_2|) |a_3|} \mu_3(\mu_1(a_3),a_1,a_2)~,
\end{aligned}
\end{equation}
where $a_i \in \frg$. These relations state that $\mu_1$ is a graded differential compatible with $\mu_2$, and $\mu_2$ is a generalization of a Lie bracket with the violation of the Jacobi identity controlled by $\mu_3$.

There is an alternative and elegant way of describing an $L_\infty$-algebra $\frg$ and its higher Jacobi relations in terms of a coalgebra and coderivations~\cite{Lada:1992wc}, cf.~also~\cite{Lada:1994mn}. To see this, consider the grade-shifted vector space $\frg[1]$, where the square bracket refers to a degree shift of all elements of $\frg$ by $-1$ and, correspondingly, of all coordinate functions by $+1$, cf.~\cite{Jurco:2018sby}. This degree shift induces a shift of the degree of the maps $\mu_i$ from $2-i$ to $1$ and allows to define a degree~$1$ coderivation
\begin{equation}
\CD:\odot^\bullet \frg[1]\to\odot^\bullet \frg[1]~,
\end{equation}
which acts on the graded symmetric coalgebra $\odot^\bullet \frg[1]$ generated by $\frg[1]$.

More explicitly, $\odot^\bullet \frg[1]$ is spanned by graded symmetric elements $a_1 \odot\dots\odot\, a_n$ and is equipped with the coproduct 
\begin{equation}
\Delta(a_1\odot\dots\odot a_n) = \sum\limits_{i+j=n}\,\,\sum\limits_{\sigma\in S_{i|j}} \epsilon(\sigma;a_1,\dots,a_n) (a_{\sigma(1)}\odot\dots\odot a_{\sigma(i)})\otimes(a_{\sigma(i+1)}\odot\dots\odot a_{\sigma(n)})~,
\end{equation}
where $S_{i|j}$ again denotes the set of $(i,j)$-unshuffles and $\epsilon$ is now the {\em graded symmetric Koszul sign}, which is related to the graded antisymmetric Koszul sign via
\begin{equation}
\epsilon(\sigma;a_1,\ldots,a_n) = \text{sgn}(\sigma)\chi(\sigma;a_1,\dots,a_n)~.
\end{equation}
A {\em coderivation} $\CD$ is given by a linear map $\CD:\odot^\bullet \frg[1]\to\odot^\bullet\frg[1]$ which satisfies the co-Leibniz rule
\begin{equation}
\Delta\circ \CD = (\CD \otimes \text{id}+\text{id} \otimes \CD) \circ \Delta~.
\end{equation}
We note that the higher products $\mu_i$ induce maps from $\odot^i \frg[1]\rightarrow \frg[1]$, which can be extended to coderivations $\CD_i$. The sum of all these codifferentials\footnote{We insert some additional signs for convenience.} combine into a coderivation $\CD$ encoding the $L_\infty$-algebra $\frg$, where the higher Jacobi identities correspond to the coderivation squaring to zero, i.e.~$\CD^2=0$.

The third way of describing $L_\infty$-algebras, which is the important one for this paper, is the dualization of the above coalgebra description. In the case of ordinary Lie algebras, this yields what is known as the Chevalley--Eilenberg algebra of a Lie algebra.

\begin{definition}
 The \underline{Chevalley--Eilenberg} algebra $\sCE(\frg)$ of an $L_\infty$-algebra $\frg$ encoded in a codifferential $\CD$ is the differential graded commutative algebra
 \begin{equation}
  \sCE(\frg)~\coloneqq~\big(\odot^\bullet(\frg[1]^*)\,,\,Q\,\big)~,
 \end{equation}
where $Q=\CD^*$ is the \underline{homological vector field}, i.e.~a vector field on $\frg[1]$ of degree~1 satisfying $Q^2=0$, which acts as a differential on $\odot^\bullet(\frg[1]^*)$. We call the graded vector space $\frg[1]$ together with $Q$ the \underline{differential graded (dg)-manifold} corresponding to the $L_\infty$-algebra $\frg$.
\end{definition}
Recall that a dg-manifold is a graded manifold with a differential on the algebra of smooth functions. These are often called $Q$-manifolds in the literature due to the homological vector field $Q$ inducing the differential. Note that $L_\infty$-algebras correspond to dg-manifolds with a vector spaces (which is often trivial) in degree~$0$. A general dg-manifold corresponds to an {\em $L_\infty$-algebroid}, and each $L_\infty$-algebroid also comes with a Chevalley--Eilenberg algebra. An important example is the grade-shifted tangent bundle $T[1]M$ of a manifold $M$, where the Chevalley--Eilenberg algebra, i.e.~the algebra of smooth functions on $T[1]M$, can be identified with the differential forms on $M$ and the differential $Q$ plays the role of the de Rham differential.

Note that we are only interested in $L_\infty$-algebras whose graded vector spaces are simple enough (e.g.~finite-dimensional) to allow for a linear dual.

As a first  example, consider a finite-dimensional Lie algebra $\frg$. Then $\frg[1]$ comes with coordinate functions $t^\alpha$ (with respect to some basis) with degree $|t^\alpha|=1$ and the homological vector field is of the form $Qt^\alpha=-\tfrac12 f^\alpha_{\beta\gamma} t^\beta t^\gamma$. The condition $Q^2=0$ is equivalent to the $f^\alpha_{\beta\gamma}$ being the structure constants of a Lie algebra.

As a second example, consider a Lie 2-algebra (or 2-term $L_\infty$-algebra) $\frg=\frg_{-1}\oplus \frg_0$. Let $(t^\alpha,r^a)$ be the generators of $\frg[1]^*$ with degrees~1 and~2. A general homological vector field $Q$ acts on the generators of $\sCE(\frg)$ according to
\begin{equation}\label{eq:CE_Lie_2_algebra}
 Q~:~t^\alpha\mapsto -\tfrac12 f^\alpha_{\beta\gamma}t^\beta t^\gamma-f^\alpha_a r^a~,~~~r^a\mapsto-f^a_{\alpha b}t^\alpha r^b+\tfrac{1}{3!}f^a_{\alpha\beta\gamma}t^\alpha t^\beta t^\gamma~,
\end{equation}
where the structure constants $f^\alpha_a, f^\alpha_{\beta\gamma}, f^a_{\alpha b}, f^a_{\alpha\beta\gamma}\in \FR$ satisfy relations corresponding to~\eqref{eq:hJacobi_Lie_2} or, equivalently, to $Q^2=0$.

To reconstruct the higher products $\mu_i$ from $Q$ in the case of a general $L_\infty$-algebra $\frg$, we introduce the tensor product $\xi=z^A\otimes \tau_A$, where the $z^A$ are the coordinate functions on $\frg[1]$, while the $\tau_A$ are the corresponding basis vectors in $\frg$, thus $|\xi|=1$. We then have the formula 
\begin{equation}
 Q\xi=-\hat \mu_1(\xi)-\tfrac12 \hat \mu_2(\xi,\xi)-\tfrac{1}{3!}\hat \mu_3(\xi,\xi,\xi)-\dots~,
\end{equation}
where $\hat \mu_i$ are the higher products $\mu_i$ on $\frg$, extended to the $L_\infty$-algebra $\odot^\bullet(\frg[1]^*)\otimes \frg$, see~\cite{Jurco:2018sby} for all the details of this construction. The ordinary products are obtained using
\begin{equation}
 \hat \mu_i(z^{A_1}\otimes\tau_{B_1},\dots,z^{A_i}\otimes\tau_{B_i})=\pm z^{A_1}\dots z^{A_i}\otimes\mu_i(\tau_{B_1},\dots,\tau_{B_i})~,
\end{equation}
where the sign $\pm$ is the combination of all Koszul signs arising from commuting coordinate functions $z^A$ past basis vectors $\tau_B$ and pulling coordinate functions $z^A$ out of the higher product $\mu_i$ of degree~$2-i$.

An immediate advantage of the dga-perspective on $L_\infty$-algebras is that the appropriate notion of morphism is immediately clear:
\begin{definition}
A \underline{morphism of $L_\infty$-algebras} $\phi:\frg\rightarrow \tilde \frg$ is (the dual of) a morphism of differential graded algebras between the corresponding Chevalley--Eilenberg algebras $\sCE(\frg)$ and $\sCE(\tilde \frg)$, 
\begin{equation}
 \Phi:\sCE(\frg)\rightarrow \sCE(\tilde\frg)~.
\end{equation}
In particular, $\Phi$ is of degree~$0$ and respects the differential, i.e.~$\Phi \circ Q = \tilde Q \circ \Phi$. If $\Phi$ is invertible, we call $\phi$ an \uline{isomorphism of $L_\infty$-algebras}.
\end{definition}

\noindent In the dual picture, this translates to a collection of totally antisymmetric, multilinear maps $\phi_i:\wedge^i\frg\to\tilde \frg$ of degree~$1-i$ satisfying
\begin{subequations}\label{eq:L_infty_morphism}
\begin{equation}
\begin{aligned}
   &\sum_{j+k=i}\sum_{\sigma\in S(j|i)}~(-1)^{k}\chi(\sigma;a_1,\ldots,a_i)\phi_{k+1}(\mu_j(a_{\sigma(1)},\dots,a_{\sigma(j)}),a_{\sigma(j+1)},\dots ,a_{\sigma(i)})\\
   \ &=\ \sum_{j=1}^i\frac{1}{j!} \sum_{k_1+\cdots+k_j=i}\sum_{\sigma\in{\rm Sh}(k_1,\ldots,k_{j-1};i)}\chi(\sigma;a_1,\ldots,a_i)\zeta(\sigma;a_1,\ldots,a_i)\,\times\\
   &\kern1cm\times \mu'_j\Big(\phi_{k_1}\big(a_{\sigma(1)},\ldots,a_{\sigma(k_1)}\big),\ldots,\phi_{k_j}\big(a_{\sigma(k_1+\cdots+k_{j-1}+1)},\ldots,a_{\sigma(i)}\big)\Big)
\end{aligned}
\end{equation}
with the sign $\zeta(\sigma;a_1,\ldots,a_i)$ given by
\begin{equation}\label{eq:zeta-sign}
 \zeta(\sigma;a_1,\ldots,a_i)\ \coloneqq\ (-1)^{\sum_{1\leq m<n\leq j}k_mk_n+\sum_{m=1}^{j-1}k_m(j-m)+\sum_{m=2}^j(1-k_m)\sum_{k=1}^{k_1+\cdots+k_{m-1}}|a_{\sigma(k)}|}~.
\end{equation}
\end{subequations}

For example, a morphism of 2-term $L_\infty$-algebras $\phi:\frg\rightarrow \tilde \frg$ consists of maps $\phi_1:\frg\rightarrow \tilde \frg$ and $\phi_2:\frg \wedge \frg\rightarrow \tilde \frg$ of degrees $0$ and $-1$, respectively. The higher products on $\frg$ and $\tilde \frg$ are then related by the following formulas:
\begin{equation}\label{eq:Lie_2_algebra_morph}
\begin{aligned}
0 &= \phi_1(\mu_1(v_1)) - \mu'_1(\phi_1(v_1))~,\\
0 &= \phi_1(\mu_2(w_1,w_2)) - \mu'_1(\phi_2(w_1,w_2))-\mu'_2(\phi_1(w_1),\phi_1(w_2))~,\\
0 &= \phi_1(\mu_2(w_1,v_1)) +\phi_2(\mu_1(v_1),w_1) - \mu'_2(\phi_1(w_1),\phi_1(v_1))~,\\
0 &= \phi_1(\mu_3(w_1,w_2,w_3)) -\phi_2(\mu_2(w_1,w_2),w_3) + \phi_2(\mu_2(w_1,w_3),w_2)\\
&\phantom{{}={}} - \phi_2(\mu_2(w_2,w_3),w_1) - \mu'_3(\phi_1(w_1),\phi_1(w_2),\phi_1(w_3)) \\
&\phantom{{}={}} + \mu'_2(\phi_1(w_1),\phi_2(w_2,w_3))- \mu'_2(\phi_1(w_2),\phi_2(w_1,w_3))\\
&\phantom{{}={}}+\mu'_2(\phi_1(w_3),\phi_2(w_1,w_2))~,
\end{aligned}
\end{equation}
where $w_i$ and $v_i$ denote elements of $\frg$ of degrees $0$ and $-1$, respectively.

We note that a morphism of $L_\infty$-algebras is invertible and thus an isomorphism of $L_\infty$-algebras if and only if $\phi_1$ is invertible. This is very clear in the above explicit formulas~\eqref{eq:Lie_2_algebra_morph} for a Lie 2-algebra morphism. Note also that an $L_\infty$-algebra isomorphism preserves the dimensions of the graded subspaces $\frg_k$ of its source $L_\infty$-algebra $\frg=\oplus_{k\in \RZ}\frg_k$. In most cases, this notion of isomorphism is too restrictive, and we shall return to this point in section~\ref{ssec:quasi_isos}.

For more details on the three descriptions of $L_\infty$-algebras in terms of higher products, differential graded coalgebras and differential graded algebras, see e.g.~\cite[Appendix A]{Jurco:2018sby}.

\subsection{Weil algebra and free algebra}\label{ssec:Weil_and_free}

Given an $L_\infty$-algebra $\frg$, it is natural to consider the corresponding $L_\infty$-algebra of inner derivations, as done e.g.~in~\cite{Sati:2008eg}. Its Chevalley--Eilenberg is known as the Weil algebra of $\frg$ and it will play a major role in our discussion.

\begin{definition}[\cite{MR0042426},\cite{Sati:2008eg}] 
 The \underline{Weil algebra} of an $L_\infty$-algebra $\frg$ is the differential graded commutative algebra
 \begin{subequations}
 \begin{equation}
  \sW(\frg)~\coloneqq~\big(\odot^\bullet(\frg[1]^*\oplus \frg[2]^*)\,,\,Q_\sW\,\big)
 \end{equation}
with the differential $Q_\sW$ defined by
 \begin{equation}
  Q_\sW|_{\frg[1]^*}\coloneqq Q_\sCE+\sigma\eand Q_\sW|_{\frg[2]^*}\coloneqq -\sigma Q_\sCE \sigma^{-1}~,
 \end{equation}
 \end{subequations}
 where $Q_\sCE$ is the Chevalley--Eilenberg differential on $\frg[1]^*$ and $\sigma:\frg[1]^*\rightarrow \frg[2]^*$ is the shift isomorphism of degree~1. Note that indeed $Q_\sW^2=0$. We denote the dual $L_\infty$-algebra by $\inn(\frg)$, that is $\sW(\frg)\eqqcolon\sCE(\inn(\frg))$.
\end{definition}

\noindent The natural embedding $i:\frg\embd \inn(\frg)$ is an $L_\infty$-algebra morphism, as one readily checks. That is, its dual yields the projection
\begin{equation}
 i^*:\sW(\frg)\twoheadrightarrow\sCE(\frg)~,
\end{equation}
which is a morphism of dgas because it satisfies $Q_{\sCE} i^*=i^*Q_{\sW}$. The kernel of $i^*$ is the ideal in $\sW(\frg)$ generated by $\frg[2]^*$. Moreover, we have an isomorphism $\sCE(\frg)\cong \sW(\frg)/\ker(i^*)$.

It is now useful to introduce the subalgebra 
\begin{equation}
 \sW_{\rm h}(\frg)\coloneqq \odot^\bullet\frg[2]^*
\end{equation}
of {\em horizontal} elements in the Weil algebra. Note that $Q_\sW$ does not necessarily close on $\sW_{\rm h}(\frg)$ and that $\sW_{\rm h}(\frg)$ is in the kernel of $i^*$.

As examples, we construct the Weil algebras of a generic Lie algebra and a generic Lie 2-algebra. Let $\frg$ be an ordinary, finite-dimensional Lie algebra $\frg$ and let $t^\alpha\in \frg[1]^*$, $\alpha=1,\dots,d$, be coordinate functions on $\frg[1]$, which are of degree~$1$. We also introduce the coordinate functions $\hat t^\alpha=\sigma t^\alpha\in\frg[2]^*$ on $\frg[2]$, which are of degree~$2$. The Weil algebra $\sW(\frg)$ is then the polynomial algebra generated by $t^\alpha$ and $\hat t^\alpha$, and the Weil differential acts as
\begin{equation}\label{eq:Weil_of_ordinary_Lie}
Q_\sW~:~t^\alpha \mapsto -\tfrac12 f^\alpha_{\beta\gamma} t^\beta t^\gamma + \hat t^\alpha \eand
\hat t^\alpha\mapsto -f^\alpha_{\beta\gamma} t^\beta \hat t^\gamma~,
\end{equation}
where $f^\alpha_{\beta\gamma}$ are again the structure constants of $\frg$.

For the case of a Lie 2-algebra $\frg=(\frg_{-1}\rightarrow \frg_0)$, recall the generators and the form of the Chevalley--Eilenberg algebra $\sCE(\frg)$ from~\eqref{eq:CE_Lie_2_algebra}. We introduce additional shifted generators $\hat t^\alpha=\sigma t^\alpha$ and $\hat r^a=\sigma r^a$, and the Weil differential acts as
\begin{equation}\label{eq:Weil_algebra_generic_Lie_2}
 \begin{aligned}
   Q_\sW~&:~&t^\alpha &\mapsto -\tfrac12 f^\alpha_{\beta\gamma} t^\beta t^\gamma-f^\alpha_a r^a+\hat t^\alpha~,\\
   &&\hat t^\alpha&\mapsto -f^\alpha_{\beta\gamma}t^\beta \hat t^\gamma+f^\alpha_a \hat r^a~,\\
   &&r^a&\mapsto\tfrac1{3!}f^a_{\alpha\beta\gamma}t^\alpha t^\beta t^\gamma-f^a_{\alpha b}t^\alpha r^b+\hat r^a~,\\
   &&\hat r^a&\mapsto -\tfrac12 f^a_{\alpha\beta\gamma}t^\alpha t^\beta \hat t^\gamma+f^a_{\alpha b}\hat t^\alpha r^b-f^a_{\alpha b}t^\alpha \hat r^b
 \end{aligned}
\end{equation}
with the same structure constants $f^\alpha_a, f^\alpha_{\beta\gamma},f^a_{\alpha b}, f^a_{\alpha\beta\gamma}\in \FR$ as appearing in~\eqref{eq:CE_Lie_2_algebra}. The relation $Q_\sW^2=0$ follows by construction.

Note that a morphism between the Chevalley--Eilenberg algebras of two $L_\infty$-algebras $\frg$ and $\tilde \frg$ readily lifts to a morphism between their Weil algebras. In particular, a morphism $\Phi:\sCE(\frg)\to\sCE(\tilde \frg)$ can be lifted to a morphism $\hat{\Phi}:\sW(\frg)\to\sW(\tilde \frg)$ using  $\sigma a \mapsto \sigma \Phi(a)$ for generators $a$ of $\sCE(\frg)$, because the following diagrams commute:
\begin{equation}\label{eq:lift_CE_map_to_W}
\begin{tikzcd}
a \arrow[d,"\hat\Phi"]\arrow[r,"Q_\sW"] & Q_\sCE a + \sigma a\arrow[d,"\hat\Phi"] & \sigma a\arrow[d,"\hat\Phi"]\arrow[r,"Q_\sW"] & -\sigma Q_\sCE a\arrow[d,"\hat\Phi"] \\
\Phi(a)\arrow[r,"Q_\sW"] & Q_{\sCE} \Phi(a) + \sigma\Phi(a) &\sigma \Phi(a)\arrow[r,"Q_{\sW}"] &-\sigma Q_{\sCE} \Phi(a)
\end{tikzcd}
\end{equation}

Closely related to the Weil algebra is the {\em free algebra}\footnote{These are also called {\em free differential algebra} in the supergravity literature, see~\cite{Castellani:1995gz} and references therein.} $\sF(\frg)$ of an $L_\infty$-algebra $\frg$, which is given by
\begin{equation}
 \sF(\frg)\coloneqq \big( \odot^\bullet(\frg[1]^*\oplus\frg[2]^*)\,,\,Q_\sF = \sigma\,\big)~,
\end{equation}
where $\sigma:\frg[1]^*\to\frg[2]^*$ is again the shift isomorphism. In fact, the Weil algebra $\sW(\frg)$ is naturally isomorphic to the corresponding free algebra $\sF(\frg)$, as we have the isomorphisms
\begin{equation}
\begin{aligned}
\Upsilon: \sF(\frg)\to\sW(\frg),\quad &a\mapsto a~,~~~~~&\Upsilon^{-1}:\sW(\frg)\to\sF(\frg),\quad & a \mapsto a~,\\
& \hat a \mapsto Q_\sW a~,~~~~~&&\hat a \mapsto \hat a - Q_\sCE a~,
\end{aligned}
\end{equation}
where $a\in \frg[1]^*$ and $\hat a\coloneqq \sigma a\in\frg[2]^*$, with $\Upsilon^{-1}\circ \Upsilon=\id_{\sF(\frg)}$ and $\Upsilon\circ \Upsilon^{-1}=\id_{\sW(\frg)}$. Note that these maps are indeed dga-morphisms, because the following diagrams commute: 
\begin{subequations}
\begin{equation}\label{eq:weil_free_iso}
\begin{tikzcd}
a \arrow[r,"Q_\sF"]\arrow[d,"\Upsilon"] & \hat a\arrow[d,"\Upsilon"] & \hat a \arrow[d,"\Upsilon"]\arrow[r,"Q_\sF"] & 0 \arrow[d,"\Upsilon"] \\
a \arrow[r,"Q_\sW"] & Q_\sW a & Q_\sW a\arrow[r,"Q_\sW"] & 0
\end{tikzcd}
\end{equation}
and
\begin{equation}
\begin{tikzcd}
a \arrow[r,"Q_\sW"]\arrow[d,"\Upsilon^{-1}"] & Q_\sW a \arrow[d,"\Upsilon^{-1}"] & \hat a \arrow[r,"Q_\sW"]\arrow[d,"\Upsilon^{-1}"] & -\sigma Q_\sCE a\arrow[d,"\Upsilon^{-1}"] \\
a \arrow[r,"Q_\sF"] & Q_\sCE a + \hat a - Q_\sCE a & \hat a -Q_\sCE a\arrow[r,"Q_\sF"] & -\sigma Q_\sCE a
\end{tikzcd}
\end{equation}
\end{subequations}

\subsection{Quasi-isomorphisms and 2-morphisms}\label{ssec:quasi_isos}

As indicated above, it turns out that in most cases, the appropriate notion of equivalences for $L_\infty$-algebras is {\em not} a bijective $L_\infty$-algebra morphism, but a generalization known as a  quasi-isomorphism. In the higher product picture, we readily extend the corresponding definition from cochain complexes:
\begin{definition}
 An $L_\infty$-algebra \uline{quasi-isomorphism} $\phi:\frg\rightarrow \frh$ is a morphism of $L_\infty$-algebras, $\phi:\frg\rightarrow \frh$, which induces an isomorphism on cohomology\footnote{Recall that $\phi_1$ is a chain map and therefore descends to cohomology.},
 \begin{equation}
  \phi_1: H^\bullet_{\mu_1}(\frg)\xrightarrow{~\cong~}H^\bullet_{\mu_1}(\frh)~.
 \end{equation}
Two $L_\infty$-algebras $\frg$ and $\frh$ are \underline{quasi-isomorphic}, if there exists a quasi-isomorphism between them and we write $\frg\approxeq\frh$.
\end{definition}
\noindent It is clear that quasi-isomorphisms form an equivalence relation. In particular, they are transitive by definition: morphisms of $L_\infty$-algebras $\phi:\frg \rightarrow \frh$ and $\psi:\frh \rightarrow \frl$ can be composed to a morphism $\psi\circ \phi: \frg\rightarrow \frl$, which descends to the composition of the isomorphisms on the cohomologies.

The definition of a quasi-isomorphisms can be reformulated as categorical equivalence, see e.g.~\cite{Baez:2003aa} for the example of 2-term $L_\infty$-algebras, and this picture is readily translated to the dga description of $L_\infty$-algebras:
\begin{proposition}[\cite{Sati:2008eg}]
 A quasi-isomorphism between $L_\infty$-algebras $\frg$ and $\frh$ is equivalent to a pair of dga-morphisms 
 \begin{subequations}
\begin{equation}
    \begin{tikzcd}
    \sCE(\frg) \arrow[r,bend left=30,"\Phi"{name=D}] &\sCE(\frh)\arrow[l,bend left=30,"\Psi"{name=U,below}]
    \end{tikzcd}
\end{equation}
with
\begin{equation}
\eta_{\Psi\circ\Phi}:\Psi\circ \Phi\xRightarrow{~\cong~} \id_{\sCE(\frg)}~,~~\eta_{\Phi\circ\Psi}:\Phi\circ \Psi\xRightarrow{~\cong~}\id_{\sCE(\frh)}~.
\end{equation}
 \end{subequations}
We call the collection $(\Phi,\Psi,\eta_{\Psi\circ\Phi},\eta_{\Phi\circ\Psi})$ a \underline{dual quasi-isomorphism} and say that $\sCE(\frg)$ and $\sCE(\frh)$ are \underline{dually quasi-isomorphic}.\footnote{Note that this nomenclature is important to distinguish from an ordinary quasi-isomorphism of the dgas $\sCE(\frg)$ and $\sCE(\frh)$, which induces an isomorphism on the $Q$-cohomologies, whereas the homotopy equivalence introduced here corresponds to the dual of a quasi-isomorphism between $\frh$ and $\frg$, which refers to $\mu_1$-cohomologies.} 
\end{proposition}

\noindent For this proposition to be meaningful, we clearly need a notion of 2-morphisms for dga-algebras. In the case of differential graded vector spaces, 2-morphisms are simply chain homotopies, but respecting the algebra product makes the definition slightly more involved. It is helpful to note that 2-morphisms between morphisms from free dgas into arbitrary dgas are again straightforward to define. Also, given $L_\infty$-algebras $\frg$ and $\frh$, together with dga-morphisms $\Phi:\sCE(\frg)\rightarrow \sCE(\frh)$ and $\Psi:\sCE(\frg)\rightarrow \sCE(\frh)$, a 2-morphism $\eta$ between $\Phi$ and $\Psi$,
 \begin{equation}\label{eq:2-morph_1}
    \begin{tikzcd}
    \sCE(\frh) &\sCE(\frg)\arrow[l,bend left=50,"\Psi"{name=U,below}]\arrow[l,bend right=50,"\Phi"{name=D},swap]
    \arrow[Rightarrow,"\eta", from=D, to=U,start anchor={[yshift=-1ex]},end anchor={[yshift=1ex]}]
    \end{tikzcd}~,
 \end{equation}
can then be extended to a 2-morphism between morphisms from $\sF(\frg)$ to $\sCE(\frh)$ as follows~\cite{Sati:2008eg}:
\begin{equation}
\begin{tikzcd}
&\sCE(\frg)\arrow[dl,bend right=30,"\Phi",swap]&&\frg[2]^*\\
\sCE(\frh) &&\sW(\frg)\arrow[ur,hookleftarrow]\arrow[ul,"i^*",swap,""{name=U,below}]\arrow[dl,"i^*"]&\sF(\frg)\arrow[l,swap,"\Upsilon"]\\
&\sCE(\frg)\arrow[ul,bend left=30,"\Psi"{name=D}]&& \arrow[Rightarrow,"\eta",from=U,to=D,start anchor={[xshift=-1ex]},end anchor={[yshift=1ex,xshift=1ex]}]
\end{tikzcd}
\end{equation}
where $\eta$ should vanish on $\frg[2]^*\hookrightarrow\sW(\frg)$ in order to compensate for the ambiguities arising in the extension from $\sCE(\frg)$ to $\sW(\frg)$. For convenience, let us also introduce the following pullbacks to $\sF$:
\begin{equation}
 \Phi_\sF\coloneqq \Phi\circ i^*\circ \Upsilon\eand \Psi_\sF\coloneqq \Psi\circ i^*\circ \Upsilon~.
\end{equation}
These considerations lead to the following definition.

\begin{definition}[\cite{Sati:2008eg}]\label{def:2-morphism}
 A \underline{2-morphism} $\eta$ from $\Phi$ to $\Psi$ as in~\eqref{eq:2-morph_1} is given by a linear map $\eta$ of degree~$-1$ on the generators of the free algebra $\sF(\frg)$,
 \begin{subequations}
    \begin{equation}
    \eta: \frg[1]^*\oplus\frg[2]^*\to \sCE(\frh)~,
    \end{equation}
which is continued to all of $\sF(\frg)$ by the formula
 \begin{equation}
\begin{aligned}\label{eq:expansion_formula}
&\eta: a_1 \dots   a_n\mapsto\tfrac{1}{n!} \sum\limits_{\sigma \in S_n} \eps(\sigma)(a_1,\dots,a_n) \times\\
&\hspace{3cm}
\sum\limits_{k=1}^n (-1)^{\sum\limits_{i=1}^{k-1}\left|a_{\sigma(i)}\right|} \Phi_\sF(a_{\sigma(1)}  \dots  a_{\sigma(k-1)}) \eta(a_{\sigma(k)})  \Psi_\sF(a_{\sigma(k+1)} \dots  a_{\sigma(n)})
\end{aligned}
\end{equation}
 \end{subequations}
for $a_i \in \frg[1]^*\oplus\frg[2]^*$ to a chain homotopy on $\sF(\frg)$, 
 \begin{equation}\label{eq:2_morph_cond}
  \Phi_\sF-\Psi_\sF\coloneqq \Phi\circ i^*\circ \Upsilon-\Psi\circ i^*\circ \Upsilon=[Q,\eta]=Q_{\sCE}\circ\eta+\eta\circ Q_\sF~,
 \end{equation}
 and which becomes trivial when restricted to the generators $\Upsilon^{-1}(\frg[2]^*)$ of $\sW_{\rm h}(\frg)$. Here, $\eps(\sigma;a_1,\dots,a_n)$ is the symmetric Koszul sign of the permutation $\sigma$ of $a_1,\dots,a_n$. 
\end{definition}

A few remarks on this definition are in order. First, we note that it suffices to ensure condition~\eqref{eq:2_morph_cond} on the generators of $\sF(\frg)$ as the continuation~\eqref{eq:expansion_formula} then extends this property to all of $\sF(\frg)$. Second, the triviality upon restriction to $\Upsilon^{-1}(\frg[2]^*)$ implies that $\eta: \sF(\frg)\to \sCE(\frh)$ induces a map $\eta_\sW:\sW(\frg)\to \sCE(\frh)$ which can be defined by its image of the generators $a\in \frg[1]^*$. The fact that $\eta_\sW$ vanishes on all $\sigma a \in \sW(\frg)$ then fixes its image of $Q_\sW a$ inside $\sW(\frg)$ and on $\sigma_\sF a$ inside $\sF(\frg)$. In particular, we have
\begin{equation}\label{eq:restiction_eta}
 \eta(\sigma_\sF a)=\eta(Q_\sCE a)
\end{equation}
on generators $a\in \sF(\frg)$. Third, for $\eta_\sW=\eta\circ \Upsilon^{-1}$ we have
\begin{equation}
(\Phi_\sW-\Psi_\sW)(a)\coloneqq (\Phi\circ i^*-\Psi\circ i^*)(a)=(Q_\sCE\circ\eta_\sW+\eta_\sW\circ Q_\sW)(a)
\end{equation}
on the generators $a$ of $\sW(\frg)$ since $\Upsilon$ is a dga-isomorphism. Very importantly, however, the continuation formula~\eqref{eq:expansion_formula} does {\em not} extend to all of $\sW(\frg)$: since $\Upsilon^{-1}(\hat a)$ for $a\in\frg[1]^*$ is not necessarily a homogeneous polynomial in the generators, the continuation formula does {\em not} have a simple analogue on $\sW(\frg)$. Fourth, let us stress that definition~\ref{def:2-morphism} naturally extends to 2-morphisms between morphisms between Weil algebras as $\sW(\frg)$ can be seen as the Chevalley--Eilenberg algebra $\sCE(\inn(\frg))$. Fifth, 2-morphisms can be composed horizontally and vertically, and details are presented in appendix~\ref{app:A:concatenation}, where also the composition of quasi-isomorphisms is discussed.

It is instructive to spell out what this definition means in the example of morphisms between Lie 2-algebras $\frg=\frg_{-1}\oplus \frg_0$ and $\tilde \frg=\tilde \frg_{-1}\oplus \tilde \frg_0$. Recall our choice of generators $(t^\alpha,r^a)$ and the action of the Chevalley--Eilenberg differential $Q$ from~\eqref{eq:CE_Lie_2_algebra}. We introduce analogous generators $(\tilde t^\mu,\tilde r^m)$ and a differential $\tilde Q$ encoded in structure constants $\tilde f^\mu_m, \tilde f^\mu_{\nu\kappa}, \tilde f^m_{\mu n}$ and $\tilde f^m_{\mu\nu\kappa}$ for $\tilde \frg$. The morphisms $\Phi$ and $\Psi$ are defined by their images of the generators of $\frg[1]^*$:
\begin{equation}
\begin{aligned}
 \Phi~&:~t^\alpha\mapsto \Phi^\alpha_\mu t^\mu~,~~~r^a\mapsto\Phi^a_m\tilde r^m+\tfrac12 \Phi^a_{\mu\nu}\tilde t^\mu\tilde t^\nu~,\\
  \Psi~&:~t^\alpha\mapsto \Psi^\alpha_\mu t^\mu~,~~~r^a\mapsto\Psi^a_m\tilde r^m+\tfrac12 \Psi^a_{\mu\nu}\tilde t^\mu\tilde t^\nu~.
\end{aligned}
\end{equation}

To fix the 2-morphism, we note that a generic map $\eta:\frg[1]^*\oplus \frg[2]^*\rightarrow \sCE(\tilde \frg)$ of degree~$-1$ has the images
\begin{equation}
\eta~:~t^\alpha\mapsto 0~,~~~
r^a\mapsto \eta^a_\mu \tilde t^\mu~,
\end{equation}
which implies that the map $\eta_\sW=\eta\circ \Upsilon^{-1}$, taking generators of $\sW(\frg)$ to $\sCE(\tilde \frg)$, satisfies
\begin{equation}
\eta_\sW~:~t^\alpha\mapsto 0~,~~~
r^a\mapsto \eta^a_\mu \tilde t^\mu~.
\end{equation}
The requirement that $\eta_\sW$ vanishes along $\frg[2]^*\subset\sW(\frg)$ together with the formula~\eqref{eq:expansion_formula} then also defines $\eta_\sW$ on $Q_\sW t^\alpha$ and $Q_\sW r^a$, which we use to calculate
\begin{equation}
\begin{aligned}
[Q,\eta] t^\alpha &= \tilde Q_\sCE(\eta_\sW(t^\alpha))-\eta_\sW(Q_\sW t^\alpha)= f^\alpha_a\eta^a_\mu \, \tilde t^{\mu}~,\\
[Q,\eta] r^a &= \tilde Q_\sCE(\eta_\sW(r^a))-\eta_\sW( Q_\sW r^a)\\
&=-\eta^a_\mu \tilde f^\mu_m\tilde r^m-\tfrac12 \eta^a_\mu\tilde f^\mu_{\nu\kappa}\tilde t^\nu\tilde t^\kappa+\tfrac12 f^a_{\alpha b}(\Psi^\alpha_\mu \eta^b_\nu+\eta^b_\mu\Psi^\alpha_\nu)\tilde t^\mu \tilde t^\nu~.
\end{aligned}
\end{equation}
The condition $\Psi_\sW-\Phi_\sW=[Q,\eta_\sW]$ then translates to
\begin{equation}
\begin{aligned}
\Phi^\alpha_\mu-\Psi^\alpha_\mu &= f^\alpha_a \eta^a_\mu~,\\
\Phi^a_m - \Psi^a_m &= - \eta^a_\mu \tilde f^\mu_m~,\\
\Phi^a_{\left[\mu\nu\right]}-\Psi^a_{\left[\mu\nu\right]} &= -\eta^a_\kappa \tilde f^\kappa_{\mu\nu}+f^a_{\alpha b} (\Psi^\alpha_\mu\eta^b_\nu+\eta_\mu^b\Psi^\alpha_\nu)~,
\end{aligned}
\end{equation}
and this agrees with the familiar condition for 2-morphisms as given in~\cite{Baez:2009:aa}, cf.\ also appendix~A of~\cite{Sati:2008eg}.

As an example of a quasi-isomorphism, let us show that the Weil algebra $\sW(\frg)$ of an $L_\infty$-algebra $\frg$ is quasi-isomorphic to the Weil algebra $\sW(*)$ of the trivial $L_\infty$-algebra. We have already shown that $\sW(\frg)$ is isomorphic to the free algebra $\sF(\frg)$, so it merely remains to show that $\sF(\frg)\approxeq\sW(*)$. The relevant morphisms are obvious,
\begin{equation}
 \begin{tikzcd}
    \sF(\frg)  \arrow[r,bend left=30]{}{\Phi} & \sW(\ast)  \arrow[l,bend left=30]{}{\Psi}
 \end{tikzcd}
\end{equation}
with 
\begin{equation}\label{eq:generic_weil_to_zero_morphism}
\Phi(-)=0 \eand \Psi:0 \mapsto 0~.
\end{equation}
Clearly, $\Phi\circ \Psi=\id_{\sW(\ast)}$, so it remains to find a 2-morphism $\eta:\Psi\circ\Phi\Rightarrow \id_{\sF(\frg)}$. There is only one generic choice, namely 
\begin{equation}\label{eq:generic_weil_contracting_homotopy}
 \eta(a)=\begin{cases}
       -\sigma_\sF^{-1}(a) &\mbox{ for generators } a\in \im(\sigma)~,\\
       0 &\mbox{ else}~.
      \end{cases}
\end{equation}
where $\sigma_\sF$ is the shift isomorphism in $\sF(\frg)$. We then have $[Q,\eta]_{\sF(\frg)}=-\id_{\sF(\frg)}=\Psi\circ \Phi-\id_{\sF(\frg)}$.

The map $\eta$ can now be used to show that the $Q$-cohomology of $\sF(\frg)$ is trivial: given an $\alpha\in \sF(\frg)$ with $Q\alpha=0$, we have $\alpha=-\id_{\sF(\frg)}(-\alpha)=[Q,\eta] (-\alpha)=Q(-\eta(\alpha))$ and therefore any $Q$-closed algebra element is $Q$-exact. The isomorphism $\Upsilon$ between $\sF(\frg)$ and $\sW(\frg)$ allows us to translate this argument to $\sW(\frg)$:
\begin{lemma}\label{lem:Q_cohomology_of_Weil_algebra_trivial}
 The $Q$-cohomology of the Weil algebra $\sW(\frg)$ of an $L_\infty$-algebra $\frg$ is trivial.
\end{lemma}
\noindent To prove this lemma, consider an $\alpha\in \sW(\frg)$ with $Q\alpha=0$. We then have $\beta=\Upsilon^{-1}(\alpha)\in \sF(\frg)$ which is exact and closed, i.e.~$\beta=Q_\sF (\eta(\beta))$. It follows that $Q_\sW\Upsilon(\eta(\beta))=\Upsilon(Q_\sF(\eta(\beta)))=\Upsilon(\beta)=\alpha$ and $\alpha$ is thus exact.

\subsection{Structural theorems for \texorpdfstring{$L_\infty$}{Linf}-algebras}

Let us briefly recall some important structural theorems for $L_\infty$-algebras which will simplify our discussion.
\begin{definition}\label{def:L_inf_algebra_types}
 Let $\frg$ be an $L_\infty$-algebra with higher products $\mu_i$, $i\in \NN^+$. We call $\frg$
 \begin{itemize}
  \setlength{\itemsep}{-1mm}
  \item[$\circ$] \uline{strict} if $\mu_i=0$ for $i\geq 3$ and $\frg$ is thus simply a differential graded Lie algebra;
  \item[$\circ$] \uline{minimal} if $\mu_1=0$;
  \item[$\circ$] \uline{linearly contractible} if $\mu_i=0$ for $i>1$ and $H^\bullet_{\mu_1}(\frg)=0$.
 \end{itemize}
\end{definition}
Fundamentally, we have the following theorem.
\begin{theorem}[Decomposition theorem, cf.~\cite{Kajiura:0306332}]
 Any $L_\infty$-algebra is isomorphic as an $L_\infty$-algebra to the direct sum of a minimal and a linearly contractible $L_\infty$-algebra.
\end{theorem}
Applying a projection to the minimal part of an $L_\infty$-algebra (which evidently induces an isomorphism on the $\mu_1$-cohomology), we immediately arrive at the following theorem, which historically predates the decomposition theorem:
\begin{theorem}[cf.~\cite{kadeishvili1982algebraic,Kajiura:0306332}]
 Any $L_\infty$-algebra is quasi-isomorphic to a minimal $L_\infty$-algebra.
\end{theorem}
We can thus endow the cohomology $H^\bullet_{\mu_1}(\frg)$ of an $L_\infty$-algebra $\frg$ with an $L_\infty$-algebra structure such that it is quasi-isomorphic to $\frg$ itself. The resulting $L_\infty$-algebra is minimal in the sense that it is a dimensionally smallest representative of the quasi-isomorphism class of $\frg$. It is therefore called a {\em minimal model} of $\frg$.

Finally, we have another extreme case, relating $L_\infty$-algebras to differential graded Lie algebras:
\begin{theorem}[\cite{igor1995,Berger:0512576}]
 Any $L_\infty$-algebra is quasi-isomorphic to a strict $L_\infty$-algebra.
\end{theorem}

Let us discuss the example of a Lie 2-algebra $\frg=\frg_{-1}\oplus \frg_0$ in more detail. First, we note that there is an exact sequence
\begin{equation}\label{eq:es_Lie2}
 0\longrightarrow\ker(\mu_1) \hooklongrightarrow \frg_{-1}\xrightarrow{~\mu_1~} \frg_0 \xrightarrow{~\pi~} {\rm coker}(\mu_1)\longrightarrow 0~,
\end{equation}
where ${\rm coker}(\mu_1)$ carries a Lie algebra structure induced by $\mu_2$. A minimal model $\frg^\circ$ of $\frg$ has underlying graded vector space
\begin{equation}
 \frg^\circ=\frg^\circ_{-1}\oplus \frg^\circ_0\cong\ker(\mu_1)\oplus {\rm coker}(\mu_1)~.
\end{equation}
Using the decomposition theorem, we can further decompose $\frg$ (non-canonically) according to
\begin{equation}\label{eq:LAsplitting}
 \frg~=~\frg_{-1}\oplus \frg_0~=\left(
\begin{tikzcd}[row sep=2pt,column sep=1.7cm]
\frg_{-1}^0=\ker(\mu_1) & \frg^0_0\cong {\rm coker}(\mu_1)  \\
\oplus & \oplus\\
\frg_{-1}^1\cong \im(\mu_1)\arrow[r,"\mu_1=\text{id}"] & \frg^1_0=\im(\mu_1)
\end{tikzcd}
 \right)
\end{equation}
with the only non-trivial higher products being 
\begin{equation}
 \mu_1:\frg^1_{-1}\rightarrow \frg^1_0~,~~~\mu_2:\frg^0_0\wedge \frg^0_0\rightarrow \frg^0_0~,~~~\mu_2:\frg^0_0\wedge \frg^0_{-1}\rightarrow \frg^0_{-1}~,~~~\mu_3:\wedge^3 \frg^0_0\rightarrow \frg^0_{-1}~.
\end{equation}
In particular, $\frg^0_0$ is a Lie algebra, $\frg^0_{-1}$ is a $\frg^0_0$-module with action induced by $\mu_2$ and $\mu_3$ is an element of the Lie algebra cohomology group $H^3(\frg^0_0,\frg^0_{-1})$.

\subsection{String Lie 2-algebra models}

In the vast category of $L_\infty$-algebras, there are particularly interesting objects which are obtained by extending metric Lie algebras by particular cocycles. As will become clear, it is these $L_\infty$-algebras that underlie truly non-abelian higher gauge theories. The simplest non-trivial one which is relevant to the application in string theory is the string Lie 2-algebra. In the following, we discuss the relevant algebraic structures, giving a minimal and a strict model.

The string group $\sString(n)$ sits in the sequence
\begin{equation}\label{eq:whitehead_tower}
\begin{tikzcd}[column sep=20pt]
\dots \arrow[r] & \sString(n)\arrow[r] & \sSpin(n)\arrow[r] &\sSpin(n)\arrow[r] & \sSO(n)\arrow[r] & \sO(n)~,
\end{tikzcd}
\end{equation}
which is known as the {\em Whitehead tower} of $\sO(n)$. It is constructed by successively removing the lowest homotopy group: $\pi_0(\sO(n))$ is removed in the step from $\sO(n)$ to $\sSO(n)$, $\pi_1(\sO(n))$ in the step to $\sSpin(n)$ and $\pi_2(\sO(n))$ is already trivial. The string group $\sString(n)$ is obtained by removing $\pi_3(\sO(n))$. That is, $\sString(n)$ is a 3-connected cover of $\sSpin(n)$~\cite{Stolz:1996:785-800}. This definition only determines $\sString(n)$ up to homotopical equivalence and consequently, there are a variety of models.

Particularly interesting models are given by Lie 2-groups, and one example is that of~\cite{Schommer-Pries:0911.2483}. As shown in~\cite{Demessie:2016ieh}, this Lie 2-group can be differentiated to a minimal $L_\infty$-algebra which we call, following the categorical nomenclature, the skeletal model. This Lie 2-algebra allows for an immediate generalization to arbitrary metric Lie algebras, cf.~also~\cite{Baez:2003aa}:
\begin{definition}
 Let $\frg$ be a Lie algebra endowed with a metric $(-,-)$. The \uline{skeletal model of the string Lie 2-algebra} or, simply, the \uline{skeletal string algebra} of $\frg$ is the 2-term $L_\infty$-algebra 
 \begin{subequations}
 \begin{equation}
\astringsk(\frg) = \big(\,\,\FR[1] \overset{0}{\longrightarrow} \frg\,\,\big)
\end{equation}
with non-trivial higher products
\begin{equation}\label{eq:skel_string_algebra_brackets}
\begin{aligned}
 \mu_2&:\frg \wedge \frg\rightarrow \frg~,~~~&\mu_2(a_1,a_2)&=[a_1,a_2]~,\\
 \mu_3&:\frg \wedge \frg \wedge \frg\rightarrow \FR~,~~~&\mu_3(a_1,a_2,a_3)&=(a_1,[a_2,a_3])~,
\end{aligned}
\end{equation}
where $[-,-]$ is the commutator in $\frg$.
\end{subequations}
\end{definition}

The Weil algebra $\sW(\astringsk(\frg))$ is generated by coordinate functions $t^\alpha, r$ of degrees~1 and~2, respectively, together with their shifted copies $\hat t^\alpha=\sigma t^\alpha$ and $\hat r=\sigma r$ of degrees~2 and~3. The differential corresponding to~\eqref{eq:skel_string_algebra_brackets} is then
\begin{equation}\label{eq:skel_string_algebra_differential}
\begin{aligned}
Q~&:~&t^\alpha &\mapsto -\tfrac12 f^{\alpha}_{\beta\gamma} t^\beta  t^\gamma + \hat t^\alpha~,~~~&r &\mapsto \tfrac1{3!} f_{\alpha\beta\gamma} t^\alpha  t^\beta   t^\gamma + \hat r~,\\
&&\hat t^\alpha &\mapsto -f^\alpha_{\beta\gamma} t^\beta   \hat t^\gamma~,~~~&\hat r &\mapsto -\tfrac12 f_{\alpha\beta\gamma} t^\alpha  t^\beta   \hat t^\gamma
\end{aligned}
\end{equation}
with $f^\alpha_{\beta\gamma}$ being the structure constants of $\frg$ and $f_{\alpha\beta\gamma}\coloneqq \kappa_{\alpha\delta}f^\delta_{\beta\gamma}$, where the components $\kappa_{\alpha\beta}$ encode the metric.

\

The string 2-group model of~\cite{Schommer-Pries:0911.2483} is rather complicated and historically, a strict model of the string 2-group therefore came first~\cite{Baez:2005sn}, which is readily obtained by integrating the following strict Lie 2-algebra model:
\begin{definition}[\cite{Baez:2005sn}]
  Let $\frg$ be again a Lie algebra endowed with a metric $(-,-)$. The \uline{loop algebra model of the string Lie 2-algebra} or, simply, the \uline{loop string algebra} is the 2-term $L_\infty$-algebra 
  \begin{subequations}
    \begin{equation}
    \astringl(\frg) = \big(\,\,\hat L_0 \frg[1]\xrightarrow{~~\mu_1~~}P_0\frg\,\,\big)\ewith \hat L_0 \frg\coloneqq L_0\frg\oplus\FR~~,
    \end{equation}
    where $P_0\frg$ and $L_0\frg$ are the spaces of based paths and loops\footnote{Our based loops are those of~\cite{Baez:2005sn} and slightly differ from the canonical definition. A based loop $\lambda\in \sL_0 \frg$ is a smooth function $\lambda:[0,1]\rightarrow \frg$ such that $\lambda(0)=\lambda(1)=0\in\frg$. In other words, they are based paths with endpoint $0\in\frg$ and we have the short exact sequence $0\rightarrow L_0\frg\embd P_0\frg \xrightarrow{\dpar} \frg \rightarrow 0$, where $\dpar$ is the endpoint evaluation. Also, composability of our paths and loops requires them to be lazy in the sense that they are constant in a neighborhood of~0 and~1. We will suppress all the technicalities related to these ``sitting instances'' etc.} in $\frg$, respectively. The non-trivial higher products are
    \begin{equation}\label{eq:loop_string_algebra_brackets}
    \begin{aligned}
    \mu_1&:\hat L_0 \frg[1] \rightarrow P_0\frg~,~~~&\mu_1\big(\lambda,r\big)&= \lambda~,\\
    \mu_2&:P_0\frg\wedge  P_0\frg\rightarrow P_0\frg~,~~~&\mu_2(\gamma_1,\gamma_2)&=[\gamma_1,\gamma_2]~,\\
    \mu_2&:P_0\frg\otimes\hat L_0 \frg[1]\rightarrow \hat L_0 \frg[1]~,~~~&\mu_2\big(\gamma,(\lambda,r)\big)&=\left([\gamma,\lambda]\; ,\; -2\int_0^1 \dd\tau \left(\gamma(\tau),\dot \lambda(\tau)\right)\right)~,
    \end{aligned}
    \end{equation}
  \end{subequations}
where here and in the following, a dot denotes the obvious derivative with respect to the loop parameter $\tau$.
\end{definition}

Note that the homogeneously graded subspace $\hat L_0 \frg\coloneqq L_0\frg\oplus\FR$ is the Lie algebra of the Kac--Moody central extension of $L_0 \sG$ for $\sG$ a Lie group integrating $\frg$.

To construct the Weil algebra $\sW(\astringl(\frg))$ we now have to somehow dualize the infinite-dimensional mapping spaces $P_0\frg$ and $L_0 \frg$. We can do this pointwise for each value of the path and loop parameters and thus introduce coordinate functions $t^{\alpha\tau}$ and $(r^{\alpha\tau},r_0)$ of degrees~1~and~2, respectively. The shifted copies are again denoted by $\hat{t}^{\alpha\tau}$ and $(\hat{r}^{\alpha\tau},\hat r_0)$. The differential corresponding to~\eqref{eq:loop_string_algebra_brackets} is then given by its action on the coordinate functions,
\begin{equation}\label{eq:Weil_string_Lie_2_loop}
\begin{aligned}
Q~&:~&t^{\alpha\tau} &\mapsto -\tfrac12 f^{\alpha}_{\beta\gamma} t^{\beta\tau} t^{\gamma\tau} - r^{\alpha\tau} +\hat{t}^{\alpha\tau}~,~~~ 
  &\hat{t}^{\alpha\tau}&\mapsto -f^\alpha_{\beta\gamma}t^{\beta\tau}\hat{t}^{\gamma\tau}+\hat{r}^{\alpha\tau}~,\\
&&r^{\alpha\tau} &\mapsto -f^\alpha_{\beta\gamma} t^{\beta\tau}r^{\gamma\tau} + \hat{r}^{\alpha\tau}~,~~~
  &\hat{r}^{\alpha\tau}&\mapsto-f^\alpha_{\beta\gamma} t^{\beta\tau}\hat{r}^{\gamma\tau} + f^\alpha_{\beta\gamma} \hat{t}^{\beta\tau}r^{\gamma\tau} ~,\\
&&r_0 &\mapsto 2\int_0^1\dd \tau\, \kappa_{\alpha\beta}t^{\alpha\tau} \dot{r}^{\beta\tau}+\hat{r}_0 ~,~~~&\hat{r}_0&\mapsto 2\int_0^1\dd\tau\,\kappa_{\alpha\beta}\left(t^{\alpha\tau} \smash{\dot{\hat{r}}}^{\beta\tau}-\hat{t}^{\alpha\tau}\dot{r}^{\beta\tau}\right)~,
\end{aligned}
\end{equation}
where $f^\alpha_{\beta\gamma}$ are again the structure constants of $\frg$.

A quasi-isomorphism between $\astringl(\frg)$ and $\astringsk(\frg)$ is readily found, cf.~also~\cite{Baez:2005sn}. We have a morphism of $L_\infty$-algebras $\psi:\astringsk(\frg)\rightarrow\astringl(\frg)$ given by the chain map
\begin{subequations}
\begin{equation}
\begin{tikzcd}[column sep=1.5cm]
\FR[1] \arrow[r,"\psi_1",hook]\arrow[d,"0"] & L_0\frg[1]\oplus\FR[1] \arrow[d,"\mu_1"] \\
\frg \arrow[r,"\psi_1\,=\,\cdot \ell(\tau)",hook] & P_0\frg 
\end{tikzcd}
\end{equation}
as well as 
\begin{equation}
 \psi_2(a_1,a_2) = \big(\,[a_1,a_2](\ell(\tau)-\ell^2(\tau)),0\,\big)~.
\end{equation}
\end{subequations}
Here, $\cdot \ell(\tau):\frg\to P_0\frg$ is the embedding of $a_0\in\frg$ as the straight line $a(\tau)=a_0\ell(\tau)$, for some function
\begin{equation}\label{eq:f-function}
 \ell\in \CC^\infty([0,1])\ewith \ell(0)=0\eand \ell(1)=1~.
\end{equation}
This morphism induces an isomorphism on the cohomologies\footnote{Note that $\ker(\mu_1)=\FR$ and $\im(\mu_1)=L_0 \frg\subset P_0\frg$ in $\astringl(\frg)$ and that two paths with the same endpoint in $P_0\frg$ differ by a loop in $L_0 \frg$.}
\begin{equation}
 H^\bullet_{\mu_1}(\astringsk(\frg))=H^\bullet_{\mu_1}(\astringl(\frg))=(\FR[1] \longrightarrow \frg)~.
\end{equation}

A quasi-isomorphism can also readily be found. That is, we introduce a second morphism $\phi$, 
\begin{equation}
 \begin{tikzcd}
    \astringsk(\frg) \arrow[r,bend left=30]{}{\psi} & \astringl(\frg) \arrow[l,bend left=30]{}{\phi}
 \end{tikzcd}~,
\end{equation}
such that $\phi\circ \psi\cong \id_{\astringsk(\frg)}$ and $\psi\circ \phi\cong \id_{\astringl(\frg)}$. Explicitly, let $\phi$ be given by the chain map  
\begin{subequations}
 \begin{equation}\label{eq:equivalence_loop_skel_part1}
\begin{tikzcd}[column sep=1.5cm]
L_0\frg[1]\oplus\FR[1] \arrow[r,"\phi_1\,=\,\text{pr}_{\FR[1]}"] \arrow[d,"\mu_1"] & \FR[1]\arrow[d,"0"] \\
P_0\frg \arrow[r,"\phi_1\,=\,\partial"] & \frg
\end{tikzcd}
\end{equation}
together with 
\begin{equation}
 \phi_2(x_1,x_2) = \int_0^1\!\dd\tau\,(\dot{x}_1,x_2)-(x_1,\dot{x}_2)~,
\end{equation}
\end{subequations}
where $\text{pr}_\FR$ is the obvious projection and $\partial: P_0\frg\to\frg$ is the endpoint evaluation. We then have $\phi\circ \psi=\id_{\astringsk(\frg)}$ and a 2-morphism $\eta^*:\psi\circ \phi\Rightarrow\id_{\astringl(\frg)}$ in the sense of~\cite{Baez:2003aa}, which is encoded in the map 
\begin{equation}\label{eq:skeletal_to_loop_2_morphism}
\eta^*: P_0\frg\to L_0\frg[1]\oplus\FR[1]~,~~~\eta^*(\gamma)=\big(\,\gamma-\ell(\tau)\partial\gamma,0\,\big)~.
\end{equation}

In the dual dga-picture, we have morphisms of differential graded algebras,
\begin{subequations}\label{eq:quasi-iso-loop-skeletal-Lie2}
\begin{equation}
 \begin{tikzcd}
    \sCE(\astringsk(\frg)) \arrow[r,bend left=30]{}{\Phi} & \sCE(\astringl(\frg)) \arrow[l,bend left=30]{}{\Psi}
 \end{tikzcd}~,
\end{equation}
which act according to\footnote{The morphism $\Phi$ is dual to the sum $\phi_1+\tfrac12 \phi_2+\dots$ degree-shifted and extended as a map to the codifferential graded commutative coalgebra underlying the $L_\infty$-algebra under consideration, cf.~also the discussion in~\cite{Jurco:2018sby}.}
\begin{equation}
\begin{aligned}
 \Phi~&:~&t^{\alpha\tau}_{\rm lp}&\mapsto \ell(\tau)t^{\alpha}_{\rm sk}
    ~,~~~&r^{\alpha\tau}_{\rm lp}&\mapsto\tfrac12 f^\alpha_{\beta\gamma}t^\beta_{\rm sk}t^\gamma_{\rm sk}(\ell(\tau)-\ell^2(\tau))~,~~~&r_{0{\rm lp}}&\mapsto -r_{\rm sk}~,\\
 \Psi~&:~&t^\alpha_{\rm sk}&=t^{\alpha 1}_{\rm lp}~,~~~&r_{\rm sk}&\mapsto -r_{0{\rm lp}}-\int_0^1\! \dd \tau~ \tfrac12\kappa_{\alpha\beta}(\dot t^{\alpha\tau}_{\rm lp}t^{\beta \tau}_{\rm lp}-t^{\alpha\tau}_{\rm lp}\dot t^{\beta\tau}_{\rm lp})~,
\end{aligned} 
\end{equation}
where we added subscripts to distinguish the generators for the skeletal model and the loop model. We note that $\Phi\circ \Psi=\id_{\sCE(\astringsk(\frg))}$ and there is a 2-morphism
\begin{equation}
 \eta:\Psi\circ \Phi\Rightarrow \id_{\sCE(\astringl(\frg))}~,
\end{equation}
which is encoded in a map $\eta: \sF(\astringl(\frg))\rightarrow \sCE(\astringl(\frg))$ non-trivial only on the generators $r^{\alpha\tau}$,
\begin{equation}
 \eta(r^{\alpha\tau}_{\rm lp})=t^{\alpha\tau}_{\rm lp}-\ell(\tau)t^{\alpha 1}_{\rm lp}~.
\end{equation}
\end{subequations}
This is indeed the dual to $\eta^*$ from~\eqref{eq:skeletal_to_loop_2_morphism} and we have in particular
\begin{equation}
[Q,\eta]\coloneqq Q_{\sCE}\circ \eta+\eta\circ \sigma=\Psi\circ \Phi\circ i^*\circ \Upsilon-\id_{\sW(\astringl(\frg))}\circ i^*\circ \Upsilon~.
\end{equation}
Explicitly, the action of $[Q,\eta]$ on the various generators reads as 
\begin{subequations}
\begin{equation}
 \begin{aligned}
  (Q_\sCE\circ \eta+\eta\circ Q_\sW)(t^{\alpha\tau}_{\rm lp})
        &=\eta(Q_\sW t^{\alpha\tau}_{\rm lp})=-\eta(r^{\alpha\tau}_{\rm lp})\\
        &=\ell(\tau)t^{\alpha1}_{\rm lp}-t^{\alpha\tau}_{\rm lp}
        =(\Psi\circ \Phi-\id)(t^{\alpha\tau}_{\rm lp})~,
 \end{aligned}
\end{equation}
\begin{equation}
 \begin{aligned}
  (Q_\sCE\circ \eta+\eta\circ Q_\sW)(r^{\alpha\tau}_{\rm lp})
        &=Q_\sCE(t^{\alpha\tau}_{\rm lp}-\ell(\tau)t^{\alpha 1}_{\rm lp})+\eta(Q_\sW r^{\alpha\tau}_{\rm lp})\\
        &=-\tfrac12f^\alpha_{\beta\gamma}(t^{\beta\tau}_{\rm lp}t^{\gamma\tau}_{\rm lp}-\ell(\tau)t^{\beta1}_{\rm lp}t^{\gamma1}_{\rm lp})-r^{\alpha\tau}_{\rm lp}-\eta(f^\alpha_{\beta\gamma}t^{\beta\tau}_{\rm lp}r^{\gamma\tau}_{\rm lp})\\
        &=\tfrac12 f^\alpha_{\beta\gamma}t^{\beta1}_{\rm lp}t^{\gamma1}_{\rm lp}(\ell(\tau)-\ell^2(\tau))-r^{\alpha\tau}_{\rm lp}
        =(\Psi\circ \Phi-\id)(r^{\alpha\tau}_{\rm lp})~,
 \end{aligned}
\end{equation}
\begin{equation}
\begin{aligned}
(Q_\sCE\circ\eta+\eta\circ Q_\sW)(r_{0{\rm lp}}) &= \eta\left(2\int_0^1\!\dd\tau ~\kappa_{\alpha\beta} t^{\alpha\tau}_{\rm lp} \dot r^{\beta\tau}_{\rm lp}\right)\\
&= \int_0^1\!\dd\tau~\tfrac12 \kappa_{\alpha\beta}(\dot t^{\alpha\tau}_{\rm lp}t^{\beta\tau}_{\rm lp}-t^{\alpha\tau}_{\rm lp}\dot t^{\beta\tau}_{\rm lp})=(\Psi\circ\Phi-\id)(r_{0{\rm lp}})~,
\end{aligned}
\end{equation}
where we used
\begin{equation}
 \begin{aligned}
  \eta(f^\alpha_{\beta\gamma}t^{\beta\tau}_{\rm lp}r^{\gamma\tau}_{\rm lp})&=\tfrac12 f^\alpha_{\beta\gamma}\left(\eta(r^{\gamma\tau}_{\rm lp})t^{\beta\tau}_{\rm lp}-(\Psi\circ \Phi)(t^{\beta\tau}_{\rm lp})\eta(r^{\gamma\tau}_{\rm lp})\right)\\
  &=-\tfrac12 f^\alpha_{\beta\gamma}\left(t^{\beta\tau}_{\rm lp}(t^{\gamma\tau}_{\rm lp}-\ell(\tau)t^{\gamma 1}_{\rm lp})+\ell(\tau)t^{\beta1}_{\rm lp}(t^{\gamma\tau}_{\rm lp}-\ell(\tau)t^{\gamma 1}_{\rm lp})\right)\\
  &=-\tfrac12 f^\alpha_{\beta\gamma} t^{\beta\tau}_{\rm lp} t^{\gamma\tau}_{\rm lp}+\tfrac12 f^\alpha_{\beta\gamma} \ell^2(\tau)t^{\beta 1}_{\rm lp} t^{\gamma 1}_{\rm lp}~,
 \end{aligned}
\end{equation}
and
\begin{equation}
\begin{aligned}
\eta\left(2\int_0^1\!\dd\tau ~\kappa_{\alpha\beta} t^{\alpha\tau}_{\rm lp}\dot r^{\beta\tau}_{\rm lp}\right) &= \int_0^1\!\dd\tau ~\kappa_{\alpha\beta} \left(\eta(\dot r^{\beta\tau}_{\rm lp})t^{\alpha\tau}_{\rm lp} -(\Psi\circ\Phi)(t^{\alpha\tau}_{\rm lp})\eta(\dot r^{\beta\tau}_{\rm lp})\right)\\
&=-\int_0^1\!\dd\tau ~ \kappa_{\alpha\beta} \left(t^{\alpha\tau}_{\rm lp} \dot t^{\beta\tau}_{\rm lp} - t^{\alpha\tau}_{\rm lp} t^{\beta 1}_{\rm lp} + \ell(\tau) t^{\alpha 1}_{\rm lp} \dot t^{\beta\tau}_{\rm lp}\right)\\
&=-\int_0^1\!\dd\tau ~\kappa_{\alpha\beta} t^{\alpha\tau}_{\rm lp} \dot t^{\beta\tau}_{\rm lp}~.
\end{aligned}
\end{equation}
\end{subequations}

\noindent Altogether, we conclude that $\astringsk(\frg)$ and $\astringl(\frg)$ are quasi-isomorphic as 2-term $L_\infty$-algebras. They form two possible extreme models of the string Lie 2-algebra: a minimal and a strict one, cf.~definition~\ref{def:L_inf_algebra_types}. Note that the simplicity of the first comes at the price of a more involved integrated version, while the simple integrated version of the second comes with the issue of having to work with infinite-dimensional spaces.

Having these two extreme examples at hand is important because, as mentioned above and stated in~\cite{Saemann:2017rjm,Saemann:2017zpd,Saemann:2019leg}, we always want to ensure that the higher gauge theories we construct are agnostic about the explicit model of the gauge $L_\infty$-algebra used to define them.

\subsection{Extended skeletal model}\label{ssec:extended_skeletal}

The string Lie 2-algebra allows for a further description which will be very useful for our discussion later. Note that $\astring(\frg)$ is essentially an extensions of $\frg$ defined by trivializing the cocycle $\mu=\frac1{3!}f_{\alpha\beta\gamma}t^\alpha t^\beta t^\gamma$. This cocycle parametrizes a map $\sCE(\FR[2])\rightarrow \sCE(\frg)$ in which the single generator of $\sCE(\FR[2])$ is mapped to $\mu$. Dually, we have a morphism $\mu:\frg\rightarrow \FR[2]$, which gives rise to a Lie 3-algebra $\aghsk$ quasi-isomorphic to $\frg$, which fits into the short exact sequence
\begin{equation}\label{eq:astringsk_ses}
 0 \longrightarrow \astringsk(\frg) \hooklongrightarrow \aghsk \xrightarrow{~~~} \FR[2] \longrightarrow 0~,~~~\aghsk \approxeq \frg~,
\end{equation}
cf.~\cite[Prop.~20]{Sati:2008eg}. An element of $\astringsk(\frg)$ can be seen as an element of $\aghsk$ in the kernel of the projection onto $\FR[2]$.\footnote{A simple analogy is the short exact sequence $0\xrightarrow{~\phantom{\tr}~} \asu(n)\xrightarrow{~\phantom{\tr}~} \au(n)\xrightarrow{~\tr~} \au(1)\xrightarrow{~\phantom{\tr}~} 0$, which justifies identifying elements of $\asu(n)$ with traceless elements of $\au(n)$.} It turns out that this description is very important for the discussion of string structures, and we shall describe the maps involved in~\eqref{eq:astringsk_ses} in the following.

We note that any $n$-term $L_\infty$-algebra possess an extension to an $n+1$-term $L_\infty$-algebra by their left-most kernel of $\mu_1$, as explained in appendix~\ref{app:C}. In the case of $\astringsk(\frg)$, this leads to the Lie 3-algebra 
\begin{equation}
 \aghsk\coloneqq \big(~\FR_q\xrightarrow{~\id~} \FR_r \longrightarrow \frg_t~\big)\coloneqq \big(~\FR[2]\xrightarrow{~\id~} \FR[1] \longrightarrow \frg_t~\big)~\approxeq~\frg~,
\end{equation}
where the subscripts $q$, $r$, and $t$ help to identify the various subspaces of $\aghsk$, in particular to distinguish between the two grade-shifted copies of $\FR$, and to suppress the grade-shifts. The Chevalley--Eilenberg algebra of $\aghsk$ is generated by coordinate functions $t^\alpha$, $r$, and $q$ of degrees~1, 2, 3, respectively. The action of the differential on generators is, cf.~again appendix~\ref{app:C},
\begin{equation}\label{eq:extended_string_sk_differential}
\begin{aligned}
Q_\sCE~&:~&t^\alpha &\mapsto -\tfrac12 f^\alpha_{\beta\gamma} t^\beta  t^\gamma ~,~~~&r &\mapsto \tfrac{1}{3!} f_{\alpha\beta\gamma} t^\alpha  t^\beta  t^\gamma +q~,&q &\mapsto 0~,
\end{aligned}
\end{equation}
where $t^\alpha\in \frg^*_t[1]$, $r\in \FR^*_r[2]$ and $q\in \FR^*_q[3]$ are the coordinate functions on the shifted graded vector space underlying $\aghsk$.

Note that we have both a projection and an embedding
\begin{equation}
 \aghsk\twoheadrightarrow \frg\eand\frg \embd \aghsk~,
\end{equation}
which yield dual maps
\begin{subequations}\label{eq:astringsk_CE-morphisms}
\begin{equation}
 \begin{tikzcd}
    \sCE(\aghsk) \arrow[r,bend left=30]{}{\Phi} & \sCE(\frg) \arrow[l,bend left=30]{}{\Psi}
 \end{tikzcd}~,
\end{equation}
\begin{equation}
\begin{aligned}
\Phi~&:~&t^\alpha&\mapsto\tilde t^\alpha~,~~~&r&\mapsto0~,~~~&q&\mapsto-\tfrac{1}{3!} f_{\alpha\beta\gamma}\tilde t^\alpha \tilde t^\beta \tilde t^\gamma~,\\
\Psi~&:~&\tilde t^\alpha&\mapsto t^\alpha~,
\end{aligned}
\end{equation}
where $\tilde t^\alpha$ are the generators of $\sCE(\frg)$. One readily checks that the differentials are respected, i.e.~$\tilde Q_\sCE\circ \Phi=\Phi\circ Q_\sCE$ and $Q_\sCE\circ \Psi=\Psi\circ \tilde Q_\sCE$.

To promote this pair of maps to a dual quasi-isomorphism, we note that $\Phi\circ\Psi$ is the identity and $\Psi\circ\Phi$ can be connected to the identity via the 2-morphism
\begin{equation}
\begin{gathered}
\eta:\sW(\aghsk)\longrightarrow\sCE(\aghsk)~,\\
\eta~:~t^\alpha\mapsto0~,~~~r\mapsto0~,~~~q\mapsto -r~.
\end{gathered}
\end{equation}
\end{subequations}

In conclusion, we can identify $\sCE(\astringsk(\frg))=\sCE(\aghsk)/\langle q\rangle$, where $\langle q\rangle$ is the differential ideal generated by $q$.

Evidently, there is a similar kernel-extension $\aghl$ leading to an analogous description of the loop model,
\begin{equation}\label{eq:astringl_ses}
 0 \longrightarrow \astringl(\frg) \hooklongrightarrow \aghl \approxeq \frg \xrightarrow{~~~} \FR[2] \longrightarrow 0\eand \sCE(\astringl(\frg))=\sCE(\aghl)/\langle q\rangle~.
\end{equation}

\section{Invariant polynomials}\label{sec:inv_polys}

Besides the Chevalley--Eilenberg and the Weil algebras of an $L_\infty$-algebra, we shall also be interested in its {\em invariant polynomials}. Again, a key point here is that all our constructions should be compatible with quasi-isomorphisms, cf.~also~\cite{Schmidt:2019pks}.

\subsection{Invariant polynomials and Chevalley--Eilenberg algebra cocycles} 

An invariant polynomial is a horizontal element $p\in\sW_{\rm h}(\frg)$, which is either required to be closed under $Q_\sW$ or, more generally, whose image under $Q_\sW$ also lies entirely in $\sW_{\rm h}(\frg)$, cf.~\cite{Sati:2008eg}. We will use the following definitions:
\begin{definition}[\cite{MR0042426},\cite{Sati:2008eg}]
 The \underline{invariant polynomials} $\inv(\frg)$ of an $L_\infty$-algebra $\frg$ form the subset of elements $p$ in $\sW_{\rm h}(\frg)$ for which $Q_\sW p\in \sW_{\rm h}(\frg)$. We also introduce the inclusion map
 \begin{equation}
  e:\inv(\frg)\embd \sW(\frg)~.
 \end{equation}
 
 The \underline{vector space of reduced invariant polynomials} $\overline{\inv}(\frg)$ is given by the $Q$-closed elements in $\sW_{\rm h}(\frg)$ modulo the equivalence relation
 \begin{equation}
  p_1\sim p_2~\Leftrightarrow~p_1-p_2\in Q_\sW \ker(i^*)~.
 \end{equation}

 This can be promoted to the \underline{free algebra of reduced invariant polynomials} $\langle \overline{\inv}(\frg)\rangle_\sF$, which is the algebra generated by the elements of $\overline{\inv}(\frg)$ and which is endowed with the trivial differential. 
\end{definition}

\noindent For Lie algebras $\frg$ both definitions correspond to the ordinary notion of invariant polynomials: a horizontal element $p\in \sW_{\rm h}(\frg)$ is a sum of terms of the form $p=\tfrac1{n!}p_{\alpha_1\dots \alpha_n}\hat t^{\alpha_1}\dots \hat t^{\alpha_n}$, in the coordinates used above in~\eqref{eq:Weil_of_ordinary_Lie} and a horizontal element $p\in \sW_{\rm h}(\frg)$ is always of even degree. Therefore,
\begin{equation}
 Q_{\sW}p\sim p_{\alpha_1\dots \alpha_n}(Q_{\sW}\hat t^{\alpha_1})\dots \hat t^{\alpha_n}=-p_{\alpha_1\dots \alpha_n}f^{\alpha_1}_{\beta\gamma}t^\beta\hat t^\gamma\dots \hat t^{\alpha_n}\in \sW_{\rm h}(\frg)~~~\Leftrightarrow~~~Q_\sW p=0~.
\end{equation}
A distinction is only apparent for higher $L_\infty$-algebras, and we shall return to this point later.

The invariant polynomials form the {\em dg-algebra of invariant polynomials} $\inv(\frg)$ that sits in the complex
\begin{equation}\label{eq:short_exact_seq_cwi}
\begin{tikzcd}
0&\sCE(\frg) \arrow[l] &\sW(\frg)\arrow[l,two heads,swap,"~~i^*"] & \inv(\frg)\arrow[l,swap,hook',"~~e"]&0 \arrow[l]~,
\end{tikzcd}
\end{equation}
which fails to be exact at $\sW(\frg)$. This complex will feature prominently in the discussion of higher connections with gauge $L_\infty$-algebra $\frg$, or $\frg$-connection objects for short, in the following.

We also note that the invariant polynomials are in the kernel of $i^*$ and that $Q_\sW$ restricts to a map $Q_\sW:\inv(\frg)\rightarrow \inv(\frg)$. Therefore, they form a dga dual to an $L_\infty$-algebra, which we denote by $\frg_\inv$ so that $\inv(\frg)\eqqcolon\sCE(\frg_\inv)$.

There is now an important relation between $L_\infty$-algebra cocycles, i.e.~$\mu\in \sCE(\frg)$ with $Q_\sCE \mu=0$, and reduced invariant polynomials $p$. Consider the following double fibration:
\begin{equation}
 \begin{tikzcd}
   & \sW(\frg) \arrow[ld,swap]{}{i^*} \arrow[rd]{}{Q_\sW}& \\
   \sCE(\frg) & & \overline{\inv}(\frg)
 \end{tikzcd}\hspace{1.5cm}
 \begin{tikzcd}
   & {\rm cs} \arrow[ld,swap]{}{i^*} \arrow[rd]{}{Q_\sW}& \\
   \mu & & p 
 \end{tikzcd}
\end{equation}
where we suppressed the projection from $Q_{\sW}(\sW(\frg))$ to $\overline{\inv}(\frg)$.
\begin{definition}[cf.~\cite{Sati:2008eg}]
 Let $\frg$ be an $L_\infty$-algebra. Given a cocycle $\mu\in \sCE(\frg)$ and a reduced invariant polynomial $p\in \overline{\inv}(\frg)$, we call an element ${\rm cs}\in \sW(\frg)$ such that
    \begin{equation}
    i^* ({\rm cs}) = \mu\eand Q_\sW {\rm cs} = p~,
    \end{equation}
 a \underline{Chern--Simons element} witnessing the transgression between $\mu$ and $p$. We say that $\mu$ \underline{transgresses} to $p$ and $p$ \underline{suspends} to $\mu$.
\end{definition}

We note that $p\in \sW_{\rm h}\subset \ker(i^*)$ implies that $\mu$ is a cocycle:
\begin{equation}
Q_\sCE \mu \,=\, Q_\sCE i^*( {\rm cs}) \,=\, i^* (Q_\sW {\rm cs}) \,=\, i^* (p) \,=\, 0~.
\end{equation}

Let us briefly examine the above correspondence. First, we note that there is always a Chern--Simons element for any invariant polynomial $p\in \overline{\inv}(\frg)$ due to  lemma~\ref{lem:Q_cohomology_of_Weil_algebra_trivial}. Second, if we modify the cocycle $\mu$ by a $Q_{\sCE}$-exact term, $\mu\mapsto \tilde \mu=\mu+Q_{\sCE} a$ for some $a\in \sCE(\frg)$, then $\tilde \mu= i^*({\rm cs} + Q_\sW b)$, so $\tilde{\rm cs}={\rm cs}+Q_\sW b$ for any $b\in \sW(\frg)$ with $i^*(b)=a$ and $\tilde\mu$ transgresses to the same invariant polynomial as $\mu$. Third, if we shifted the representative $p\in \overline{\inv}(\frg)$ according to $p\mapsto p'=p+Q_\sW q$ with $q\in \ker(i^*)$, then the Chern--Simons element will be modified to $\tilde{\rm cs}={\rm cs}+q$, but the cocycle $\mu$ remains the same as $i^*(\tilde{\rm cs})=i^*({\rm cs})$. This motivates the definition of the reduced invariant polynomials.

\subsection{Invariant polynomials and quasi-isomorphisms}\label{ssec:inv_and_quasi_isos}

One of our guiding principles is that all our constructions should be agnostic about the representative of the quasi-isomorphism class of the $L_\infty$-algebra we are using. In particular, we would expect that a quasi-isomorphism $\phi:\tilde \frg\rightarrow \frg$ should lead to the following commutative diagram:
\begin{equation}\label{eq:compatibility_diagram}
\begin{tikzcd}
0&\sCE(\tilde \frg) \arrow[l] &\sW(\tilde \frg)\arrow[l,two heads,swap,"~~\tilde i^*"] & \inv(\tilde \frg)\arrow[l,swap,hook',"~~\tilde e"]&0 \arrow[l]\\
0\arrow[u]&\sCE(\frg)\arrow[u,swap,"\Phi"] \arrow[l] &\sW(\frg)\arrow[l,two heads,swap,"~~i^*"] \arrow[u,swap,"\hat\Phi"] & \inv(\frg)\arrow[l,swap,hook',"~~e"] \arrow[u,swap,"\bar\Phi"]&0 \arrow[l]\arrow[u]
\end{tikzcd}
\end{equation}
Here, $\Phi:\sCE(\frg)\rightarrow \sCE(\tilde \frg)$ is the dual of $\phi$, $\hat \Phi$ is its lift to the Weil algebras and $\bar\Phi$ the restriction to the algebras of invariant polynomials. As $\phi$ is a quasi-isomorphism, $\Phi$ forms part of a dual quasi-isomorphism $(\Phi,\Psi,\eta_{\Psi\circ\Phi},\eta_{\Phi\circ\Psi})$, which can always be lifted to a dual quasi-isomorphism $(\hat \Phi,\hat \Psi,\eta_{\hat\Psi\circ\hat\Phi} , \eta_{\hat\Phi\circ\hat\Psi})$ between the Weil algebras, whereas the restrictions to $\bar\Phi$ and $\bar\Psi$ do not necessarily form parts of a dual quasi-isomorphism.

However, in the case that the above commutative diagram does induce a dual quasi-isomorphism $(\bar\Phi,\bar\Psi,\eta_{\bar\Psi\circ\bar\Phi},\eta_{\bar\Phi\circ\bar\Psi})$ between $\inv(\frg)$ and $\inv(\tilde\frg)$ we would expect that there exists an induced isomorphism between the vector spaces $\overline{\inv}(\frg)$ and $\overline{\inv}(\tilde \frg)$. The latter is indeed the case as we shall show in the following. First, we note that $\hat \Phi$ and $\hat \Psi$ induce vector space morphisms 
\begin{equation}
 \hat \Phi:\overline{\inv}(\frg)\to \overline{\inv}(\tilde \frg)\eand
 \hat \Psi:\overline{\inv}(\tilde \frg)\to \overline{\inv}(\frg)
\end{equation}
on the reduced invariant polynomials: $\hat \Phi$ and $\hat \Psi$ restrict to $\bar\Phi$ and $\bar\Psi$ and thus, as dga-morphisms, they map closed invariant polynomials to closed invariant polynomials. Furthermore, as these maps are lifts of $\Phi$ and $\Psi$, respectively, they respect the kernels of the projections $i^\ast$ and $\tilde i^\ast$, i.e.
\begin{equation}
 \hat \Phi:\ker(i^*)\to \ker(\tilde i^*)\eand
 \hat \Psi:\ker(\tilde i^*)\to \ker(i^*)~.
\end{equation}
This, together with the fact that, as dga-morphisms, $\hat\Phi$ and $\hat\Psi$ map exact elements to exact elements, ensures that they are well-defined on the equivalence classes of $\overline{\inv}(\frg)$ and $\overline{\inv}(\tilde\frg)$.

It remains to show that the induced vector space morphisms are isomorphisms. We do this by showing that both $\hat \Phi$ and $\hat\Psi$ have a trivial kernel on the reduced invariant polynomials. The argument is the same in both cases, so we focus on $\hat\Phi$. Thus, consider the 2-morphism $\eta_{\hat\Psi\circ \hat\Phi}:\hat \Psi\circ \hat \Phi\Rightarrow \id$,
\begin{equation}
\begin{tikzcd}
&\sW(\frg)\arrow[dl,bend right=30,"\hat \Psi \circ \hat \Phi",swap]&&\inn(\frg)[2]^*\\
\sW(\frg) &&\sW(\inn(\frg))\arrow[ur,hookleftarrow]\arrow[ul,"i^*",swap,""{name=U,below}]\arrow[dl,"i^*"]&\sF(\inn(\frg))\rlap{~.}\arrow[l,swap,"\Upsilon"]\\
&\sW(\frg)\arrow[ul,bend left=30,"\id"{name=D}]&& \arrow[Rightarrow,"\eta_{\hat \Psi\circ\hat \Phi}",from=U,to=D,start anchor={[xshift=-1ex]},end anchor={[yshift=1ex,xshift=1ex]}]
\end{tikzcd}
\end{equation}
Let $p\in \sW(\frg)$ be a representative of some class $[p]\in \overline{\inv}(\frg)$ with $\hat\Phi(p)=0$. Let $p_\sW$ be $p$ seen as an element of $\sW(\inn(\frg))$, and define $p_\sF=\Upsilon^{-1}(p_\sW)$. On $\sF(\inn(\frg))$, we then have the identity
\begin{equation}\label{eq:p_reduced}
 -p=-(\id\circ i^*\circ \Upsilon)(p_\sF)=Q_\sW(\eta_{\hat \Psi\circ\hat \Phi}(p_\sF))+\eta_{\hat \Psi\circ\hat \Phi}(\sigma_\sF p_\sF)~.
\end{equation}
Note that, for a general 2-morphism, neither does $\eta_{\hat\Psi\circ\hat\Phi}(\sigma_\sF p_\sF)$ vanish nor is $\eta_{\hat\Psi\circ\hat\Phi}(p_\sF)$ in $\ker(i^*)$. However, by assumption, there is another 2-morphism $\eta_{\bar\Psi\circ\bar\Phi}:\bar\Psi\circ\bar\Phi\Rightarrow\id$,
\begin{equation}
\begin{tikzcd}
&\inv(\frg)\arrow[dl,bend right=30,"\bar \Psi \circ \bar \Phi",swap]&&\frg_\inv[2]^*\\
\inv(\frg) &&\sW(\frg_\inv)\arrow[ur,hookleftarrow]\arrow[ul,"i^*",swap,""{name=U,below}]\arrow[dl,"i^*"]&\sF(\frg_\inv)\rlap{~,}\arrow[l,swap,"\Upsilon"]\\
&\inv(\frg)\arrow[ul,bend left=30,"\id"{name=D}]&& \arrow[Rightarrow,"\eta_{\bar \Psi\circ\bar \Phi}",from=U,to=D,start anchor={[xshift=-1ex]},end anchor={[yshift=1ex,xshift=1ex]}]
\end{tikzcd}
\end{equation}
where $\eta_{\bar\Psi\circ\bar\Phi} \eqqcolon \eta_{\hat\Psi\circ\hat\Phi}\circ e$. Both $p_\sF$ and $\sigma_{\sF_\inv} p_\sF$ are now generators of $\sF(\frg_\inv)$ and together with equation~\eqref{eq:restiction_eta} it follows that
\begin{equation}
\eta_{\bar\Psi\circ\bar\Phi}(\sigma_{\sF_\inv}p_\sF) = \eta_{\bar\Psi\circ\bar\Phi}(Q_{\inv(\frg)} p_\sW) = 0~,
\end{equation}
which implies $\eta_{\hat\Psi\circ\hat\Phi}(\sigma_\sF p_\sF)=\eta_{\bar\Psi\circ\bar\Phi}(\sigma_{\sF_\inv}p_\sF)=0$. Additionally, as $\eta_{\bar\Psi\circ\bar\Phi}(p_\sF)$ is an invariant polynomial it follows that $\eta_{\hat\Psi\circ\hat\Phi}(p_\sF)\in\inv(\frg)\subset\ker i^\ast$. Thus, on $\sW(\frg)$, equation~\eqref{eq:p_reduced} reduces to
\begin{equation}
 p=Q_\sW(-\eta_{\hat \Psi\circ\hat \Phi}(p_\sF))\in Q_\sW(\ker i^\ast)~,
\end{equation}
i.e.~$[p]=[0]\in\overline{\inv}(\frg)$ and we showed the following theorem:
\begin{theorem}\label{thm:quasi-iso_inv_inv}
 A quasi-isomorphism of $L_\infty$-algebras $\phi:\tilde \frg\xrightarrow{~\approxeq~}\frg$ for which there exists an induced dual quasi-isomorphism $(\hat \Phi,\hat \Psi,\eta_{\hat\Psi\circ\hat\Phi}, \eta_{\hat\Phi\circ\hat\Psi})$ between the corresponding Weil algebras which restricts to a dual quasi-isomorphism between the invariant polynomials $\inv(\frg)$ and $\inv(\tilde \frg)$ induces a vector space isomorphism between $\overline{\inv}(\frg)$ and $\overline{\inv}(\tilde \frg)$.
\end{theorem}

\

Let us consider a simple example of all of the above in which our compatibility condition is indeed satisfied, namely the $L_\infty$-algebra $\frg=\inn(\frh)$ of some Lie algebra $\frh$,
\begin{equation}\label{eq:g_as_h_to_h}
\begin{tikzcd}
\frg~:~\frh[1] \arrow[r,"\text{id}"] & \frh
\end{tikzcd}~,
\end{equation}
which is concentrated in degrees $-1$ and $0$. This $L_\infty$-algebra is readily seen to be quasi-isomorphic to the trivial $L_\infty$-algebra $*$: the cohomology $H^\bullet_{\mu_1}(\frg)$ is trivial and there is a morphism $\phi:\frg\rightarrow *$ with $\phi_1: \frg \rightarrow *$ and $\phi_2$ trivial, cf.~\eqref{eq:Lie_2_algebra_morph}.\footnote{Note also that $\sCE(\frg)=\sW(\frh)\approxeq \sW(*)$, where the dual isomorphism merely inverts the signs of the coordinate functions.} The corresponding Chevalley--Eilenberg algebra $\sCE(\frg)$ is generated by coordinate functions $t^\alpha$ and $r^\alpha$ of degree~$1$ and~$2$, respectively, with the differential acting according to
\begin{equation}
\begin{aligned}\label{eq:QCE_inv_example_h_h}
Q_\sCE&:&t^\alpha &\mapsto -\tfrac12 f^\alpha_{\beta\gamma} t^\beta t^\gamma - r^\alpha~,~~~&r^\alpha &\mapsto -f^\alpha_{\beta\gamma} t^\beta r^\gamma~,
\end{aligned}
\end{equation}
where $f^\alpha_{\beta\gamma}$ are the structure constants of the Lie algebra structure on $\frh$. The dual of the above morphism $\phi$ corresponds to
\begin{subequations}\label{eq:quasi-iso_h_to_h}
\begin{equation}
 \begin{tikzcd}
    \sCE(\frg) \arrow[r,bend left=30]{}{\Phi} & \sCE(\ast) \arrow[l,bend left=30]{}{\Psi}
 \end{tikzcd}
\end{equation}
with 
\begin{equation}
\Phi: t^\alpha,r^\alpha\mapsto 0\eand \Psi:0 \mapsto 0~.
\end{equation}
\end{subequations}
A 2-morphism $\eta:\Psi\circ\Phi\Rightarrow\id$ completing the quasi-isomorphism is given by
\begin{equation}\label{eq:CE_contracting_homotopy}
\begin{aligned}
\eta~&:~&t^\alpha&\mapsto 0~,~~~&r^\alpha &\mapsto t^\alpha~.
\end{aligned}
\end{equation}
Lifting to the Weil algebra $\sW(\frg)$ yields the differential
\begin{equation}\label{eq:QW_inv_example_h_h}
\begin{aligned}
Q_\sW&:&t^\alpha &\mapsto -\tfrac12 f^\alpha_{\beta\gamma} t^\beta t^\gamma -r^\alpha+ \hat t^\alpha~,~~~&r^\alpha &\mapsto -f^\alpha_{\beta\gamma} t^\beta r^\gamma + \hat r^\alpha~,\\
&&\hat t^\alpha &\mapsto -f^\alpha_{\beta\gamma} t^\beta \hat t^\gamma + \hat r^\alpha~,~~~&\hat r^\alpha &\mapsto -f^\alpha_{\beta\gamma} t^\beta \hat r^\gamma + f^\alpha_{\beta\gamma} \hat t^\beta r^\gamma~,
\end{aligned}
\end{equation}
and morphisms
\begin{equation}
\hat\Phi: t^\alpha,r^\alpha,\hat t^\alpha, \hat r^\alpha\mapsto 0\eand \Psi:0 \mapsto 0~,
\end{equation}
as well as the 2-morphism $\eta:\hat\Psi\circ\hat\Phi\Rightarrow\id$ given by the map
\begin{equation}\label{eq:eta-h-to-h}
\begin{aligned}
\eta~&:~&t^\alpha&\mapsto 0~,~~~&r^\alpha &\mapsto t^\alpha~,\\
&&\hat t^\alpha&\mapsto 0~,~~~&\hat r^\alpha &\mapsto -\hat t^\alpha~.
\end{aligned}
\end{equation}
One readily checks that
\begin{equation}
 (Q_\sW\circ\eta_\sW+\eta_\sW\circ Q_{\sF})(a)=-a=(\hat\Psi\circ\hat\Phi-\id)(a)
\end{equation}
for elements $a\in\sF(\inn(\frg))$.

To conclude that this dual quasi-isomorphism restricts to a dual quasi-isomorphism on $\inv(\frg)$ we need to check that $\eta(\inv(\frg))\subset\inv(\frg)$. Comparing degrees, it is easy to see that a generic invariant polynomial consists of sums of invariant polynomials of the form
\begin{equation}
 p=p_{\alpha_1\dots \alpha_n\beta_1\dots \beta_m}\hat t^{\alpha_1}\dots \hat t^{\alpha_n}\hat r^{\beta_1}\dots \hat r^{\beta_m}~,
\end{equation}
where $p_{\alpha_1\dots \alpha_n\beta_1\dots \beta_m}\in \FR$, $n,m\in \NN$. The condition that $Q_\sW p \in \sW_{\rm h}(\frg)$ together with~\eqref{eq:QW_inv_example_h_h} implies that
\begin{equation}\label{eq:f_p_identity}
\sum\limits_{i=1}^{n} p_{\alpha_1\dots\hat\alpha_i\dots\alpha_n\beta_1\dots\beta_m}f^{\rho}_{\alpha_i\nu}+\sum\limits_{i=1}^m p_{\alpha_1\dots\alpha_n\beta_1\dots\hat\beta_i\dots\beta_m}f^{\rho}_{\beta_i\nu} = 0~,
\end{equation}
such that we have
\begin{equation}\label{eq:Qp_h_h}
Q_\sW p= np_{\alpha_1\dots \alpha_n\beta_1\dots \beta_m}\hat t^{\alpha_1}\dots \hat t^{\alpha_{n-1}}\hat r^{\alpha_n}\hat r^{\beta_1}\dots \hat r^{\beta_m}~.
\end{equation}
Applying the 2-morphism in~\eqref{eq:eta-h-to-h} to $p$ we obtain 
\begin{equation}\label{eq:etap_h_h}
\eta(p) = -\tfrac{m}{n+m}p_{\alpha_1\dots \alpha_n\beta_1\dots \beta_m}\hat t^{\alpha_1}\dots \hat t^{\alpha_n}\hat t^{\beta_1}\hat r^{\beta_2 }\dots \hat r^{\beta_m}~,
\end{equation}
which due to the identity~\eqref{eq:f_p_identity} again forms an invariant polynomial. Thus, we arrive at the following proposition:
\begin{proposition}
The dual quasi-isomorphism~\eqref{eq:quasi-iso_h_to_h} induces a dual quasi-isomorphism between the dga of invariant polynomials $\inv(\frg)$, where $\frg$ is defined in~\eqref{eq:g_as_h_to_h}, and the dga of invariant polynomials $\inv(*)=*$. 
\end{proposition}
\noindent Together with theorem~\ref{thm:quasi-iso_inv_inv}, we then have an expected corollary:
\begin{corollary}
 The vector space $\overline{\inv}(\frg)$ is the trivial vector space.
\end{corollary}
Explicitly, one can show that $\eta$ in~\eqref{eq:etap_h_h} acts as the inverse of $Q_\sW$ on $\inv(\frg)$. Thus, any $Q_\sW$-closed element $p$ in $\sW_{\rm h}(\frg)$ is automatically $Q_\sW$-exact in $\sW_{\rm h}(\frg)$, 
\begin{equation}
\begin{aligned}
Q_\sW\eta(p)= -\tfrac{m}{n+m}p_{\alpha_1\dots\alpha_n\beta_1\dots\beta_m} \hat t^{\alpha_1}\dots \hat t^{\alpha_n}\hat r^{\beta_1}\dots \hat r^{\beta_m} \propto p~,
\end{aligned}
\end{equation}
rendering $\overline{\inv}(\frg)$ the trivial vector space.

\subsection{Adjusted Weil algebras of \texorpdfstring{$\aghsk$}{ghsk} and \texorpdfstring{$\astringsk(\frg)$}{stringsk(g)}}\label{ssec:Weil_adj_sk}

Unfortunately, dual quasi-isomorphisms between Chevalley--Eilenberg algebras do not induce dual quasi-isomorphisms between the dgas of invariant polynomials in general. This may not be surprising because the Weil algebra, which sits between both dgas in the complex~\eqref{eq:short_exact_seq_cwi}, is always dually quasi-isomorphic to the trivial one. There is, in fact, some freedom in constructing a suitable Weil algebra sitting above the Chevalley--Eilenberg algebra, leading to what we will call an adjusted Weil algebra.

As an example, let $\frg$ be some Lie algebra and consider the quasi-isomorphic Lie 3-algebra $\aghsk$ introduced in section~\ref{ssec:extended_skeletal}. The Weil algebra of $\aghsk$ is given by the differential acting on generators as
\begin{equation}\label{eq:extended_string_sk_differential_weil}
\begin{aligned}
Q_\sW~&:~&t^\alpha &\mapsto -\tfrac12 f^\alpha_{\beta\gamma} t^\beta  t^\gamma + \hat t^\alpha~,~~~&r &\mapsto \tfrac{1}{3!} f_{\alpha\beta\gamma} t^\alpha  t^\beta  t^\gamma +q+ \hat r~,&q &\mapsto \hat q~,\\
&&\hat t^\alpha &\mapsto -f^\alpha_{\beta\gamma} t^\beta  \hat t^\gamma~,~~~&\hat r&\mapsto -\tfrac12 f_{\alpha\beta\gamma} t^\alpha  t^\beta  \hat t^\gamma - \hat q~,~~~& \hat q&\mapsto0~,
\end{aligned}
\end{equation}
where $t^\alpha\in \frg^*_t[1]$, $r\in \FR^*_r[1]$ and $q\in \FR^*_q[1]$ are the coordinate functions on the shifted graded vector space underlying $\aghsk$ and $\hat t^\alpha$, $\hat r$, $\hat q$ are the additional copies introduced for the Weil algebra.

The dual quasi-isomorphism~\eqref{eq:astringsk_CE-morphisms} between the Chevalley--Eilenberg algebras of $\aghsk$ and $\frg$ induces the dual quasi-isomorphism
\begin{subequations}\label{eq:quasi-iso_ordinary_Weil}
\begin{equation}\label{eq:Weil_algebras}
 \begin{tikzcd}
    \sW(\aghsk) \arrow[r,bend left=30]{}{\hat\Phi} & \sW(\frg)\arrow[l,bend left=30]{}{\hat\Psi}
 \end{tikzcd}~,
\end{equation}
\begin{equation}
\begin{aligned}
\hat\Phi~&:~&t^\alpha&\mapsto\tilde t^\alpha~,~~~&r&\mapsto0~,~~~&q&\mapsto-\tfrac{1}{3!} f_{\alpha\beta\gamma}\tilde t^\alpha \tilde t^\beta \tilde t^\gamma~,\\
&&\hat t^\alpha&\mapsto\hat{\tilde{t}}^\alpha~,~~~&\hat r&\mapsto0~,~~~&\hat q&\mapsto-\tfrac12 f_{\alpha\beta\gamma}\tilde t^\alpha \tilde t^\beta \hat{\tilde{t}}^\gamma~,\\
\hat\Psi~&:~&\tilde t^\alpha&\mapsto t^\alpha~,~~~&\hat{\tilde{t}}^\alpha&\mapsto \hat t^\alpha
\end{aligned}
\end{equation}
with $\eta_{\hat\Phi\circ \hat\Psi}:\hat\Phi\circ\hat\Psi\Rightarrow \id$ trivial and the connecting 2-morphism $\eta_{\hat\Psi\circ\hat\Phi}:\hat\Psi\circ\hat\Phi\Rightarrow \id$ fixed by 
\begin{equation}
\begin{aligned}
\eta_{\hat\Psi\circ\hat\Phi}~&:~&
t^\alpha&\mapsto 0~,~~~&r&\mapsto 0~,~~~&q&\mapsto -r~,\\
&&\hat t^\alpha&\mapsto 0~,~~~&\hat r&\mapsto 0~,~~~&\hat q&\mapsto \hat r~.
\end{aligned}
\end{equation}
\end{subequations}
Here, $\tilde t^\alpha$ and $\hat{\tilde{t}}^\alpha$ are the evident generators of $\sW(\frg)$, cf.~\eqref{eq:Weil_of_ordinary_Lie}.

Clearly, the dual quasi-isomorphism~\eqref{eq:quasi-iso_ordinary_Weil} is problematic since it maps the invariant polynomial $\hat q$ to the element $\Phi(\hat q)=-\tfrac12 f_{\alpha\beta\gamma}\tilde t^\alpha \tilde t^\beta \hat{\tilde{t}}^\gamma\notin \sW_{\rm h}$. A quick computation\footnote{This and the following computations are best done using a computer algebra program.} reveals that the dual quasi-isomorphism between $\sCE(\aghsk)$ and $\sCE(\frg)$ does not allow for a deformation that solves the issue.

We therefore have to deform the Weil algebra $\sW(\aghsk)$ to an adjusted Weil algebra $\sW_{\rm adj}(\aghsk)$ such that~\eqref{eq:compatibility_diagram} with $\tilde \frg=\aghsk$ becomes a commutative diagram. Note that we cannot change the Chevalley--Eilenberg algebras on which the Weil algebras in~\eqref{eq:Weil_algebras} project or the relating morphisms between them, since this would amount to changing the underlying Lie 3-algebra $\aghsk$. 

In general, such an adjustment is not unique, but we impose a number of simplifying constraints which fix it. First, we choose to preserve the embedding $\sW(\FR[2])\embd \sW(\aghsk)$ induced by the sequence~\eqref{eq:astringsk_ses}, which fixes
\begin{equation}
 Q_{\sW_{\rm adj}} q=\hat q~,~~~Q_{\sW_{\rm adj}} \hat q=0~.
\end{equation}
This is essentially a choice of coordinates. Second, we choose $\Phi(\hat q)=\tfrac12 \mathsf{p}_1$, where $\tfrac12 \mathsf{p}_1$ is the first fractional Pontryagin class to which the Chevalley--Eilenberg cocycle $\mu$ transgresses:
\begin{equation}\label{eq:transgression_data}
 \begin{tikzcd}
   & {\rm cs}\coloneqq -\tfrac1{3!}f_{\alpha\beta\gamma}\tilde t^\alpha\tilde t^\beta\tilde t^\gamma+\kappa_{\alpha\beta}\tilde{t}^\alpha\hat{\tilde{t}}^\beta \arrow[ld,swap]{}{i^*} \arrow[rd]{}{Q_\sW}& \\
    \mu\coloneqq -\tfrac1{3!}f_{\alpha\beta\gamma}\tilde t^\alpha\tilde t^\beta\tilde t^\gamma & & \tfrac12 \mathsf{p}_1\coloneqq \kappa_{\alpha\beta}\hat{\tilde{t}}^\alpha\hat{\tilde{t}}^\beta
 \end{tikzcd}
\end{equation}
This induces the adjustment
\begin{equation}
\begin{tikzcd}[row sep=1.5cm,ampersand replacement=\&]
  \sCE(\aghsk) \arrow[d]{}{(i^*\circ \Phi)(q)=\mu} \& \sW(\aghsk) \arrow[l, swap]{}{i^*}\arrow[d]{}{
  \begin{aligned}
  \Phi(q)&=\mu\\[-0.2cm]
  \Phi(\hat q)&=-\tfrac12 f_{\alpha\beta\gamma}\tilde t^\alpha \tilde t^\beta \hat{\tilde t}^\gamma
  \end{aligned}
  }
  \\
  \sCE(\frg) \& \sW(\frg) \arrow[l,swap]{}{i^*}
\end{tikzcd}
~~\longrightarrow~~
\begin{tikzcd}[row sep=1.5cm,ampersand replacement=\&]
  \sCE(\aghsk) \arrow[d]{}{(i^*\circ \Phi)(q)=\mu} \& \sW_{\rm adj}(\aghsk) \arrow[l, swap]{}{i^*}\arrow[d]{}{
  \begin{aligned}
  \Phi_{\rm adj}(q)&={\rm cs}\\[-0.1cm]
  \Phi_{\rm adj}(\hat q)&=\tfrac12 \mathsf{p}_1
  \end{aligned}
  }
  \\
  \sCE(\frg) \& \sW(\frg) \arrow[l,swap]{}{i^*}
\end{tikzcd}
\end{equation}
which ensures that invariant polynomials in $\sW_{\rm adj}(\aghsk)$ are mapped to invariant polynomials in $\sW(\frg)$ and vice versa. We note, however, that $\kappa_{\alpha\beta}\hat t^\alpha\hat t^\beta-\hat q$ is now in the kernel of $\Phi_{\rm adj}$, so it should trivialize in $\sW_{\rm adj}(\aghsk)$, which fixes
\begin{equation}
 Q_{\sW_{\rm adj}} \hat r=\kappa_{\alpha\beta}\hat t^\alpha \hat t^\beta-\hat q~,
\end{equation}
up to an isomorphic choice of the generator $\hat r$. Finally, we also demand that $Q_{\sW_{\rm adj}} t^\alpha =Q_\sW t^\alpha$. This is enough to completely fix $Q_{\sW_{\rm adj}}$:
\begin{definition}
 The \underline{adjusted Weil algebra} $\sW_{\rm adj}(\aghsk)$ has the same generators as $\sW(\aghsk)$ with the differential $Q_{\sW_{\rm adj}}$ acting as 
\begin{equation}\label{eq:ext_twt_string_sk_differential}
\begin{aligned}
Q_{\sW_{\rm adj}}~&:~&t^\alpha &\mapsto -\tfrac12 f^\alpha_{\beta\gamma} t^\beta  t^\gamma + \hat t^\alpha~,~&r &\mapsto \tfrac{1}{3!} f_{\alpha\beta\gamma} t^\alpha  t^\beta  t^\gamma -\kappa_{\alpha\beta}t^\alpha\hat t^\beta+q+ \hat r~,~&q &\mapsto \hat q~,\\
&&\hat t^\alpha &\mapsto -f^\alpha_{\beta\gamma} t^\beta  \hat t^\gamma~,~
 &\hat r&\mapsto \kappa_{\alpha\beta}\hat t^\alpha\hat t^\beta - \hat q~,~
 & \hat q&\mapsto0~.
\end{aligned}
\end{equation}
\end{definition}

\noindent In the adjusted case, the dual quasi-isomorphism reads as 
\begin{subequations}\label{eq:quasi-iso_adjusted_Weil}
\begin{equation}
 \begin{tikzcd}
    \sW_{\rm adj}(\aghsk) \arrow[r,bend left=30]{}{\Phi_{\rm adj}} & \sW(\frg) \arrow[l,bend left=30]{}{\Psi_{\rm adj}}
 \end{tikzcd}~,
\end{equation}
\begin{equation}
\begin{aligned}
\Phi_{\rm adj}~&:~&t^\alpha&\mapsto\tilde t^\alpha~,~~~&r&\mapsto0~,~~~&q&\mapsto{\rm cs}~,~~~\\
&&
\hat t^\alpha&\mapsto\hat{\tilde{t}}^\alpha~,~~~&\hat r&\mapsto0~,~~~&\hat q&\mapsto \tfrac12 \mathsf{p}_1~,\\
\Psi_{\rm adj}~&:~&\tilde t^\alpha&\mapsto t^\alpha~,~~~&\hat{\tilde{t}}^\alpha&\mapsto \hat t^\alpha~,
\end{aligned}
\end{equation}
where $\tilde Q_\sW\circ \Phi_{\rm adj}=\Phi_{\rm adj}\circ Q_{\sW_{\rm adj}}$ and $Q_{\sW_{\rm adj}}\circ \Psi_{\rm adj}=\Psi_{\rm adj}\circ \tilde Q_\sW$ as well as $\Phi_{\rm adj}\circ\Psi_{\rm adj}=\id$ and unmodified $\eta_{\rm adj}:\Psi_{\rm adj}\circ\Phi_{\rm adj}\rightarrow \id$:
\begin{equation}
\begin{aligned}
\eta_{\rm adj}~&:~&
t^\alpha&\mapsto 0~,~~~&r&\mapsto 0~,~~~&q&\mapsto -r~,\\
&&\hat t^\alpha&\mapsto 0~,~~~&\hat r&\mapsto 0~,~~~&\hat q&\mapsto \hat r~.
\end{aligned}
\end{equation}
\end{subequations}

We note that the dual quasi-isomorphism $(\Phi_{\rm adj},\Psi_{\rm adj},\eta_{\rm adj},0):\sW_{\rm adj}(\aghsk)\rightarrow \sW(\frg)$ implies that also $\sW_{\rm adj}(\aghsk)$ and $\sW(\aghsk)$ are dually quasi-isomorphic. Moreover, the projection $i^*$ to the Chevalley--Eilenberg algebra is the same for both ordinary and adjusted Weil algebra.

Let us now show that with the adjusted Weil algebra $\sW_{\rm adj}(\aghsk)$, the diagram~\eqref{eq:compatibility_diagram} is indeed commutative and $\inv_{\rm adj}(\aghsk)\approxeq \inv(\frg)$. We first note that 
\begin{equation}
 \begin{tikzcd}
    \sW_{{\rm adj},{\rm h}}(\aghsk)\arrow[r,bend left]{}{\Phi_{\rm adj}} & \sW_{\rm h}(\frg) \arrow[l,bend left]{}{\Psi_{\rm adj}}
 \end{tikzcd}
 ~~~\Rightarrow~~~
 \begin{tikzcd}
    \inv_{\rm adj}(\aghsk)\arrow[r,bend left]{}{\Phi_{\rm adj}} & \inv(\frg) \arrow[l,bend left]{}{\Psi_{\rm adj}}
 \end{tikzcd}~~,
\end{equation}
that is, $\Phi_{\rm adj}$ and $\Psi_{\rm adj}$ indeed restrict to morphisms between the dgas of invariant polynomials. Moreover, $\Phi_{\rm adj}\circ \Psi_{\rm adj}$ restricts to $\id_{\inv(\frg)}$ and therefore it remains to show that $\eta_{\rm adj}$ restricts to a 2-morphism $\eta_{\rm adj}:\Psi_{\rm adj}\circ \Phi_{\rm adj}\Rightarrow \id_{\inv_{\rm adj}(\aghsk)}$ on the generators of $\inv_{\rm adj}(\aghsk)$. Obviously, $\inv_{\rm adj}(\aghsk)\cong \inv(\frg)[\hat r,\hat q]$, that is, $\inv_{\rm adj}(\aghsk)$ consists of polynomials in $\hat r$ and $\hat q$ with coefficients in $\inv(\frg)$. We note that
\begin{equation}
 \begin{aligned}
  (\Psi_{\rm adj}\circ \Phi_{\rm adj}-\id_{\inv(\aghsk)})(p)=0=(Q_{\sW_{\rm adj}}\circ \eta_{\rm adj}+\eta_{\rm adj}\circ Q_{\sW_{\rm adj}})(p)~,\\
  (\Psi_{\rm adj}\circ \Phi_{\rm adj}-\id_{\inv(\aghsk)})(\hat r)=-\hat r=(Q_{\sW_{\rm adj}}\circ \eta_{\rm adj}+\eta_{\rm adj}\circ Q_{\sW_{\rm adj}})(\hat r)~,\\
  (\Psi_{\rm adj}\circ \Phi_{\rm adj}-\id_{\inv(\aghsk)})(\hat q)=\tfrac12 \kappa_{\alpha\beta}\hat t^\alpha \hat t^\beta-\hat q=(Q_{\sW_{\rm adj}}\circ \eta_{\rm adj}+\eta_{\rm adj}\circ Q_{\sW_{\rm adj}})(\hat q)~,
 \end{aligned}
\end{equation}
where $p$ denotes invariant polynomials not containing $\hat r$ or $\hat q$, i.e.~all other generators of $\inv_{\rm adj}(\aghsk)$. Together with theorem~\ref{thm:quasi-iso_inv_inv}, we then have the following result:
\begin{theorem}
 The dgas $\inv(\frg)$ and $\inv_{\rm adj}(\aghsk)$ are quasi-isomorphic and the vector spaces $\overline{\inv}(\frg)$ and $\overline{\inv}_{\rm adj}(\aghsk)$ are isomorphic.
\end{theorem}

Note that this is important if we want to model the string Lie 2-algebra as an $L_\infty$-subalgebra of $\aghsk$ as in equation~\eqref{eq:astringsk_ses}. It shows in particular, that $\sW(\aghsk)$ is problematic, while $\sW_{\rm adj}(\aghsk)$, when factored by the differential ideal generated by $q$ and $\hat q$, becomes a good model for the Weil algebra of $\astringsk(\frg)$.

\subsection{Adjusted Weil algebras of \texorpdfstring{$\aghl$}{ghl} and \texorpdfstring{$\astringl(\frg)$}{stringl(g)}}\label{ssec:Weil_adj_l}

Because we shall require it later on, let us also give the explicit formulas for the Lie 3-algebra $\aghl$ involving path and loop spaces quasi-isomorphic to $\frg$. Here we have 
\begin{equation}
 \aghl\coloneqq \big(~\FR_q\xhookrightarrow{~~~} \hat  L_0\frg_r \longrightarrow P_0\frg~\big)\coloneqq \big(~\FR[2]\xhookrightarrow{~~~} \hat  L_0\frg[1] \longrightarrow P_0\frg~\big)~\approxeq~\big(~ L_0\frg[1] \longrightarrow P_0\frg~\big)~\approxeq~\frg~.
\end{equation}
The Weil algebra $\sW(\aghl)$ has the generators of $\sW(\astringl(\frg))$, cf.~\eqref{eq:Weil_string_Lie_2_loop}, as well as the coordinate functions $q$ and $\hat q$ of degrees~3 and~4 and the differential acts according to 
\begin{equation}\label{eq:extended_string_l_differential}
\begin{aligned}
Q_\sW~&:&t^{\alpha\tau} &\mapsto -\tfrac12 f^{\alpha}_{\beta\gamma} t^{\beta\tau} t^{\gamma\tau} - r^{\alpha\tau} +\hat{t}^{\alpha\tau}~,~~~ 
  &\hat{t}^{\alpha\tau}&\mapsto -f^\alpha_{\beta\gamma}t^{\beta\tau}\hat{t}^{\gamma\tau}+\hat{r}^{\alpha\tau}~,\\
 &&r^{\alpha\tau} &\mapsto -f^\alpha_{\beta\gamma} t^{\beta\tau}r^{\gamma\tau} + \hat{r}^{\alpha\tau}~,~~~
  &\hat{r}^{\alpha\tau}&\mapsto-f^\alpha_{\beta\gamma} t^{\beta\tau}\hat{r}^{\gamma\tau} + f^\alpha_{\beta\gamma} \hat{t}^{\beta\tau}r^{\gamma\tau} ~,\\
 &&r_0 &\mapsto 2\int_0^1\dd \tau\, \kappa_{\alpha\beta}t^{\alpha\tau} \dot{r}^{\beta\tau}+q +\hat{r}_0~,~&\hat{r}_0&\mapsto 2\int_0^1\dd\tau\,\kappa_{\alpha\beta}\left(t^{\alpha\tau} \smash{\dot{\hat{r}}}^{\beta\tau}-\hat{t}^{\alpha\tau}\dot{r}^{\beta\tau}\right)-\hat q~,\\
 &&q &\mapsto \hat q~,& \hat q&\mapsto0~.
\end{aligned}
\end{equation}

The endpoint evaluation map $\dpar$ and the smooth function $\ell$ from~\eqref{eq:f-function} yield projections and embeddings,
\begin{equation}
 \aghl\xrightarrow{~\phantom{\dpar}~} P_0\frg\xrightarrow {~\dpar~}\frg\eand \frg \xhookrightarrow{~\cdot \ell(\tau)~}P_0\frg\xhookrightarrow{~\phantom{\cdot \ell(\tau)}}\aghl~.
\end{equation}
Dually, we have embeddings and projections on the Chevalley--Eilenberg algebras of\linebreak $\aghl$ and $\frg$, which form a dual quasi-isomorphism. Lifted to the Weil algebra, it reads as
\begin{subequations}\label{eq:quasi-iso_ordinary_Weil_loop}
\begin{equation}
 \begin{tikzcd}
    \sW(\aghl) \arrow[r,bend left=30]{}{\Phi} & \sW(\frg) \arrow[l,bend left=30]{}{\Psi}
 \end{tikzcd}~,
\end{equation}
\begin{equation}
\begin{aligned}
\Phi~&:~&
t^{\alpha\tau}&\mapsto \ell(\tau)\tilde t^\alpha~,&r^{\alpha\tau}&\mapsto (\ell(\tau)-\ell^2(\tau))\tfrac12 f^\alpha_{\beta\gamma}\tilde t^\beta \tilde t^\gamma~,&r_0&\mapsto0~,&q&\mapsto\tfrac{1}{3!} f_{\alpha\beta\gamma}\tilde t^\alpha \tilde t^\beta \tilde t^\gamma~,~\\
&&\hat t^{\alpha\tau}&\mapsto \ell(\tau)\hat{\tilde{t}}^\alpha~,&\hat r^{\alpha\tau}&\mapsto -(\ell(\tau)-\ell^2(\tau))f^\alpha_{\beta\gamma} \tilde t^\beta \hat{\tilde{t}}^\gamma~,&\hat r_0&\mapsto0~,~&\hat q&\mapsto\tfrac12 f_{\alpha\beta\gamma}\tilde t^\alpha \tilde t^\beta \hat{\tilde{t}}^\gamma~,\\
\Psi~&:~&\tilde t^\alpha&\mapsto t^{\alpha 1}~,&\hat{\tilde{t}}^\alpha&\mapsto \hat t^{\alpha 1}~,
\end{aligned}
\end{equation}
where $\tilde t^\alpha$ and $\hat{\tilde{t}}^\alpha$ are again the generators of $\sW(\frg)$. A 2-morphism $\eta:\Psi\circ \Phi\Rightarrow \id$ is given by the map $\eta:\sW(\inn(\aghl))\Rightarrow\sW(\aghl)$ with 
\begin{equation}
\begin{aligned}
\eta~:~
t^{\alpha\tau},\, r_0&\mapsto0~,~~~
&r^{\alpha\tau}&\mapsto t^{\alpha\tau}-\ell(\tau)t^{\alpha 1}~,~~~
&q&\mapsto -r_0-\int_0^1\dd\tau\,\kappa_{\alpha\beta} \dot t^{\alpha\tau} t^{\beta\tau}~,\\
\hat t^{\alpha\tau},\,\hat r_0&\mapsto0~,~~~
&\hat r^{\alpha\tau}&\mapsto \ell(\tau)\hat t^{\alpha 1}- \hat t^{\alpha\tau}~,~~~
&\hat q&\mapsto \hat r_0-\int_0^1\dd\tau\,\kappa_{\alpha\beta} \left(\dot t^{\alpha\tau} \hat t^{\beta\tau}-\dot{\hat{t}}^{\alpha\tau} t^{\beta\tau}\right)~.
\end{aligned}
\end{equation}
\end{subequations}
Note that this dual quasi-isomorphism is merely a composition of the dual quasi-isomor\-phism~\eqref{eq:quasi-iso_ordinary_Weil} and the lift of~\eqref{eq:quasi-iso-loop-skeletal-Lie2} to the Weil algebras of $\aghsk$ and $\aghl$. This observation allows us to derive the corresponding adjusted Weil algebra. Up to trivial isomorphisms, we arrive at the following definition:
\begin{definition}
 The \underline{adjusted Weil algebra} $\sW_{\rm adj}(\aghl)$ has the same generators as $\sW(\aghl)$ with the differential $Q_{\sW_{\rm adj}}$ acting as 
\begin{subequations}\label{eq:ext_twt_string_l_differential}
\begin{equation}
\begin{aligned}
    Q_{\sW_{\rm adj}}:~t^{\alpha\tau} &\mapsto -\tfrac12 f^{\alpha}_{\beta\gamma} t^{\beta\tau} t^{\gamma\tau} - r^{\alpha\tau} +\hat{t}^{\alpha\tau}~,~~~&\hat{t}^{\alpha\tau}&\mapsto -f^\alpha_{\beta\gamma}t^{\beta\tau}\hat{t}^{\gamma\tau}+\chi^{\alpha\tau}(t,\hat{t})+\hat{r}^{\alpha\tau}~,\\
r^{\alpha\tau} &\mapsto -f^\alpha_{\beta\gamma} t^{\beta\tau}r^{\gamma\tau} +\chi^{\alpha\tau}(t,\hat t)+ \hat{r}^{\alpha\tau}~,~
&\hat{r}^{\alpha\tau}&\mapsto0~,\\
r_0 &\mapsto 2\int_0^1\dd\tau\,\kappa_{\alpha\beta}t^{\alpha\tau} \smash{\dot{{r}}}^{\beta\tau}+\chi(t,\hat{t})+q+\hat{r}_0&\hat{r}_0&\mapsto -\chi(\hat t,\hat t)-\hat q~,\\
q &\mapsto \hat q~,~~~&\hat q&\mapsto0~,
\end{aligned}
\end{equation}
where
\begin{equation}\label{eq:chi_loop_case}
 \chi^{\alpha\tau}(t,\hat t)\coloneqq f^\alpha_{\beta\gamma}(t^{\beta\tau}\hat t^{\gamma\tau}-\ell(\tau)t^{\beta 1}\hat t^{\gamma 1})\eand \chi(t,\hat{t})\coloneqq 2\int_0^1\dd\tau\,\kappa_{\alpha\beta} \dot t^{\alpha\tau} \hat t^{\beta\tau}~.
\end{equation}
\end{subequations}
\end{definition}

\noindent The quasi-isomorphism reads as 
\begin{subequations}\label{eq:quasi-iso_adjusted_Weil_loop}
\begin{equation}
 \begin{tikzcd}
    \sW_{\rm adj}(\aghl) \arrow[r,bend left]{}{\Phi_{\rm adj}} & \sW(\frg) \arrow[l,bend left]{}{\Psi_{\rm adj}}
 \end{tikzcd}
\end{equation}
\begin{equation}
\begin{aligned}
\Phi_{\rm adj}&:&t^{\alpha\tau}&\mapsto \ell(\tau)\tilde t^\alpha~,
&r^{\alpha\tau}&\mapsto (\ell(\tau)-\ell^2(\tau))\tfrac12 f^\alpha_{\beta\gamma}\tilde t^\beta \tilde t^\gamma~,
&r_0&\mapsto 0~,
&q&\mapsto -{\rm cs}~,\\
&
&\hat t^{\alpha\tau}&\mapsto \ell(\tau)\hat{\tilde{t}}^\alpha~,
&\hat r^{\alpha\tau}&\mapsto 0~,
&\hat r_0&\mapsto 0~,~
&\hat q&\mapsto -\tfrac12 \mathsf{p}_1~,\\
\Psi_{\rm adj}&:& \tilde t^\alpha&\mapsto  t^{\alpha 1}~,~~~&\hat{\tilde{t}}^\alpha&\mapsto  \hat t^{\alpha 1}~,
\end{aligned}
\end{equation}
where $\tilde Q_\sW\circ \Phi_{\rm adj}=\Phi_{\rm adj}\circ Q_{\sW_{\rm adj}}$ and $Q_{\sW_{\rm adj}}\circ \Psi_{\rm adj}=\Psi_{\rm adj}\circ \tilde Q_\sW$ as well as $\Phi_{\rm adj}\circ\Psi_{\rm adj}=\id$ and $\eta_{\rm adj}:\Psi_{\rm adj}\circ\Phi_{\rm adj}\Rightarrow \id$:
\begin{equation}
\begin{aligned}
\eta_{\rm adj}~:~
t^{\alpha\tau},\,r_0&\mapsto0~,~~~
&r^{\alpha\tau}&\mapsto t^{\alpha\tau}-\ell(\tau)t^{\alpha 1}~,~~~
&q&\mapsto -r_0-\int_0^1\dd\tau\,\kappa_{\alpha\beta} \dot t^{\alpha\tau} \hat t^{\beta\tau}~,\\
\hat t^{\alpha\tau},\,\hat r_0&\mapsto0~,~~~
&\hat r^{\alpha\tau}&\mapsto \ell(\tau)\hat t^{\alpha 1}-\hat t^{\alpha\tau}~,~~~
&\hat q&\mapsto \hat r_0~.
\end{aligned}
\end{equation}
\end{subequations}

Note that $\Phi_{\rm adj}$, $\Psi_{\rm adj}$ and $\eta_{\rm adj}$ in the quasi-isomorphism~\eqref{eq:quasi-iso_adjusted_Weil_loop} restrict to maps between horizontal elements and therefore the same arguments as in the skeletal case apply:
\begin{proposition}
 The differential graded algebras $\inv(\frg)$ and $\inv(\aghl)$ are quasi-isomorphic and the vector spaces $\overline{\inv}(\frg)$ and $\overline{\inv}(\aghl)$ are isomorphic.
\end{proposition}
Finally, we note that the above adjustment in the loop case mirrors the adjustment of the skeletal case in that it consists of replacing
\begin{equation}
\begin{aligned}
Q_\sW r &\mapsto Q_{\sW_{\rm adj}}r \coloneqq Q_\sW r \mp \chi(t,\hat t)~,\\
Q_\sW \hat r &\mapsto Q_{\sW_{\rm adj}}\hat r \coloneqq \pm\chi(\hat t,\hat t) - \hat q~,
\end{aligned}
\end{equation}
where in the skeletal case $\chi$ corresponds to the Killing form and in the loop case $\chi$ is given in~\eqref{eq:chi_loop_case}. The corresponding functions $\chi_{\rm sk}$ and $\chi_{\rm lp}$ on the dual $L_\infty$-algebras are related to the cocycles that are trivialized in $\astringsk(\frg)$ and $\astringl(\frg)$, respectively, via $\chi_{\rm sk} \circ \mu_2 = \mu_3$ and $\chi_{\rm lp} \circ \mu_1 = \mu_2$.

\subsection{Adjusted Weil algebra of a general Lie 2-algebra}

An obvious question is now whether the string Lie 2-algebra plays a special role, or whether one can define adjusted Weil algebras leading to compatibility of invariant polynomials with quasi-isomorphisms in general. The evident generalization of~\eqref{eq:astringsk_ses} for a general Lie 2-algebra $\frg=(\frg_{-1}\rightarrow \frg_0)$ is to use the decomposition~\eqref{eq:LAsplitting} and consider the short exact sequence of $L_\infty$-algebras
\begin{equation}
 0~\longrightarrow~(\frg^0_{-1}\rightarrow \frg^0_0)~\hooklongrightarrow~\hat \frg^0_0\coloneqq (\frg_{-1}^0[1]\,\embd\,\frg_{-1}\rightarrow \frg_0)~\longrightarrow~ (\frg_{-1}^0\rightarrow * \rightarrow *)~\longrightarrow~0
\end{equation}
with $\hat \frg^0_0\approxeq \frg^0_0$. Explicitly, we have the following decomposition:
\begin{equation}
 \begin{tikzcd}[row sep=0cm,column sep=2cm]
  \frg_{-1}^0[1] \arrow[r]{}{\mu_1=\id} & \frg_{-1}^0 & \frg_0^0\\
  & \oplus & \oplus\\
  & \frg_{-1}^1 \arrow[r]{}{\mu_1=\id} & \frg_0^1
 \end{tikzcd}
\end{equation}
with corresponding coordinate functions $t^\alpha\in\frg_0^0$, $t^a\in\frg_0^1$, $r^i\in \frg^0_{-1}$, $r^a\in\frg^1_{-1}$ and $q^i\in\frg_{-1}^0[1]$ of degrees $0$, $0$, $1$, $1$ and $2$, respectively. The (unadjusted) Weil algebra reads as 
\begin{equation}
 \begin{aligned}
  Q_\sW~&:~&t^\alpha&\mapsto-\tfrac12 \mu^\alpha_{\beta\gamma}t^\beta t^\gamma+\hat t^\alpha~,~~~&t^a&\mapsto -r^a+\hat t^a~,\\
  &&r^i&\mapsto \tfrac1{3!} \mu^i_{\alpha\beta\gamma}t^\alpha t^\beta t^\gamma-\mu^i_{\alpha j}t^\alpha r^j+q^i+\hat r^i~,~~~&r^a&\mapsto\hat r^a~,\\
  &&q^i&\mapsto -\mu^i_{\alpha j}t^\alpha q^j+\hat q^i~,\\
  &&\hat t^\alpha&\mapsto-\mu^\alpha_{\beta\gamma}t^\beta \hat t^\gamma~,~~~&\hat t^a&\mapsto \hat r^a~,\\
  &&\hat r^i&\mapsto -\tfrac1{2} \mu^i_{\alpha\beta\gamma}t^\alpha t^\beta \hat t^\gamma-\mu^i_{\alpha j}t^\alpha \hat r^j+\mu^i_{\alpha j}\hat t^\alpha r^j-\hat q^i~,~~~&\hat r^a&\mapsto0~,\\
  &&\hat q^i&\mapsto -\mu^i_{\alpha j} t^\alpha \hat q^j+\mu^i_{\alpha j}\hat t^\alpha q^j~,
 \end{aligned}
\end{equation}
and the dual quasi-isomorphism to $\sW(\frg^0_0)$ is given by maps 
\begin{subequations}\label{eq:quasi-iso_unadjusted_general_Weil}
\begin{equation}
 \begin{tikzcd}
    \sW(\hat \frg_0^0) \arrow[r,bend left]{}{\hat\Phi} & \sW(\frg_0^0) \arrow[l,bend left]{}{\hat\Psi}
 \end{tikzcd}~,~~~\eta_{\hat\Psi\circ\hat\Phi}:\hat\Psi\circ \hat\Phi\Rightarrow \id_{\sW(\hat \frg_0^0)}~,
\end{equation}
\begin{equation}
\begin{aligned}
\hat\Phi&:&t^{\alpha}&\mapsto \tilde t^\alpha~,
&t^a,r^i,r^a&\mapsto 0~,~
&q^i&\mapsto -\tfrac{1}{3!} \mu^i_{\alpha\beta\gamma}\tilde t^\alpha \tilde t^\beta \tilde t^\gamma~,\\
&
&\hat t^{\alpha}&\mapsto \hat{\tilde{t}}^\alpha~,
&\hat t^a, \hat r^i, \hat r^a&\mapsto 0~,~
&\hat q^i&\mapsto -\tfrac12 \mu_{\alpha\beta\gamma}^i \tilde t^\alpha \tilde t^\beta \hat{\tilde t}^\gamma~,\\
\hat\Psi&:& \tilde t^\alpha&\mapsto  t^{\alpha}~,~~~&\hat{\tilde{t}}^\alpha&\mapsto  \hat t^{\alpha}~,\\
\eta_{\hat\Psi\circ\hat\Phi}&:&t^{\alpha},t^a,r^i&\mapsto 0~,~&r^a&\mapsto t^a~,&q^i&\mapsto -r^i~,\\
&&\hat t^\alpha, \hat t^a,\hat r^i &\mapsto 0~,~&\hat r^a&\mapsto -\hat t^a~,~~~&\hat q^i&\mapsto \hat r^i~.
\end{aligned}
\end{equation}
\end{subequations}
As in the example~\eqref{eq:quasi-iso_ordinary_Weil}, the quasi-isomorphism $\Phi$ does not restrict to a map $\Phi:\sW_{\rm h}(\hat \frg_0^0)\to \sW_{\rm h}(\frg_0^0)$ and the unadjusted Weil algebra is not suitable for a definition of invariant polynomials.

An adjusted form $\sW_{\rm adj}(\hat \frg_0^0)$ of the Weil algebra  is readily found:
\begin{equation}
 \begin{aligned}
  Q_{\sW_{\rm adj}}~&:~&t^\alpha&\mapsto-\tfrac12 \mu^\alpha_{\beta\gamma}t^\beta t^\gamma+\hat t^\alpha~,~~~&t^a&\mapsto -r^a+\hat t^a~,\\
  &&r^i&\mapsto \tfrac1{3!} \mu^i_{\alpha\beta\gamma}t^\alpha t^\beta t^\gamma-\mu^i_{\alpha j}t^\alpha r^j+q^i+\hat r^i~,~~~&r^a&\mapsto\hat r^a~,\\
  &&q^i&\mapsto -\mu^i_{\alpha j}t^\alpha q^j-\tfrac12 \mu^i_{\alpha\beta\gamma} t^\alpha t^\beta \hat t^\gamma-\mu^i_{\alpha j}\hat t^\alpha r^j+\hat q~,~~~\\
  &&\hat t^\alpha&\mapsto-\mu^\alpha_{\beta\gamma}t^\beta \hat t^\gamma~,~~~&\hat t^a&\mapsto \hat r^a~,\\
  &&\hat r^i&\mapsto -\mu^i_{\alpha j}t^\alpha \hat r^j-\hat q^i~,~~~&\hat r^a&\mapsto0~,\\
  &&\hat q^i&\mapsto -\mu^i_{\alpha j} t^\alpha \hat q^j+\mu^i_{\alpha j}\hat t^\alpha \hat r^j~,
 \end{aligned}
\end{equation}
and the dual quasi-isomorphism to $\sW(\frg^0_0)$ is given by maps 
\begin{subequations}\label{eq:quasi-iso_adjusted_general_Weil}
\begin{equation}
 \begin{tikzcd}
    \sW_{\rm adj}(\hat \frg_0^0) \arrow[r,bend left]{}{\Phi_{\rm adj}} & \sW(\frg_0^0) \arrow[l,bend left]{}{\Psi_{\rm adj}}
 \end{tikzcd}~,~~~\eta_{\rm adj}:\Psi_{\rm adj}\circ \Phi_{\rm adj}\Rightarrow \id_{\sW_{\rm adj}(\hat \frg_0^0)}~,
\end{equation}
\begin{equation}
\begin{aligned}
\Phi_{\rm adj}&:&t^{\alpha}&\mapsto \tilde t^\alpha~,
&t^a,r^i,r^a&\mapsto 0~,~
&q^i&\mapsto -\tfrac{1}{3!} \mu^i_{\alpha\beta\gamma}\tilde t^\alpha \tilde t^\beta \tilde t^\gamma~,\\
&
&\hat t^{\alpha}&\mapsto \hat{\tilde{t}}^\alpha~,
&\hat t^a, \hat r^i, \hat r^a&\mapsto 0~,~
&\hat q^i&\mapsto 0~,\\
\Psi_{\rm adj}&:& \tilde t^\alpha&\mapsto  t^{\alpha}~,~~~&\hat{\tilde{t}}^\alpha&\mapsto  \hat t^{\alpha}~,\\
\eta_{\rm adj}&:&t^{\alpha},t^a,r^i&\mapsto 0~,~&r^a&\mapsto t^a~,&q^i&\mapsto -r^i~,\\
&&\hat t^\alpha, \hat t^a,\hat r^i &\mapsto 0~,~&\hat r^a&\mapsto -\hat t^a~,~~~&\hat q^i&\mapsto \hat r^i~.
\end{aligned}
\end{equation}
\end{subequations}
Here, we indeed have the restriction $\Phi_{\rm adj}:\sW_{\rm adj,h}(\hat \frg_0^0)\to \sW_{\rm h}(\frg_0^0)$, a necessary condition for compatibility of the quasi-isomorphism with the dga of invariant polynomials. Note that contrary to the adjusted Weil algebra $\sW_{\rm adj}(\aghsk)$, however, $q^i$ and $\hat q^i$ do not generate a differential ideal, in general. In particular, $Q_{\sW_{\rm adj}}$ does not close on the subspace generated by $q^i$ and $\hat q^i$. The reduction of $\hat \frg^0_0$ to the Lie 2-algebra $\frg$, which would correspond to quotienting the Weil algebra of $\hat \frg^0_0$ by this non-existing differential ideal is therefore not possible.

Without additional structure on $\frg^0_0$, there is no other adjustment or quasi-isomorphism that leads to the desired differential ideal. We thus recognize that the string Lie 2-algebra is special as a Lie 2-algebra in that it admits an adjusted Weil algebra.

\section{Higher gauge theory}\label{sec:HGT}

We now come to the discussion of higher gauge theory based on the structures introduced in the previous sections. The framework we use is based on ideas due to H.~Cartan~\cite{Cartan:1949aaa,Cartan:1949aab} and it is closely related to the Atiyah algebroid~\cite{Atiyah:1957} and the Alexandrov, Kontsevich, Schwarz, Zaboronsky (AKSZ) construction~\cite{Alexandrov:1995kv}. It was rediscovered in various forms and extended to higher form gauge potentials in the context of high-energy physics~\cite{D'Auria:1982nx,Castellani:1991et,Bojowald:0406445,Kotov:2007nr,Gruetzmann:2014ica}, and then extended to a rather full picture for higher gauge theory in~\cite{Sati:2008eg}, cf.~also~\cite{Fiorenza:2011jr}.

\subsection{Basic idea}\label{ssec:HGT_basic_ideas}

The local kinematical data of a gauge theory on some contractible patch $U$ of a space-time manifold with structure or gauge Lie algebra $\frg$ 
consists of a $\frg$-valued one-form $A\in \Omega^1(U,\frg)$ called the {\em gauge potential}, its {\em curvature} $F\in \Omega^2(U,\frg)$, which satisfies a {\em Bianchi identity}, and a Lie algebra of (infinitesimal) {\em gauge transformations} which act on $A$ and $F$. In higher gauge theories, we have correspondingly higher differential forms taking values in higher Lie algebras modeled by $L_\infty$-algebras.

The language of Chevalley--Eilenberg algebras introduced above presents a gauge $L_\infty$-algebras in terms of a differential graded algebra, putting them on equal footing with the other key ingredient in the kinematical data of a gauge theory, the differential forms. This unifying framework leads to a formulation of the local description of a gauge theory in terms of morphisms of differential graded algebras, which can be generalized vastly.

We now go through a reformulation of the kinematical data of ordinary gauge theories, which is readily extended to higher gauge theories. Explicitly, let $U$ be a contractible patch of our space-time and let $\frg$ be an ordinary Lie algebra, as above. We use again coordinate functions $t^\alpha$ and $\hat t^\alpha$ on the Weil algebra $\sW(\frg)$, cf.~equation~\eqref{eq:Weil_of_ordinary_Lie}. The local kinematical data of a gauge theory with structure Lie algebra $\frg$ on $U$ is given by a dga-morphism\footnote{The set $\Omega^\bullet(U)$ is the Chevalley--Eilenberg algebra of the tangent algebroid $T[1]U$ and it can also be regarded as the Weil algebra of the trivial Lie algebroid $U$ so that $\CA:\sW(\frg)\rightarrow \sW(U)$.}
\begin{equation}
\begin{gathered}
\begin{tikzcd}[column sep=2cm]
\Omega^\bullet(U) & \sW(\frg)\arrow[l,"{\CA}",swap]
\end{tikzcd}~,\\
 \CA~:~t^\alpha \mapsto A^\alpha\in \Omega^1(U)~,~~~\hat t^\alpha \mapsto F^\alpha\in \Omega^2(U)~.
\end{gathered}
\end{equation}
Compatibility with the differentials in both dgas, $\dd\circ \CA=\CA\circ Q$, implies that
\begin{equation}
F^\alpha = \dd A^\alpha + \tfrac12 f^\alpha_{\beta\gamma} A^\beta   A^\gamma\eand
\dd F^\alpha = -f^\alpha_{\beta\gamma} A^\beta F^\gamma~.
\end{equation}
We thus recovered the gauge potential, its curvature and the corresponding Bianchi identity. Note that if had we merely considered morphisms $\CA:\sCE(\frg)\rightarrow \Omega^\bullet(U)$, we would have only recovered {\em flat} connections.

Gauge transformations are encoded in flat homotopies between two such gauge configurations~\cite{Fiorenza:2010mh}, i.e.~in dga-morphisms
\begin{equation}
\begin{tikzcd}[column sep=2cm]
\Omega^\bullet(U\times I) & \sW(\frg)\rlap{~,} \arrow[l,"\check\CA",swap]
\end{tikzcd}
\end{equation}
where $I=[0,1]$ denotes the interval and flatness means that $\check{F}$ vanishes when contracted with vector fields along $I$. The differential on $\Omega^\bullet(U\times I)$ is given by $\dd_{U\times I} = \dd_U + \dd \tau \der{\tau}$, where $\tau$ is the additional coordinate on $I$, and we have the morphism
\begin{equation}
  \check{\CA}~:~t^\alpha \mapsto A^\alpha + \Lambda_0^\alpha \dd \tau\in \Omega^1(U)\oplus\Omega^1(I)~,~~~\hat t^\alpha \mapsto F^\alpha_U\in \Omega^2(U)~.
\end{equation}
The infinitesimal gauge transformations are then parametrized by $\Lambda_0 \in \Omega^0(U)\otimes \frg$ and act according to
\begin{equation}
 \delta A^\alpha=\der{\tau} \check A^\alpha|_{\tau=0}\eand \delta F^\alpha=\der{\tau} \check F^\alpha|_{\tau=0}~.
\end{equation}
The compatibility of $\check\CA$ with $\dd_{U\times I}$ and $Q$ yields
\begin{equation}\label{eq:old_gauge_trafo}
\delta A^\alpha = \dd \Lambda^\alpha_0 + f^\alpha_{\beta\gamma} A^\beta \Lambda^\gamma_0\eand
\delta F^\alpha = f^\alpha_{\beta\gamma}F^\beta \Lambda_0^\gamma~.
\end{equation}
Therefore, the gauge transformations of $A$ and $F$ are determined by the expression for $F$ and the form of the Bianchi identity.

Finally, note that an invariant polynomial $p\in \inv(\frg)\subset \sW_{\rm h}(\frg)$ is mapped to a polynomial $\CA(p)$ in the curvature $F$, and the condition $Q_\sW p=q\in \sW_{\rm h}(\frg)$ ensures that $\dd \CA(p)$ is again a polynomial $\CA(q)$ in $F$, which implies that $\CA(p)$ is gauge invariant:
\begin{equation}
 \delta \CA(p)=\der{\tau} \check \CA(p)|_{\tau=0}=\iota_{\der{\tau}}\check \CA(q)|_{\tau=0}=0~.
\end{equation}
That is, the topological observables are the images of the invariant polynomials $\inv(\frg)$ under $\CA$. The topological invariants are the images of $\overline{\inv}(\frg)$.

\subsection{BRST complex from an AKSZ-like construction}\label{ssec:BRST_complex}

Let us provide a convenient and closely related description of local $\frg$-connection objects which follows the Alexandrov, Kontsevich, Schwarz, Zaboronsky (AKSZ) construction~\cite{Alexandrov:1995kv}. The advantage of this construction is that we obtain the full gauge algebroid rather directly, in the form of a BRST complex. Mathematically, we construct the Chevalley--Eilenberg algebra of the gauge $L_\infty$-algebroid of the kinematical data, cf.~e.g.~the discussion in~\cite{Jurco:2018sby,Jurco:2019bvp}.

Recall that the input data of the AKSZ construction consists of two differential graded manifolds, the {\em source} $(\Sigma,\dd_\Sigma)$ and the {\em target} $(X,\dd_X)$, where the source is endowed with an additional measure $\mu$ and the target is endowed with a symplectic form $\omega$. The space of fields is the space of {\em not necessarily grade preserving} maps $\CCA\in\sMap(\Sigma,X)$, which carry a $\RZ$-grading and contain the morphisms of graded manifolds $\sHom(\Sigma,X)$ in degree~0, see e.g.~\cite[Section~5.1]{Cattaneo:2010re} for more details on this point. To restore compatibility with the grading, one introduces the ghost degree, as discussed in more detail in the example below. The measure $\mu$ and the symplectic form $\omega$ induce a symplectic structure on $\sMap(\Sigma,X)$ and the differentials on $\Sigma$ and $X$ linearly combine to a differential on the graded space $\sMap(\Sigma,X)$, which is Hamiltonian with respect to the Poisson bracket induced by $\omega$ on $\sMap(\Sigma,X)$. We thus obtain the BV complex of a field theory. This structure induces an action for a topological field theory and different choices of $\Sigma$ and $X$ lead e.g.~to Chern--Simons theory and its higher variants, to BF-theories, Poisson-sigma models, etc. Since we are merely interested in the kinematical data, this part of the AKSZ construction is irrelevant for us.

Instead, let us construct the Chevalley--Eilenberg algebra of the gauge $L_\infty$-algebroid of local $\frg$-connection objects for some $L_\infty$-algebra $\frg$. This is usually called the BRST complex. Recall that the gauge $L_\infty$-algebroid consists of the fields, the (higher) Lie algebra of (higher) gauge transformations and the higher products encode the actions of gauge transformations on the fields as well as the compositions of (higher) gauge transformations.

Since we are merely interested in local gauge theories, we choose $\Sigma=T[1]U$ for $U$ some contractible patch of our manifold $M$. As target $X$ for topological $\frg$-connections, one would usually take the dg-manifold $\frg[1]$ corresponding to the $L_\infty$-algebra $\frg$. This will yield flat connections, as mentioned previously in section~\ref{ssec:HGT_basic_ideas}. For our purposes, we enlarge\footnote{The same enlarged target was considered in~\cite{Fiorenza:2011jr} but with a focus on maps of degree~0.} $X$ to the dg-manifold $\inn(\frg)[1]$ corresponding to $\inn(\frg)$. The morphisms of dg-manifolds $\Sigma\rightarrow X$ are now precisely the duals of the morphisms $\sW(\frg)\rightarrow \Omega^\bullet(U)$ and encode $\frg$-connection objects, as discussed above.

If $X=\frg[1]$ was the dg-manifold corresponding to $\frg$, we would recover the BV complex of a corresponding AKSZ model from the general morphisms $\sMap(\Sigma,X)$. Recall that we are only interested in the kinematical data and its gauge transformation, i.e.~the BRST complex contained in the BV complex. We therefore put all generators corresponding to antifields (as well as antifields of ghosts and higher ghosts) to zero, as they encode equations of motion and corresponding Noether identities, cf.~\cite{Jurco:2018sby}. We also have to put to zero the generators corresponding to additional gauge parameters arising from doubling $\frg[1]$ to $\inn(\frg)[1]$. The result is the BRST complex encoding the kinematical data of our gauge theory.

Let us illustrate this for ordinary gauge theory with $\frg$ an ordinary Lie algebra. We thus consider maps
\begin{equation}
 \CCA:\sW(\frg)\rightarrow \Omega^\bullet(U)~,
\end{equation}
which describe $\frg$-valued differential forms on $U$, and we decompose them according to the form degree of the image:
\begin{equation}
\begin{aligned}
 \CCA(t^\alpha)&=\Lambda^\alpha_0+A^\alpha+A^{+\alpha}+\Lambda^{+\alpha}_0+\dots~,\\
 \CCA(\hat t^\alpha)&=\vartheta^\alpha_0+\vartheta^\alpha_1+F^\alpha+F^{+\alpha}+\dots~.
\end{aligned}
\end{equation}
To preserve the grading, we associate an additional degree called {\em ghost degree} to each component and call the resulting space $\sMap(\Sigma,X)$. The components then have the following interpretation:
\begin{center}
\begin{tabular}{@{}lccl@{}}
\toprule
   component & form degree & ghost degree & interpretation \\
\midrule
   $\Lambda_0$ & 0 & 1 & ghost or gauge parameter\\
   $A$ & 1 & 0 & local gauge potential 1-form\\
   $A^+$ & 2 & $-1$ & antifield of $A$, put to zero\\
   $\Lambda^+_0$ & 3 & $-2$ & antifield of $\Lambda_0$, put to zero\\
   $\vartheta^\alpha_0$ & 0 & 2 & additional gauge parameter, put to zero\\
   $\vartheta^\alpha_1$ & 1 & 1 & additional gauge parameter, put to zero\\
   $F$ & 2 & 0 & curvature of $A$\\
   $F^+$ & 3 & $-1$ & antifield of $F$, put to zero\\
\bottomrule
\end{tabular}
\end{center}
As usual in the AKSZ-formalism, $Q_{\rm BRST}$ on $\sMap(\Sigma,X)$ is induced by a linear combination of the precomposition of the map with $Q_\sW$ and the postcomposition of the map with the de Rham differential $\dd$:
\begin{equation}\label{eq:Q_BRST}
 Q_{\rm BRST}\CCA\coloneqq \dd\circ \CCA-\CCA\circ Q_\sW~,~~~\CCA\in \sMap(\Sigma,X)~.
\end{equation}
Decomposing again by form degree, we obtain
\begin{equation}
\begin{aligned}
 Q_{\rm BRST}~&:~&\Lambda_0^\alpha&\mapsto \tfrac12 f^\alpha_{\beta\gamma}\Lambda_0^\beta \Lambda_0^\gamma-\vartheta_0^{\alpha}~,\\
 &&A^\alpha &\mapsto \dd \Lambda_0^\alpha+f^\alpha_{\beta\gamma}A^\beta\Lambda_0^\gamma-\vartheta_1^\alpha~,\\
 &&A^{+\alpha}&\mapsto \dd A^\alpha+f^\alpha_{\beta\gamma}(\Lambda_0^\beta A^{+\gamma}+\tfrac12 A^\beta A^\gamma)-F^\alpha~,\\
 &&\Lambda_0^{+\alpha}&\mapsto \dd A^{+\alpha}+f^\alpha_{\beta\gamma}(\Lambda_0^\beta \Lambda_0^{+\gamma}+A^\beta A^{+\gamma})-F^{+\alpha}~,\\
 &&\vartheta^\alpha_0&\mapsto f^\alpha_{\beta\gamma}\Lambda_0^\beta \vartheta^\gamma_0~,\\
 &&\vartheta^\alpha_1&\mapsto \dd \vartheta^\alpha_0+f^\alpha_{\beta\gamma}(\Lambda_0^\beta\vartheta_1^\gamma+A^\beta \vartheta_0^\gamma)~,\\
 &&F^\alpha&\mapsto \dd \vartheta^\alpha_1+f^\alpha_{\beta\gamma}(A^\beta\vartheta_1^\gamma+A^{+\beta}\vartheta_0^\gamma-F^\beta\Lambda_0^\gamma)~,\\
 &&F^{+\alpha}&\mapsto \dd F^\alpha+f^\alpha_{\beta\gamma}(\Lambda_0^\beta F^{+\alpha}+A^\beta F^\gamma+A^{+\beta}\vartheta_1^\gamma+\Lambda_0^{+\beta}\vartheta_0^\gamma)~.
\end{aligned}
\end{equation}
Putting all antifields and additional gauge parameters to zero, we recover the fields and gauge parameters of the kinematical data of ordinary gauge theory as well as
\begin{equation}
\begin{gathered}
 F=\dd A+\tfrac12 [A,A]~,~~~\dd F+[A,F]=0~,\\
 \delta A=Q_{\rm BRST}A=\dd \Lambda_0+[A,\Lambda_0]~,~~~\delta F=Q_{\rm BRST} F= -[F,\Lambda_0]~.
\end{gathered}
\end{equation}
Note that here $\delta A$ and $\delta F$ are of ghost degree one and, thus, may differ in signs from the expressions found in~\eqref{eq:old_gauge_trafo}. To recover these more familiar expressions, one can pull out the ghost degree to either side which fixes the discrepancies. We will refrain from doing so in the following and provide the infinitesimal gauge transformations as found from the BRST complex.

\subsection{Equivalence of gauge theory kinematical data}

Our construction in the last section also allows us to explore possible equivalences of general gauge theories. As discussed in~\cite{Jurco:2018sby}, any Lagrangian field theory gives rise to a BV-complex, which is the dual of an $L_\infty$-algebra. Two field theories are then physically equivalent if there is a quasi-isomorphism between them. This requires, in particular, that gauge equivalent field configurations are mapped to gauge equivalent field configurations and thus that the $L_\infty$-algebras encoding the kinematical data of equivalent field theories are quasi-isomorphic. In other words, if closed, the BRST complexes of equivalent field theories are dually quasi-isomorphic.

Consider now two quasi-isomorphic $L_\infty$-algebras $\frg$ and $\tilde \frg$. Their Chevalley--Eilenberg algebras are dually quasi-isomorphic, which lifts to a dual quasi-isomorphism of the Weil algebras $(\Phi,\Psi,\eta_{\Psi\circ \Phi},\eta_{\Phi\circ \Psi}): \sW(\frg)\approxeq \sW(\tilde \frg)$. A morphism $\CCA:\sW_{\rm adj}(\frg)\rightarrow \Omega^\bullet(U)$ encoding kinematical data induces a morphism $\tilde \CCA:\sW_{\rm adj}(\tilde \frg)\rightarrow \Omega^\bullet(U)$ by pullback:
\begin{equation}\label{eq:pullbacks_g_connection_objects}
 \begin{tikzcd}[row sep=0.5cm,column sep=2cm]
  \sW_{\rm adj}(\frg) \arrow[dd,bend left]{}{\Phi} \arrow[dr]{}{\CCA}\\
  & \Omega^\bullet(U) \\
  \sW_{\rm adj}(\tilde \frg)\arrow[uu,bend left]{}{\Psi} \arrow[ur,dashed,swap]{}{\tilde \CCA=\Psi^*\CCA =\CCA\circ \Psi}
 \end{tikzcd}
\end{equation}
Moreover, the dual quasi-isomorphism of Weil algebras $(\Phi,\Psi,\eta_{\Psi\circ \Phi},\eta_{\Phi\circ \Psi})$ induces a full quasi-isomorphism between the BRST $L_\infty$-algebroids. 

One may now think that a general dual quasi-isomorphism of Weil algebras $(\Phi,\Psi,\eta_{\Psi\circ \Phi},$ $\eta_{\Phi\circ \Psi}): \sW(\frg)\approxeq \sW(\tilde \frg)$ leads to a quasi-isomorphism between the two BRST $L_\infty$-algebroids. This, however, is not true: recall that in our construction of the BRST complex, we put certain gauge parameters equal to zero. General dual quasi-isomorphisms of Weil algebras will mix gauge parameters with gauge fields and therefore change, in general, the structure of gauge transformations. This can also be seen from another perspective: general dual quasi-isomorphisms of the Weil algebras will mix gauge fields with curvatures, which induces a change of the notion of flatness. Since gauge transformations are partially flat homotopies, this leads to inequivalent gauge transformations. We have, in fact, seen an example of this in our discussion of unadjusted and adjusted Weil algebras. 

To preserve the gauge structure, the dual quasi-isomorphism between Weil algebras has to restrict to a dual quasi-isomorphism of the underlying Chevalley--Eilenberg algebras, which is equivalent to a quasi-isomorphism of the gauge $L_\infty$-algebras. We thus have the following statement:
\begin{theorem}\label{thm:equivalence_of_field_theories}
    The kinematical data of two gauge field theories are equivalent, if and only if their gauge $L_\infty$-algebras are quasi-isomorphic.
\end{theorem}

\subsection{Unadjusted Weil algebras lead to fake flatness}\label{ssec:unadjusted_lead_to_fake_flatness}

Let us now consider the kinematical data of higher gauge theory for a generic Lie 2-algebra $\frg=(\frg_{-1}\rightarrow \frg_0)$ with Chevalley--Eilenberg algebra~\eqref{eq:CE_Lie_2_algebra} and Weil algebra~\eqref{eq:Weil_algebra_generic_Lie_2}, as derived by the truncated enlarged AKSZ-construction presented in the previous section. That is, we have a map
\begin{equation}
\begin{aligned}
 \CCA(t^\alpha)&=\Lambda^\alpha_0+A^\alpha+A^{+\alpha}+\Lambda^{+\alpha}_0+\dots~,\\
 \CCA(\hat t^\alpha)&=\vartheta^\alpha_0+\vartheta^\alpha_1+F^\alpha+F^{+\alpha}+\dots~,\\
 \CCA(r^a)&=\Sigma_0^a+\Lambda_1^a+B^a+B^{+a}+\Lambda_1^{+a}+\dots~,\\
 \CCA(\hat r^a)&=\Theta^a_0+\Theta^a_1+\Theta^a_2+H^a+H^{+a}+\dots~,
\end{aligned}
\end{equation}
where we decompose $\CCA$ now in components and subspaces of $\frg$. The action of the differential~\eqref{eq:Q_BRST} is readily computed and we recover the kinematical data of higher gauge theory, as expected:
\begin{equation}
 \begin{aligned}
  \delta A&=\dd \Lambda_0+\mu_2(A,\Lambda_0)+\mu_1(\Lambda_1)~,~&\delta B&=\dd \Lambda_1+\mu_2(A,\Lambda_1)+\mu_2(\Lambda_0,B)-\tfrac12\mu_3(A,A,\Lambda_0)~,\\
  F&=\dd A+\tfrac12 \mu_2(A,A)+\mu_1(B)~,~&H&=\dd B+\mu_2(A,B)-\tfrac1{3!}\mu_3(A,A,A)~,\\
  \dd F&=-\mu_2(A,F)+\mu_1(H)~,~&\dd H&=-\mu_2(A,H)+\mu_2(F,B)-\tfrac12\mu_3(A,A,F)~.
 \end{aligned}
\end{equation}
The truncation $\Theta_2^a=0$, however, also produces the equation
\begin{equation}
 -\tfrac12 \mu_3(\Lambda_0,\Lambda_0,F)+\mu_2(F,\Sigma_0)=0~.
\end{equation}
As $\Lambda_0$ and $\Sigma_0$ are independent gauge parameters it follows that both $\mu_2(F,\Sigma_0)=0$ and $\mu_3(\Lambda_0,\Lambda_0,F)=0$, which requires a restriction of the gauge transformations or of $F$ in order to ensure that $Q_{\rm BRST}$ is well-defined. The nilpotency $Q_{\rm BRST}^2=0$ then follows from the definition of $Q_{\rm BRST}$. This point is familiar from the quantization of field theories with 2-form potentials, where such a BRST algebra is called {\em open} and considered to close only up to equations of motion $F=0$.

The curvature $F$ was called {\em fake curvature} in the original paper on non-abelian\linebreak gerbes~\cite{Breen:math0106083}, and the requirement of vanishing fake curvature is ubiquitous. One might think that a transition to strict Lie 2-algebras, for which $\mu_3$ vanishes, together with ignoring the higher gauge transformations parameterized by $\Sigma_0$ might solve the issue. This, however, is not the case. The fake curvature condition $F=0$ then arises generically in the composition of finite gauge transformations. Moreover, it has been noted that the fake curvature condition is a requirement for the parallel transport described by the connection $(A,B)$ to be reparametrization invariant~\cite{Baez:2004in}.

This is in fact a rather generic feature of higher gauge theories constructed from unadjusted Weil algebras, cf.~\cite{Jurco:2018sby}. The BRST complex of a higher gauge theory based on a Lie $n$-algebra for $n>1$ is open and requires the fake curvatures, which are all curvature forms except for the top one, to vanish.

We can now finally define the notion of adjusted Weil algebra:
\begin{definition}\label{def:adjusted_Weil_algbra}
 We call a Weil algebra \underline{adjusted}, if the resulting BRST complex for $\frg$-connection objects is closed. That is, $Q_{\rm BRST}^2=0$ without invoking any equations of motion.
\end{definition}
We will find in section~\ref{ssec:local_connections_ordinary_string_structures} that the Weil algebras introduced in sections~\ref{ssec:Weil_adj_sk} and~\ref{ssec:Weil_adj_l} are indeed adjusted in this sense.

\subsection{Fake flat higher gauge theories are locally abelian}\label{ssec:fake_flat_trivial}

While vanishing of the fake curvatures is clearly not an issue for higher versions of Chern--Simons theories, it is highly problematic in the context of higher gauge theories with locally non-vanishing curvatures, cf.~\cite{Demessie:2016ieh,Saemann:2019leg}. To underline this point, let us give an explicit geometrical proof that fake-flat kinematical data of a higher gauge theory based on 2-groups is locally abelian. A detailed analytical proof has been given before in~\cite{Gastel:2018joi}. For simplicity, we assume that we can use categorical equivalence to strictify the gauge 2-group. That is, we work with the non-abelian gerbes first introduced in the literature in~\cite{Breen:math0106083,Aschieri:2003mw} and ignore the (categorically equivalent) generalizations, as e.g.~the ones of~\cite{Jurco:2014mva}.

We thus start from a strict gauge Lie 2-group modeled by a crossed module of Lie groups $\CG=(\sH\xrightarrow{~\dpar~}\sG)$. That is, we have two Lie groups $\sH$ and $\sG$, a group homomorphism $\dpar:\sH\rightarrow \sG$, and an action $\acton:\sG\times \sH\rightarrow \sH$ such that
\begin{equation}\label{eq:xmrelations}
 \dpar(g\acton h_1)=g\dpar(h_1)g^{-1}\eand \dpar(h_1)\acton h_2=h_1h_2h_1^{-1}
\end{equation}
for all $g\in \sG$ and $h_{1,2}\in \sH$. The second relation implies that $\ker(\dpar)$ is an abelian subgroup of $\sH$, because the group commutator reads as
\begin{equation}
 [h_1,h_2]\coloneqq h_1 h_2 h_1^{-1} h_2^{-1}=\big(\dpar(h_1)\acton h_2\big) h_2^{-1}=h_1\big(\dpar(h_2)\acton h_1^{-1}\big)~,~~~h_i\in\sH~.
\end{equation}
The first relation in~\eqref{eq:xmrelations} implies that $\im(\dpar)$ is a normal subgroup of $\sG$ and therefore $\sG$ is a principal $\im(\dpar)$-bundle over $\sG^0\coloneqq {\rm coker}(\dpar)=\sG/\im(\dpar)$, which is a group with product induced by that on $\sG$:
\begin{equation}
g_1\dpar(h_1)g_2\dpar(h_2)=g_1g_2\dpar(g^{-1}_2\acton h_1)\dpar(h_2)=g_1g_2\dpar((g^{-1}_2\acton h_1)h_2)\sim g_1 g_2~,
\end{equation}
where $g_i\in \sG$ and $h_i\in \sH$. We also note that a group commutator with an element in $\im(\dpar)$ takes values in $\im(\dpar)$:
\begin{equation}\label{eq:group_commutator}
 [\dpar(h),g]=\dpar(h)g\dpar(h^{-1})g^{-1}=\dpar(h)\dpar(g\acton h^{-1})=\dpar(h(g\acton h^{-1}))\in \im(\dpar)
\end{equation}
for all $g\in \sG$ and $h\in \sH$.

A crossed module of Lie groups differentiates to a crossed module of Lie algebras which in turn corresponds to a strict Lie 2-algebra
\begin{equation}
 \frg~=~(\,\frg_{-1}\rightarrow \frg_0\,)~=~(\,\sLie(\sH)\rightarrow \sLie(\sG)\,)~,
\end{equation}
where $\mu_1$ is the differential of $\dpar$ and $\mu_2$ arises from the commutator on $\sLie(\sG)$ and the action of $\sLie(\sG)$ on $\sLie(\sH)$. We again have the exact sequence
\begin{equation}\label{eq:es_cm}
 0\longrightarrow\ker(\mu_1) \hooklongrightarrow \frg_{-1}\xrightarrow{~\mu_1~} \frg_0 \xrightarrow{~\pi~} {\rm coker}(\mu_1)\longrightarrow 0~,
\end{equation}
cf.~\eqref{eq:es_Lie2}, where ${\rm coker}(\mu_1)$ is a Lie algebra.

Recall from our discussion above that a $\frg$-connection object over a contractible patch $U$ of some manifold $M$ is locally given by a 1- and a 2-form,
\begin{equation}
 A\in \Omega^1(U)\otimes \frg_0\eand B\in \Omega^2(U)\otimes \frg_{-1}
\end{equation}
with curvatures
\begin{equation}
 F=\dd A+\tfrac12 \mu_2(A,A)+\mu_1(B)=0\eand H=\dd B+\mu_2(A,B)~.
\end{equation}
Finite gauge transformations are parametrized by $g\in \Omega^0(U,\sG)$ and $\Lambda_1 \in \Omega^1(U)\otimes \frg_{-1}$ and act on the gauge potential 1- and 2-forms according to 
\begin{equation}
\begin{aligned}
 A&\mapsto \tilde A=g^{-1}A g+g^{-1}\dd g+\mu_1(\Lambda_1)~,\\
 B&\mapsto \tilde B=g^{-1}\acton B+\dd \Lambda_1+\tilde A\acton \Lambda_1+\tfrac12 [\Lambda_1,\Lambda_1]~,
\end{aligned}
\end{equation}
cf.~e.g.~\cite{Baez:0511710}.

To analyze these transformations further, note that when writing a group element infinitesimally as\footnote{Here and in the following, we use suggestive matrix group notation, adding Lie algebra elements to Lie group elements and obtaining Lie group elements. This is just done for simplicity and our discussion can be made rigorous, cf.~\cite{Jurco:2014mva}.} $g=\unit_\sG + A$ we have
\begin{equation}
 \unit_\sG+\dpar(h^{-1})A\dpar(h)-A=[\dpar(h),\unit_\sG+A]\in\im(\dpar)
\end{equation}
by equation~\eqref{eq:group_commutator} and differentiating yields
\begin{equation}
\dpar(h)A\dpar(h^{-1})-A\in \im(\mu_1)=\ker(\pi)~.
\end{equation} 
This implies that gauge transformations $(g,\Lambda_1)$ of $A$ descend to gauge transformations of $\pi(A)$ parametrized by the element $[g]\in \sG^0$: let $h\in \sH$ and $g\in \sG$, then 
\begin{equation}
\begin{aligned}
 \pi(\tilde A)&=\pi\Big(\dpar(h^{-1})g^{-1}Ag\dpar(h)+\dpar(h^{-1})g^{-1}\dd \big(g\dpar(h)\big)+\mu_1(\Lambda_1^1)\Big)\\
 &=\pi\big(g^{-1}Ag+g^{-1}\dd g\big)\\
 &=[g^{-1}]\pi(A)[g]+[g^{-1}]\dd [g]~.
\end{aligned}
\end{equation}
Because of $\pi(F)=\pi(\dd A+\tfrac12 [A,A])=0$, we have an element $g_0\in \Omega^0(U,\sG^0)$ such that
\begin{equation}
 \pi(A)=g_0^{-1}\dd g_0~.
\end{equation}

The pullback bundle $g^*_0 \sG$ is topologically trivial over $U$ because $U$ is contractible. Therefore, $g^*_0\sG$ admits a global section, which induces a lift $\hat g_0\in \Omega^0(U,\sG)$ of $g_0\in \Omega^0(U,\sG^0)$. Acting with the corresponding gauge transformation $(g,\Lambda_1)=(\hat g_0,0)$, we obtain gauge potential 1- and 2-forms
\begin{equation}
 \tilde A\in \Omega^1(U)\otimes \frg_0\eand \tilde B\in \Omega^2(U)\otimes \frg_{-1}\ewith \pi(\tilde A)=0~.
\end{equation}
That is, $\tilde A\in \ker(\pi)$ and there is an element $(g,\Lambda_1)=(\unit_\sG,\Lambda_1)$ in the 2-group of gauge transformation  such that $\tilde A=\mu_1(\Lambda_1)$. Acting with this gauge transformation on the potential yields the gauge potentials
\begin{equation}
 \tilde{\tilde{A}}\in \Omega^1(U)\otimes \frg_0\eand \tilde{\tilde B}\in \Omega^2(U)\otimes \frg_{-1}\ewith \tilde{\tilde A}=0~,~~\mu_1(\tilde{\tilde B})=0~,
\end{equation}
where the former relation is implied by the gauge invariant equation $F=0$. We are thus left with a 2-form $B'\in \Omega^2(U)\otimes \ker(\mu_1)$, the connective structure on an abelian gerbe over $U$. Altogether, we arrive at the following theorem.
\begin{theorem}\label{thm:fake_flat_trivial}
 A connection on a non-abelian principal 2-bundle is locally gauge equivalent to a connection on an abelian gerbe.
\end{theorem}

A generalization to principal 3-bundles based on 2-crossed modules of Lie groups with trivial Peiffer lifting is straightforward, cf.~also the analytical discussion in~\cite{Gastel:2018joi}. A proof for even higher bundles should also be possible using a reformulation in terms of simplicial principal bundles~\cite{May:book:1993}, cf.~also~\cite{Jurco20122389}, after extending the Poincar\'e lemma beyond what has been done in~\cite{Demessie:2014ewa}.

\section{String structures and their metric extensions}\label{sec:metric_string_structures}

The problems that the fake curvature condition creates in higher gauge theory are overcome by employing adjusted Weil algebras in the construction of $\frg$-connection objects. For the string Lie 2-algebra, this leads to {\em string structures} and we shall discuss them in the following.

\subsection{String structures and string theory}

The local connection data for string structures was first discovered in the context of supergravity~\cite{Bergshoeff:1981um,Chapline:1982ww}. The full topological picture first arose in the work of Killingback~\cite{Killingback:1986rd} and~\cite{Witten:1987cg} and we shall briefly review the relevant points in the following.

Recall that a point particle with world-line supersymmetry leads to a global anomaly of the corresponding quantum mechanics whose cancellation requires the $n$-dimensional target space manifold $M$ to be spin, i.e.~admit a {\em spin structure}. The latter is a lift of the frame bundle, which is a principal $\sSO(n)$-bundle, to a $\sSpin(n)$-bundle along the projection in the short exact sequence
\begin{equation}
 1 \longrightarrow \RZ_2 \longrightarrow \sSpin(n) \longrightarrow \sSO(n) \longrightarrow 1~.
\end{equation}
A spin structure exists if the second Stiefel--Whitney class $w_2(M)\in H^2(M,\RZ_2)$ vanishes, in which case inequivalent spin structures are characterized by the group $H^1(M,\RZ_2)$.

Similarly, a supersymmetric sigma model with target space an $n$-dimensional manifold $M$ leads to a quantum theory with global anomalies~\cite{Freed:2004yc}. As shown in~\cite{Freed:1986:483-513}, the anomaly can be canceled if the first fractional Pontryagin class $\tfrac12 \mathsf{p}_1(M)\in H^4(M,\RZ)$ vanishes and if a trivialization of the relevant cocycle is provided. The condition $\tfrac12 \mathsf{p}_1(M)=0$ can be interpreted as a lift of the spin bundle $\CL \sSpin(n)$ over the free loop $\CL M\coloneqq \sMap(S^1,M)$ along the projection in the short exact sequence
\begin{equation}
 1 \longrightarrow S^1 \longrightarrow \hat\CL\sSpin(n) \longrightarrow \CL \sSpin(n) \longrightarrow 1~,
\end{equation}
cf.~\cite{Killingback:1986rd,mclaughlin1992orientation}. Such a lift exists if $H^3(\CL M,\RZ)$ is trivial and inequivalent lifts are characterized by the group $H^1(\CL M,S^1)\cong H^2(\CL M,\RZ)\cong H^3(M,\RZ)$. We say that $M$ is a {\em string manifold}.\footnote{There is actually a nice sequence behind this: vanishing of the first and second Stiefel--Whitney classes of a manifold $M$ correspond to $M$ being orientable and spin, respectively. Recall that a spin structure on loop space is a lift of the looped frame bundle to its $S^1$-central extension. The free loop space $\CL M$ is now orientable if $M$ is spin and it is spin if $M$ is string, cf.~also~\cite{mclaughlin1992orientation,Waldorf:2012yv,Bunk:2018xwb}.}.

More generally, we can have a cancellation between a principal fiber bundle $P$ over $M$ and the tangent bundle of $M$ so that the condition $\mathsf{p}_1(M)=\mathsf{p}_1(TM)=0$ is replaced by the condition
\begin{equation}
 \mathsf{p}_1(P)-\mathsf{p}_1(TM)=0~,
\end{equation}
which is the condition for the cancellation of global and perturbative anomalies in string theory~\cite{Killingback:1986rd}.

Let $P$ be a principal $\sSpin(n)$-bundle. We then have the following two equivalent definitions of a string structure:
\begin{itemize}
 \setlength{\itemsep}{-1mm}
  \item[$\circ$] A string structure on $P$ is a lift of the structure group of $P$ to $\sString(n)$~\cite{Stolz:2004aa}.
 \item[$\circ$] A string structure on $P$ is a trivialization of the Chern--Simons gerbe of $P$~\cite{Waldorf:2009uf}, i.e.~of the 2-gerbe with topological class $\tfrac12 \mathsf{p}_1(P)$.
\end{itemize}
These topological string structures can be endowed with connections. In this paper, we shall be exclusively interested in the local description of these connections.

\subsection{Local connection data for string structures}\label{ssec:local_connections_ordinary_string_structures}

As mentioned above, the local structure of a string connection was first discovered in the context of supergravity~\cite{Bergshoeff:1981um,Chapline:1982ww}. This data can be obtained as a $\frg$-connection object using the adjusted Weil algebra $\sW_{\rm adj}(\hat \frg)$ introduced in section~\ref{ssec:Weil_adj_sk} as first observed in~\cite{Sati:2008eg,Sati:2009ic}.

We describe a $\aghsk$-connection object on an contractible patch $U$ as a triple
\begin{equation}
 (A,B,C)~\in~(\Omega^1(U)\otimes \frg)~\oplus~\Omega^2(U)~\oplus~\Omega^3(U)
\end{equation}
with curvature 2-, 3- and 4-forms
\begin{equation}
\begin{gathered}
 F\coloneqq \dd A+\tfrac12 [A,A]~,\\
 H\coloneqq \dd B-\tfrac{1}{3!}\mu_3(A,A,A)+ \chi_{\rm sk}(A,F)-C= \dd B+(A,\dd A)+\tfrac13 (A,[A,A])-C~,\\
 G\coloneqq \dd C~,
\end{gathered}
\end{equation}
which satisfy the Bianchi identities
\begin{equation}
 \dd F+[A,F]=0~,~~~\dd H-\chi_{\rm sk}(F,F)+G=\dd H-(F,F)+G=0~,~~~\dd G=0~.
\end{equation}
Gauge transformations act on the $\aghsk$-connection object according to
\begin{equation}
 \delta A=\dd \Lambda_0+\mu_2(A,\Lambda_0)~,~~~\delta B=\dd \Lambda_1+(\Lambda_0,F)-\tfrac12 \mu_3(A,A,\Lambda_0)-\Sigma~,~~~\delta C=\dd \Sigma~,
\end{equation}
which induces the following transformations of the curvatures:
\begin{equation}
 \delta F=-\mu_2(F,\Lambda_0)~,~~~\delta H=0~,~~~\delta G=0~.
\end{equation}
To restrict to a $\astringsk(\frg)$-connection object, we simply impose the condition $C=G=0$. This reproduces the kinematical data in~\cite{Bergshoeff:1981um,Chapline:1982ww}, if we let $\frg=\frh\oplus \aspin(1,9)$ with metrics of opposite signs on the two summands, where $\frh=\fre_8\times \fre_e$ or $\frh=\aso(32)$ the additional gauge algebra in heterotic supergravity and $\aspin(1,9)$ the structure Lie algebra of the frame bundle lifted to a spin structure.

The loop model case follows analogously. Here, we describe a $\aghl$-connection object on a contractible patch $U$ as a triple
\begin{equation}
 (A,B,C)~\in~(\Omega^1(U)\otimes P_0\frg)~\oplus~(\Omega^2(U)\otimes \hat \Omega\frg)~\oplus~\Omega^3(U)
\end{equation}
with curvature 2-, 3- and 4-forms
\begin{equation}
 F\coloneqq \dd A+\tfrac12 [A,A]+\mu_1(B)~,~~~H\coloneqq \dd B+\mu_2(A,B)-\chi_{\rm lp}(A,F)-\mu_1(C)~,~~~G\coloneqq \dd C~,
\end{equation}
where $\chi_{\rm lp}(-,-)$ is introduced in equation~\eqref{eq:chi_loop_case}. In this case, the curvature forms satisfy the Bianchi identities
\begin{equation}
\begin{gathered}
 \dd F+[A,F]=\mu_1(\chi_{\rm lp}(A,F))+\mu_1(H)~,\\
 \dd H+\chi_{\rm lp}(F,F)+\mu_1(G)=0~,~~~\dd G=0~.
\end{gathered}
\end{equation}
Gauge transformations act on the $\aghl$-connection object according to
\begin{equation}
\begin{gathered}
\delta A=\dd \Lambda_0+\mu_2(A,\Lambda_0)+\mu_1(\Lambda_1)~,\\
\delta B=\dd \Lambda_1+\mu_2(A,\Lambda_1)+\mu_2(\Lambda_0,B)-\chi_{\rm lp}(\Lambda_0,F)-\mu_1(\Sigma)~,~~~\delta C=\dd \Sigma~,
\end{gathered}
\end{equation}
which induces the following transformations of the curvatures:
\begin{equation}
\delta F=-\chi_{\rm lp}(\Lambda_0,F)-\mu_2(F,\Lambda_0)~,~~~\delta H=0~,~~~\delta G=0~.
\end{equation}
Again, to restrict to a $\astringl(\frg)$-connection object we can impose the condition $C=G=0$.

We note that in both models, the self-duality equation $H=*H$ for $U=\FR^{1,5}$ is gauge covariant for arbitrary 2-form curvature $F$. This is {\em not} the case for curvatures obtained from the unadjusted Weil algebra.

The two BRST complexes from which the above sets of gauge transformations are obtained indeed close without any further condition on the 2-form curvatures $F$. This confirms that $\sW_{\rm adj}(\aghsk)$ and $\sW_{\rm adj}(\aghl)$ are adjusted Weil algebras in the sense of definition~\ref{def:adjusted_Weil_algbra}. Note that the above constructions are readily truncated by simply using adjusted Weil algebras for both string Lie 2-algebras. However, the extended picture better fits the definition of string structures in terms of trivializations of a Chern--Simons gerbe. Moreover, it will be useful in the metric extension discussed next.

Let us also recall the morphism between local string structures, as derived in~\cite{Saemann:2017rjm}. We have induced maps
\begin{equation}
 \begin{tikzcd}
    (A_{\rm sk},B_{\rm sk},C_{\rm sk})\arrow[r,bend left=30]{}{\psi} & (A_{\rm  lp},B_{\rm  lp},C_{\rm  lp})\arrow[l,bend left=30]{}{\phi}
 \end{tikzcd}~,
\end{equation}
\begin{equation}
 \begin{aligned}
  \phi~&:~&A_{\rm  lp}&\mapsto A_{\rm sk}=\dpar A_{\rm  lp}~,~~~&B_{\rm  lp}&\mapsto B_{\rm sk}=\pr_{\FR}B_{\rm  lp}+\int_0^1\dd \tau\,(\dot A_{\rm  lp}, A_{\rm  lp})~,\\
  &&C_{\rm  lp}&\mapsto C_{\rm sk}=C_{\rm  lp}~,\\
  \psi~&:~&A_{\rm sk}&\mapsto A_{\rm  lp}=A_{\rm sk}\ell(\tau)~,~~~&B_{\rm sk}&\mapsto B_{\rm  lp}=B_{\rm sk}+\tfrac12 [A_{\rm sk},A_{\rm sk}](\ell(\tau)-\ell^2(\tau))~,\\
  &&C_{\rm sk}&\mapsto C_{\rm  lp}=C_{\rm sk}~.
 \end{aligned}
\end{equation}
These are, of course, the pullbacks of $\frg$-connection objects as discussed in~\eqref{eq:pullbacks_g_connection_objects}.

Finally, we note that using string structures replaces the problematic fake curvature condition $F=0$ with the Bianchi identity, e.g.~$\nabla F=\mu_1(H)$ in the skeletal case, which still ensures that quasi-isomorphic Lie 2-algebras give rise to the same gauge-equivalence classes of kinematical data.

\subsection{Metric string Lie 4-algebra}

To write down an action for a field theory involving an $L_\infty$-algebra, we need to introduce the analogue of a metric or inner product. The appropriate notion of a metric $L_\infty$-algebra is the following, cf.~\cite{Kontsevich:1992aa,Penkava:9512014} or also~\cite{Jurco:2018sby}.
\begin{definition}
 A \underline{cyclic structure} on an $L_\infty$-algebra $\frg$ is a non-degenerate bilinear graded symmetric map $\langle-,-\rangle:\frg\times \frg\rightarrow \FR$ such that 
\begin{equation}\label{eq:cyclicity}
 \langle a_1,\mu_i(a_2,\ldots,a_{i+1})\rangle\ =\ (-1)^{i+i(|a_1|+|a_{i+1}|)+|a_{i+1}|\sum_{j=1}^{i}|a_j|}\langle a_{i+1},\mu_i(a_1,\ldots,a_{i})\rangle
\end{equation}
for $a_1,\dots,a_{i+1}\in \frg$. An $L_\infty$-algebra endowed with a cyclic structure is called a \underline{cyclic} \underline{$L_\infty$-algebra}.
\end{definition}

In the dual, dga-picture, a cyclic structure on an $L_\infty$-algebra $\frg$ corresponds to a homogeneously graded symplectic form $\omega$ on the grade-shifted vector space $\frg[1]$. Here, the differential $Q$ becomes symplectic and can be written as $Q=\{\CQ,-\}$, where the Poisson bracket is the one induced by $\omega$.

Clearly, neither the Lie 3-algebras $\aghsk$ nor $\aghl$ are cotangent spaces and therefore they do not admit a cyclic structure. The solution to this problem for a general $3$-term $L_\infty$-algebra $\frg$ is to minimally extend the grade-shifted vector space $\frg[1]$ to $T^*[4]\frg[1]$, which carries a canonical symplectic form, and to endow it with a minimal Hamiltonian $\CQ$ such that the restriction of $\{\CQ,-\}$ to $\frg[1]$ reproduces the differential in the dga $\sCE(\frg)$~\cite{Saemann:2017rjm}. This is in fact a slight generalization of what is done in the BV formalism when introducing antifields, cf.~appendix~\ref{app:B:symplectic_extension}.

Let us first discuss $\aghsk$. The extension to the cotangent space reads as~\cite{Saemann:2017rjm}
\begin{equation}\label{eq:graded_skeletal}
\begin{aligned}
 T^*[4]\aghsk[1] &= \Big(\,(\frg^*\oplus \FR)[3]\,\oplus\,(\FR^*\oplus\FR)[2]\,\oplus\, (\FR^*\oplus \frg)[1]\,\Big)\\
 &=: \Big(\,(\frg_u^*\oplus \FR_q)[1]\,\oplus\,(\FR^*_s\oplus\FR_r)[1]\,\oplus\, (\FR^*_p\oplus \frg_t)[1]\,\Big)~,
\end{aligned}
\end{equation}
where the subscripts, again, merely help to assign the coordinate functions $t^\alpha, p, r,s,q,u_\alpha$ of degrees~$1,1,2,2,3,3$, respectively, and allow us to drop the grade-shift. In terms of these, the canonical symplectic form reads as
\begin{equation}\label{eq:symplectic_cyclic}
 \omega=\dd t^\alpha\wedge \dd u_\alpha+\dd q\wedge \dd p+\dd r\wedge \dd s~,
\end{equation}
and the symplectic completion discussed in appendix~\ref{app:B:symplectic_extension} yields the homological vector field
\begin{equation}
 \hat Q=-\tfrac12 f^\alpha_{\beta\gamma}t^\beta t^\gamma\der{t^\alpha}+\left(\tfrac{1}{3!}f_{\alpha\beta\gamma}t^\alpha t^\beta t^\gamma +q\right)\der{r}-s\der{p}+\left(-f^\gamma_{\alpha\beta}t^\beta u_\gamma -\tfrac12 f_{\alpha\beta\gamma}t^\beta t^\gamma s\right)\der{u_\alpha}~.
\end{equation}
Note that $\hat Q$ restricts indeed to the homological vector field of $\aghsk$ for $p=s=u=0$.

It turns out that we still wish to preserve the quasi-isomorphism of the extended $L_\infty$-algebra to $\frg$, and this is achieved by kernel-extending the $L_\infty$-algebra with dg-manifold $T^*[4]\aghsk[1]$ as explained in appendix~\ref{app:C}. This means that the cyclic structure only exists on a subset of the extension, which is sufficient for our purposes~\cite{Saemann:2017zpd}:
\begin{definition}
 The metric extension of $\aghsk$ is the Lie 4-algebra
 \begin{equation}\label{eq:string_cyclic_Lie_4_gv}
 \aghsk^\omega=\left(
 \begin{tikzcd}[row sep=0cm,column sep=2cm]
    \frg^*_v\arrow[r]{}{\mu_1=\id} & \frg^*_u & \FR^*_s \arrow[r]{}{\mu_1=\id} & \FR_p^*\\
    & \oplus & \oplus & \oplus\\
    & \FR_q \arrow[r]{}{\mu_1=\id} & \FR_r & \frg_t
 \end{tikzcd}\right)
\end{equation}
with higher products
\begin{equation}\label{eq:brackets_skeletal_metric}
\begin{aligned}
 &\mu_2:\frg_t\wedge \frg_t\rightarrow \frg_t~,~~~&&\mu_2(t_1,t_2)=[t_1,t_2]~,\\
 &\mu_2:\frg_t\wedge\frg^*_u\rightarrow \frg^*_u~,~~~&&\mu_2(t,u)=u\big([-,t]\big)~,\\
 &\mu_2:\frg_t\wedge\frg_v^*\rightarrow \frg_v^*~,~~~&&\mu_2(t,v)=v\big([-,t]\big)~,\\
 &\mu_3:\frg_t\wedge \frg_t \wedge \frg_t\rightarrow \FR_r~,~~~&&\mu_3(t_1,t_2,t_3)=(t_1,[t_2,t_3])~,\\
 &\mu_3:\frg_t\wedge \frg_t \wedge \FR_s\rightarrow \frg^*_u~,~~~&&\mu_3(t_1,t_2,s)= s\big(\,(-,[t_1,t_2])\,\big)~,
\end{aligned}
\end{equation}
and obvious pairings
\begin{equation}
 \langle u,t\rangle= u(t)~,~~~\langle q,p\rangle= q(p)=qp~,~~~\langle s,r\rangle=s(r)=sr~.
\end{equation}
\end{definition}
It is rather clear that the $\mu_1$-cohomologies of $\aghsk^\omega$ and $\frg$ agree, but let us give the explicit dual quasi-isomorphism for the (ordinary) Weil algebra $\sW(\aghsk^\omega)$. Its generators read as\footnote{We slightly abuse notation here by using the same letters for elements of subspaces as well as the coordinate functions on the grade-shifted versions of these subspaces.}
\begin{equation}\label{eq:generators_full_Weil}
\begin{tabular}{@{}lccccc@{}}
\toprule
     degrees & 5 & 4 & 3 & 2 & 1 \\
\midrule
   generators & $\hat v_\alpha$ & $\hat u_\alpha$ & $\hat s$ & $\hat p$ & \\
   & & $\hat q$ & $\hat r$ & $\hat t^\alpha$ \\
   & & $v_\alpha$ & $u_\alpha$ & $s$ & $p$ \\
   & & & $q$ & $r$ & $t^\alpha$\\
\bottomrule
 \end{tabular}
\end{equation}
and the differential acts as
\begin{equation}
\begin{aligned}
Q_\sW~&:~&t^\alpha &\mapsto  -\tfrac12 f^\alpha_{\beta\gamma} t^\beta t^\gamma+\hat t^\alpha~,~~~& p&\mapsto -s+\hat p~,\\
 &&\hat t^\alpha &\mapsto  -f^\alpha_{\beta\gamma} t^\beta \hat t^\gamma~,~~~& \hat p&\mapsto \hat s~,\\
 &&r&\mapsto \tfrac{1}{3!} f_{\alpha\beta\gamma} t^\alpha t^\beta t^\gamma+q+\hat r~,~~~& s&\mapsto \hat s~,\\
 &&\hat r&\mapsto -\tfrac{1}{2} f_{\alpha\beta\gamma} t^\alpha t^\beta \hat t^\gamma-\hat q~,~~~& \hat s&\mapsto 0~,\\
 &&u_\alpha &\mapsto -f^\gamma_{\alpha\beta}t^\beta u_\gamma-\tfrac12 f_{\alpha\beta\gamma}t^\beta t^\gamma s-v_\alpha+\hat u_\alpha ~,~~~&
 q &\mapsto \hat q~,\\
 &&\hat u_\alpha &\mapsto -f^\gamma_{\alpha\beta} t^\beta\hat u_\gamma+f^\gamma_{\alpha\beta}\hat t^\beta u_\gamma+f_{\alpha\beta\gamma}\hat t^\beta t^\gamma s+\tfrac12 f_{\alpha\beta\gamma}t^\beta t^\gamma \hat s+\hat v_\alpha~,~~~&
 \hat q &\mapsto 0~,\\
&&v_\alpha&\mapsto -f^\gamma_{\alpha\beta} t^\beta v_\gamma+\hat v_\alpha~,\\
&&\hat v_\alpha&\mapsto -f^\gamma_{\alpha\beta} t^\beta\hat v_\gamma+f^\gamma_{\alpha\beta} \hat  t^\beta v_\gamma~.
\end{aligned}
\end{equation}
We have a dual quasi-isomorphism $(\hat\Phi,\hat\Psi,\eta_{\hat\Psi\circ\hat\Phi},0):\sW(\aghsk^\omega)\approxeq \sW(\frg)$, which reads as 
\begin{subequations}
\begin{equation}\label{eq:skeletal_metric_quasi_iso}
\begin{aligned}
\hat \Phi:~~&t^\alpha\mapsto \tilde t^\alpha~,~~~\hat t^\alpha\mapsto \tilde{\hat{t}}^\alpha~,~~~q\mapsto -\tfrac{1}{3!}f_{\alpha\beta\gamma}\tilde t^\alpha \tilde t^\beta \tilde t^\gamma~,~~~\hat q\mapsto -\tfrac{1}{2}f_{\alpha\beta\gamma}\tilde t^\alpha \tilde t^\beta \tilde{\hat t}^\gamma~,\\
\hat \Psi:~~&\tilde t^\alpha \mapsto t^\alpha~,~~~\tilde{\hat{t}}^\alpha\mapsto \hat t^\alpha
\end{aligned} 
\end{equation}
with the remaining generators in the kernel of $\hat\Phi$ and
\begin{equation}
\begin{aligned}
\eta_{\hat\Psi\circ\hat\Phi}:~~&s\mapsto p~,~~~\hat s\mapsto -\hat p~,~~~q\mapsto -r~,~~~\hat q\mapsto \hat r~,\\
&v_\alpha\mapsto u_\alpha-\tfrac12 f_{\alpha\beta\gamma}t^\beta t^\gamma p~,~~~\hat v_\alpha \mapsto -\hat u_\alpha+f_{\alpha\beta\gamma}\hat t^\beta t^\gamma p+\tfrac12 f_{\alpha\beta\gamma}t^\alpha t^\beta \hat p~,
\end{aligned}
\end{equation}
\end{subequations}
and all other generators of $\sW(\aghsk^\omega)$ are in the kernel of $\eta_{\hat\Psi\circ\hat\Phi}$, which also defines the extension of this 2-morphism to all of $\sW(\inn(\aghsk^\omega))$.

Consider now the corresponding constructions for the extended loop string algebra $\aghl$:
\begin{definition}
 The metric extension of $\aghl$ is the Lie 4-algebra
\begin{equation}\label{eq:string_cyclic_Lie_4_gv_loop}
 \aghl^\omega=\left(
 \begin{tikzcd}[row sep=0cm,column sep=2cm]
     \frg^*_v\arrow[r]{}{\dpar^*} & (P_0\frg)^*_u \arrow[r,twoheadrightarrow] & ( L_0 \frg)^*_s\\
     & & \oplus \\
     & \oplus & \FR^*_{s_0}\arrow[r]{}{\id} & \FR_p^* \\
     & & \oplus \\
     & \FR_q \arrow[r]{}{\id} &\FR_{r_0} & \oplus \\
     & & \oplus \\
     & & ( L_0 \frg)_r \arrow[r,hook]{}{} & (P_0\frg)_t
 \end{tikzcd}\right)
\end{equation}
with higher products
\begin{equation}\label{eq:brackets_loop_metric}
\begin{aligned}
 &\mu_2:P_0\frg_t\wedge P_0\frg_t\rightarrow P_0 \frg_t~,~~~&&\mu_2(t_1,t_2)=[t_1,t_2]~,\\
 &\mu_2:P_0\frg_t\wedge  L_0\frg_r\rightarrow  L_0\frg_r~,~~~&&\mu_2(t,r)=[t,r]~,\\
 &\mu_2:P_0\frg_t\wedge  L_0\frg_r\rightarrow \FR_{r_0}~,~~~&&\mu_2(t,r)=-2\int_0^1\dd\tau\,(t,\dot r)~,\\
 &\mu_2:P_0\frg_t\wedge (P_0\frg)^*_u\rightarrow (P_0\frg)^*_u~,~~~&&\mu_2(t,u)=u\big([-,t]\big)~,\\
 &\mu_2:P_0\frg_t\wedge ( L_0\frg)^*_s\rightarrow ( L_0\frg)^*_s~,~~~&&\mu_2(t,s)=s\big([-,t]\big)~,\\
 &\mu_2:( L_0\frg)_r \wedge ( L_0\frg)^\ast_s\rightarrow (P_0\frg)_u^*~,~~~&&\mu_2(r,s)=s\big([-,r]\big)~,\\
 &\mu_2:P_0\frg_t\wedge\FR^*_{s_0}\rightarrow ( L_0\frg)^*_s~,~~~&&\mu_2(t,s_0)=-2s_0\int_0^1\dd \tau\,(\dot t,-)~,\\
 &\mu_2: L_0\frg_r\wedge\FR^*_{s_0}\rightarrow (P_0\frg)_u^*~,~~~&&\mu_2(r,s_0)=-2s_0\int_0^1\dd \tau\,(\dot  r,-)~,\\
 &\mu_2:P_0\frg_t\wedge \frg_v^*\rightarrow \frg_v^*~,~~~&&\mu_2(t,v)=v\big([-,\dpar t]\big)~,\\
\end{aligned}
\end{equation}
and obvious pairings
\begin{equation}
 \langle u,t\rangle= u(t)~,~~~\langle s,r\rangle =s(r)~,~~~\langle s_0,r_0\rangle=s_0r_0~,~~~\langle p,q\rangle=pq~.
\end{equation}
\end{definition}

The unadjusted Weil algebra of $\aghl^\omega$ has generators
\begin{equation}\label{eq:generators_full_Weil_loop}
\begin{tabular}{@{}lccccc@{}}
\toprule
     degrees & 5 & 4 & 3 & 2 & 1 \\
\midrule
generators & $\hat v_\alpha$ & $\hat u_{\alpha\tau}$ & $\hat s_{\alpha\tau}$, $\hat s_0$ & $\hat p$ & \\
& & $\hat q$ & $\hat r^{\alpha\tau}$, $\hat r_0$ & $\hat t^{\alpha\tau}$ \\
   & & $v_\alpha$ & $u_{\alpha\tau}$ & $s_{\alpha\tau}$, $s_0$ & $p$ \\
   & & & $q$ & $r^{\alpha\tau}$, $r_0$ & $t^{\alpha\tau}$\\
\bottomrule
 \end{tabular}
\end{equation}
and the differential acts according to
\begin{equation}
    \def\arraystretch{1.5}
    \begin{array}{llrlrl}
        Q_\sW&:& t^{\alpha\tau} &\mapsto  -\tfrac12 f^\alpha_{\beta\gamma} t^{\beta\tau} t^{\gamma\tau}-r^{\alpha\tau}+\hat t^{\alpha\tau}~,~~~&p&\mapsto -s_0+\hat p~,\\
        &&\hat t^{\alpha\tau} &\mapsto  -f^\alpha_{\beta\gamma} t^{\beta\tau} \hat t^{\gamma\tau}+\hat r^{\alpha\tau}~,~~~
        &\hat p&\mapsto \hat s_0~,\\
        &&r^{\alpha\tau} &\mapsto -f^\alpha_{\beta\gamma} t^{\beta\tau}r^{\gamma\tau} + \hat{r}^{\alpha\tau}~,~~~&s_0&\mapsto \hat s_0~,\\
        &&\hat{r}^{\alpha\tau}&\mapsto-f^\alpha_{\beta\gamma} t^{\beta\tau}\hat{r}^{\gamma\tau} + f^\alpha_{\beta\gamma} \hat{t}^{\beta\tau}r^{\gamma\tau}~,~~~
        &\hat s_0&\mapsto 0~,\\[0.2cm]
        &&r_0 &\displaystyle \mapsto 2\int_0^1\dd\tau\,\kappa_{\alpha\beta}t^{\alpha\tau} \smash{\dot{{r}}}^{\beta\tau}+q+\hat{r}_0 ~,~~~
        &q&\mapsto \hat q~,\\[0.3cm]
        &&\hat{r}_0&\mapsto \displaystyle 2\int_0^1\dd\tau\,\kappa_{\alpha\beta}\left(t^{\alpha\tau} \smash{\dot{\hat{r}}}^{\beta\tau}-\hat{t}^{\alpha\tau}\dot{r}^{\beta\tau}\right)-\hat q~,~~~
        &\hat q &\mapsto 0~,\\
        && s_{\alpha \tau}&\multicolumn{3}{l}{\mapsto -f^\gamma_{\alpha\beta}t^{\beta\tau}s_{\gamma\tau}+2\kappa_{\alpha\beta}\dot t^{\beta\tau}s_0+u_{\alpha\tau}+\hat s_{\alpha\tau}~,}\\
        && \hat s_{\alpha\tau}&\multicolumn{3}{l}{\mapsto f^\gamma_{\alpha\beta}(\hat t^{\beta\tau}s_{\gamma\tau}-t^{\beta\tau}\hat s_{\gamma\tau})-2\kappa_{\alpha\beta}(\dot{\hat{t}}^{\beta\tau}s_0-\dot t^{\beta\tau}\hat s_0)-\hat u_{\alpha\tau}~,}\\
        &&u_{\alpha\tau}&\multicolumn{3}{l}{\mapsto -f^\gamma_{\alpha\beta}(t^{\beta\tau}u_{\gamma\tau}+r^{\beta\tau}s_{\gamma\tau})+2\kappa_{\alpha\beta}\dot r^{\beta\tau}s_0-v_\alpha\delta(\tau-1)+\hat u_{\alpha\tau}~,}\\
        &&\hat u_{\alpha\tau}&\multicolumn{3}{l}{\mapsto f^\gamma_{\alpha\beta}(\hat t^{\beta\tau}u_{\gamma\tau}-t^{\beta\tau}\hat u_{\gamma\tau}+\hat r^{\beta\tau}s_{\gamma\tau}+r^{\beta\tau}\hat s_{\gamma\tau})+\hat v_\alpha \delta(\tau-1)-}\\
            ~~~&&&\phantom{\mapsto}-2\kappa_{\alpha\beta}(\dot{\hat{r}}^{\beta\tau}s_0+\dot r^{\beta\tau}\hat s_0)~,\\
        && v_\alpha &\multicolumn{3}{l}{\mapsto -f^\gamma_{\alpha\beta} t^{\beta 1}v_{\gamma}+\hat v_\alpha~,}\\
        && \hat v_\alpha &\multicolumn{3}{l}{\mapsto f^\gamma_{\alpha\beta} (\hat t^{\beta 1}v_{\gamma}-t^{\beta 1}\hat v_\gamma)~.}
    \end{array}
\end{equation}

Again, the $\mu_1$-cohomologies of $\aghl^\omega$ and $\frg$ evidently agree. We refrain from writing out the dual quasi-isomorphism and turn directly to the necessary adjustments.

\subsection{Adjusted Weil algebras for the metric extensions}

The generators of the adjusted Weil algebra $\sW_{\rm adj}(\aghsk^\omega)$ satisfying the condition of definition~\ref{def:adjusted_Weil_algbra} are again as in~\eqref{eq:generators_full_Weil}. As usual, the adjustment is not unique but a possible choice is given by
\begin{equation}
\def\arraystretch{1.5}
\begin{aligned}
Q_{\sW_{\rm adj}}~&:~&t^\alpha &\mapsto  -\tfrac12 f^\alpha_{\beta\gamma} t^\beta t^\gamma+\hat t^\alpha~,~~~& p&\mapsto -s+\hat p~,\\
 &&\hat t^\alpha &\mapsto  -f^\alpha_{\beta\gamma} t^\beta \hat t^\gamma~,~~~& \hat p&\mapsto \hat s~,\\
 &&r&\mapsto \tfrac{1}{3!} f_{\alpha\beta\gamma} t^\alpha t^\beta t^\gamma-\kappa_{\alpha\beta}t^\alpha \hat t^\beta+q+\hat r~,~~~& s&\mapsto \hat s~,\\
 &&\hat r&\mapsto \kappa_{\alpha\beta}\hat t^\alpha \hat t^\beta-\hat q~,~~~& \hat s&\mapsto 0~,\\
 &&u_\alpha &\mapsto -f^\gamma_{\alpha\beta}t^\beta u_\gamma-\tfrac12 f_{\alpha\beta\gamma}t^\beta t^\gamma s-v_\alpha+\hat u_\alpha~,~~~&
 q &\mapsto \hat q~,\\
 &&\hat u_\alpha &\mapsto -f^\gamma_{\alpha\beta}t^\beta \hat u_\gamma+\hat v_\alpha~,~~~&
 \hat q &\mapsto 0~,\\
&&v_\alpha&\mapsto -f^\gamma_{\alpha\beta}t^\beta v_\gamma - f^\gamma_{\alpha\beta}\hat t^\beta u_\gamma +f_{\alpha\beta\gamma}t^\beta\hat t^\gamma s-\tfrac12 f_{\alpha\beta\gamma} t^\beta t^\gamma \hat s+\hat v_\alpha~,\\
&&\hat v_\alpha&\mapsto -f^\gamma_{\alpha\beta}t^\beta \hat v_\gamma+f^\gamma_{\alpha\beta}\hat t^\beta \hat u_\gamma~.
\end{aligned}
\end{equation}
As required, this adjustment avoids any dynamical constraints on the curvatures in the BRST complex. While the choice of adjustment is not unique, it requires the presence of a modified $\hat v$ and cannot be accomplished by modifying $\hat u$ alone. The above choice is a minimal one which allows for simple expressions and covariant Bianchi identities.

The dual quasi-isomorphism in~\eqref{eq:skeletal_metric_quasi_iso} can be slightly amended to a dual quasi-isomor\-phism $(\Phi_{\rm adj},\Psi_{\rm adj},\eta_{\rm adj},0):\sW_{\rm adj}(\aghsk^\omega)\approxeq \sW(\frg)$, which reads as 
\begin{subequations}
\begin{equation}
\begin{aligned}
 \Phi_{\rm adj}:~~&t^\alpha\mapsto \tilde t^\alpha~,~~~\hat t^\alpha\mapsto \tilde{\hat{t}}^\alpha~,~~~q\mapsto {\rm cs}~,~~~\hat q\mapsto \tfrac12 \mathsf{p}_1 ~,\\
  \Psi_{\rm adj}:~~&\tilde t^\alpha \mapsto t^\alpha~,~~~\tilde{\hat{t}}^\alpha\mapsto \hat t^\alpha
\end{aligned} 
\end{equation}
with the remaining generators again living in the kernel of $\Phi_{\rm adj}$ and
\begin{equation}
\begin{aligned}
\eta:~~&s\mapsto p~,~~~\hat s\mapsto -\hat p~,~~~q\mapsto -r~,~~~\hat q\mapsto \hat r~,\\
&v_\alpha\mapsto u_\alpha-\tfrac12 f_{\alpha\beta\gamma}t^\beta t^\gamma p~,~~~\hat v_\alpha \mapsto -\hat u_\alpha~.
\end{aligned}
\end{equation}
\end{subequations}

A quick computation now shows that this dual quasi-isomorphism for the adjusted Weil algebra is indeed compatible with the invariant polynomials in the sense of diagram~\eqref{eq:compatibility_diagram}.

Let us also describe the adjustment for the loop model. The generators of the adjusted Weil algebra $\sW_{\rm adj}(\aghl^\omega)$ are as in~\eqref{eq:generators_full_Weil_loop}, and an adjustment that satisfies the condition of definition~\ref{def:adjusted_Weil_algbra} is the following:
\begin{equation}
    \def\arraystretch{1.3}
    \begin{array}{llrlrl}
        Q_{\sW_{\rm adj}}&:& t^{\alpha\tau} &\mapsto  -\tfrac12 f^\alpha_{\beta\gamma} t^{\beta\tau} t^{\gamma\tau}-r^{\alpha\tau}+\hat t^{\alpha\tau}~,~~~&p&\mapsto -s_0+\hat p~,\\
        &&\hat t^{\alpha\tau} &\mapsto  -f^\alpha_{\beta\gamma} t^{\beta\tau} \hat t^{\gamma\tau}+\chi^{\alpha\tau}(t,\hat t)+\hat r^{\alpha\tau}~,~~~
        &\hat p&\mapsto \hat s_0~,\\
        &&r^{\alpha\tau} &\mapsto -f^\alpha_{\beta\gamma} t^{\beta\tau}r^{\gamma\tau} + \chi^{\alpha\tau}(t,\hat t)+\hat{r}^{\alpha\tau}~,~~~&s_0&\mapsto \hat s_0~,\\
        &&\hat{r}^{\alpha\tau}&\mapsto 0~,~~~
        &\hat s_0&\mapsto 0~,\\
        &&r_0 &\mapsto 2\int_0^1\dd\tau\,\kappa_{\alpha\beta}t^{\alpha\tau} \smash{\dot{{r}}}^{\beta\tau}+\chi(t,\hat{t})+q+\hat{r}_0 ~,~~~
        &q&\mapsto \hat q~,\\
        &&\hat{r}_0&\mapsto -\chi(\hat t,\hat t)-\hat q~,~~~
        &\hat q &\mapsto 0~,\\
        && s_{\alpha \tau}&\multicolumn{3}{l}{\mapsto -f^\gamma_{\alpha\beta}t^{\beta\tau}s_{\gamma\tau}+2\kappa_{\alpha\beta}\dot t^{\beta\tau}s_0+u_{\alpha\tau}+\hat s_{\alpha\tau}~,}\\
        && \hat s_{\alpha\tau}&\mapsto-\hat u_{\alpha\tau}~,\\
        &&u_{\alpha\tau}&\multicolumn{3}{l}{\mapsto -f^\gamma_{\alpha\beta}(t^{\beta\tau}u_{\gamma\tau}+r^{\beta\tau}s_{\gamma\tau})+2\kappa_{\alpha\beta}\dot r^{\beta\tau}s_0-v_\alpha\delta(\tau-1)+}\\
        &&&\multicolumn{3}{l}{\phantom{\mapsto}+\chi_{\alpha\tau}(\hat t, s)-\chi_{\alpha\tau}(t,\hat s)+\chi_{\alpha\tau}(\dot{\hat{t}},s_0)-\chi_{\alpha\tau}(\dot{t},\hat s_0)+\hat u_{\alpha\tau}~,}\\
        &&\hat u_{\alpha\tau}&\mapsto \hat v_\alpha \delta(\tau-1)~,\\
        && v_\alpha &\multicolumn{3}{l}{\mapsto -f^\gamma_{\alpha\beta} t^{\beta 1}v_{\gamma}-\chi_\alpha(\hat t,u)+\chi_\alpha(t,\hat u)+\chi_\alpha(\hat t,\chi(t,s))+}\\
        &&&\multicolumn{3}{l}{\phantom{\mapsto}+\chi_\alpha(\hat t,\chi(\dot t,s_0))+\hat v_\alpha~,}\\
        && \hat v_\alpha &\multicolumn{3}{l}{\mapsto 0~,}
    \end{array}
\end{equation}
where the additional $\chi(-,-)$ are defined in~\eqref{eq:loop_additional_structure_maps}.

\subsection{Local differential metric string structures}\label{ssec:local_connections_metric_string_structures}

The above adjusted Weil algebras were derived by constructing a consistent BRST complex using the method given in section~\ref{ssec:BRST_complex}. This complex contains the full local information of differential string structures, and below we summarize the results.

A $\aghsk^\omega$-connection object on a patch $U$ of some manifold is given by potential forms
\begin{subequations}\label{eq:metric_string_structures_sk}
\begin{equation}
\begin{aligned}
 A&\in\Omega^1(U)\otimes (\frg\oplus \FR^*)~,~~~&B&\in \Omega^2(U)\otimes (\FR\oplus \FR^*)[1]~,\\
 C&\in \Omega^3(U)\otimes (\frg^*\oplus\FR)[2]~,~~~&D&\in \Omega^4(U)\otimes \frg^*[3]~,
\end{aligned}
\end{equation}
from which the curvatures
\begin{equation}\label{eq:higher_curvs_sk}
\begin{aligned}
 F&=\dd A+\tfrac12 \mu_2(A,A)+\mu_1(B)&&\in~\Omega^2(U)\otimes (\frg\oplus\FR^*)~,\\
 H&=\dd B-\tfrac{1}{3!}\mu_3(A,A,A)+\chi_{\rm sk}(A,F)-\mu_1(C)\\
  &=\dd B+(A,\dd A)+\tfrac13(A,\mu_2(A,A))-\mu_1(C)&&\in~\Omega^3(U)\otimes (\FR\oplus\FR^*)[1]~,\\
 G&=\dd C+\mu_2(A,C)+\tfrac12\mu_3(A,A,B)+\mu_1(D)&&\in~\Omega^4(U)\otimes (\frg^*\oplus\FR)[2]~,\\
 I&=\dd D+\mu_2(A,D)+\chi_{\rm sk}(F,C)+\tfrac12 \chi_{\rm sk}(A,A,H)&&\\
 &\phantom{{}={}}+\chi_{\rm sk}(F,A,B)&&\in~\Omega^5(U)\otimes \frg^*[3]
\end{aligned}
\end{equation}
are constructed, where, in addition to the higher brackets defined in~\eqref{eq:brackets_skeletal_metric}, the adjustment gives rise to the additional structure maps
\begin{equation}
\begin{aligned}
 &\chi_{\rm sk}:\frg \wedge \frg \rightarrow \FR[1]~,~~~&&\chi_{\rm sk}(a_1,a_2)=(a_1,a_2)~,\\
 &\chi_{\rm sk}:\frg \wedge \frg^*[2]\rightarrow \frg^*[3]~,~~~&&\chi_{\rm sk}(a_1,a^*_2)=a^*_2\big([-,a_1]\big)~,\\
 &\chi_{\rm sk}:\frg \wedge \frg\wedge \FR^*[1]\rightarrow \frg^*[3]~,~~~&&\chi_{\rm sk}(a_1,a_2,s)=s\big(-,[a_1,a_2]\big)~.
\end{aligned}
\end{equation}
The curvatures satisfy the Bianchi identities
\begin{equation}\label{eq:bianchi_higher_curvs_sk}
\begin{aligned}
 \dd F+\mu_2(A,F)-\mu_1(H)&=0~,~~~&\dd H-\chi_{\rm sk}(F,F)+\mu_1(G)&=0~,\\
 \dd G+\mu_2(A,G)-\mu_1(I)&=0~,~~~&\dd I+\mu_2(A,I)-\chi_{\rm sk}(F,G)&=0~.
\end{aligned}
\end{equation}

The infinitesimal gauge transformations (read-off from the BRST transformations of the gauge potentials and curvatures) are given by 
\begin{equation}
\begin{aligned}
 \delta A&=\dd \Lambda_0+\mu_2(A,\Lambda_0)+\mu_1(\Lambda_1)~,\\
 \delta B&=\dd \Lambda_1+\chi_{\rm sk}(F,\Lambda_0)-\tfrac12 \mu_3(A,A,\Lambda_0)-\mu_1(\Lambda_2)~,\\
 \delta C&=\dd \Lambda_2+\mu_2(A,\Lambda_2)+\mu_2(C,\Lambda_0)+\tfrac12 \mu_3(A,A,\Lambda_1)+\mu_3(A,B,\Lambda_0)+\mu_1(\Lambda_3)~,\\
 \delta D&=\dd \Lambda_3+\mu_2(A,\Lambda_3)-\mu_2(D,\Lambda_0)+\chi_{\rm sk}(F,\Lambda_2)-\chi_{\rm sk}(A,F,\Lambda_1)+\\
 &\hspace{1cm}+\chi_{\rm sk}(A,H,\Lambda_0)+\chi_{\rm sk}(B,F,\Lambda_0)~,\\
\end{aligned}
\end{equation}
and 
\begin{equation}
 \begin{aligned}
  \delta F&=-\mu_2(F,\Lambda_0)~,~~~&\delta H&=0~,\\
  \delta G&=-\mu_2(G,\Lambda_0)~,~~~&\delta I&=\mu_2(I,\Lambda_0)
 \end{aligned}
\end{equation}
with gauge parameters (still carrying ghost degrees)
\begin{equation}
    \begin{aligned}
         \Lambda_0&\in \Omega^0(U)\otimes(\frg\oplus \FR^*)~,~~~&\Lambda_1&\in\Omega^1(U)\otimes(\FR\oplus \FR^*)[1]~,\\
         \Lambda_2&\in \Omega^2(U)\otimes(\frg^*\oplus \FR)[2]~,~~~&\Lambda_3&\in\Omega^3(U)\otimes\frg^*[3]~.
    \end{aligned}
\end{equation}
\end{subequations}

In the case of the loop model $\aghl^\omega$, we have the potential forms
\begin{subequations}\label{eq:metric_string_structures_lp}
\begin{equation}
\begin{aligned}
 A&\in\Omega^1(U)\otimes (P_0\frg\oplus \FR^*)~,~~~&B&\in \Omega^2(U)\otimes (L_0\frg\oplus  (L_0\frg)^*\oplus\FR\oplus \FR^*)[1]~,\\
 C&\in \Omega^3(U)\otimes ((P_0\frg)^*\oplus\FR)[2]~,~~~&D&\in \Omega^4(U)\otimes \frg^*[3]
\end{aligned}
\end{equation}
with curvatures
\begin{equation}\label{eq:higher_curvs_lp}
\begin{aligned}
 F&=\dd A+\tfrac12 \mu_2(A,A)+\mu_1(B)&&\in~\Omega^2(U)\otimes (P_0\frg\oplus\FR^*)~,\\
 H&=\dd B+\mu_2(A,B)-\chi_{\rm lp}(A,F)-\mu_1(C)&&\in~\Omega^3(U)\otimes (L_0\frg\oplus  (L_0\frg)^*\oplus\FR\oplus \FR^*)[1]~,\\
 G&=\dd C+\mu_2(A,C)+\mu_2(B,B)+\mu_1(D)+&&\\
 &\phantom{{}={}}+\chi_{\rm lp}(A,H)-\chi_{\rm lp}(F,B)&&\in~\Omega^4(U)\otimes ((P_0\frg)^*\oplus\FR)[2]~,\\
 I&=\dd D+\mu_2(A,D)+\chi_{\rm lp}(F,C)-&&\\
 &\phantom{{}={}}-\chi_{\rm lp}(A,H)-\chi_{\rm lp}(F,\chi_{\rm lp}(A,B))&&\in~\Omega^5(U)\otimes \frg^*[3]~,
\end{aligned}
\end{equation}
where, again, due to the adjustment, we introduce additional structure maps complementing the higher brackets defined in~\eqref{eq:brackets_loop_metric}:
\begin{equation}\label{eq:loop_additional_structure_maps}
\begin{aligned}
 &\chi_{\rm lp}: P_0\frg \wedge P_0\frg \rightarrow (L_0\frg\oplus\FR)[1]~,~&&\chi_{\rm lp}\big(\gamma_1,\gamma_2\big)=\bigg([\gamma_1,\gamma_2]-\ell(\tau)[\partial\gamma_1,\partial\gamma_2]~,\\
 &&&\hspace{2.5cm}\; 2\int_0^1 \dd\tau \left(\dot\gamma_1(\tau),\gamma_2(\tau)\right)\bigg)~,\\
 &\chi_{\rm lp}:P_0\frg \wedge ((L_0\frg)^*\oplus\FR^*)[1]\rightarrow (P_0\frg)^*[2]~,~&&\chi_{\rm lp}\big(\gamma,(\lambda^*,s)\big)=\lambda^*\big([-,\gamma]-\ell(\tau)\partial[-,\gamma]\big)-\\
 &&&\hspace{2.6cm}\; -2s\int_0^1\dd\tau\, (\dot\gamma(\tau),-)~,\\
 &\chi_{\rm lp}:P_0\frg \wedge (P_0\frg)^*[2]\rightarrow \frg^*[3]~,~&&\chi_{\rm lp}(\gamma_1,\gamma^*_2)=\gamma_2^*(\ell(\tau)[-,\partial\gamma_1])~.
\end{aligned}
\end{equation}
The curvatures satisfy the Bianchi identities
\begin{equation}\label{eq:bianchi_higher_curvs}
\begin{aligned}
 \dd F+\mu_2(A,F)-\mu_1(\chi_{\rm lp}(A,F))-\mu_1(H)&=0~,~~~&\dd H+\chi_{\rm lp}(F,F)+\mu_1(G)&=0~,\\
 \dd G-\mu_1(I)&=0~,~~~&\dd I&=0~.
\end{aligned}
\end{equation}
Here, the infinitesimal gauge transformations read as
\begin{equation}
\begin{aligned}
 \delta A&=\dd \Lambda_0+\mu_2(A,\Lambda_0)+\mu_1(\Lambda_1)~,\\
 \delta B&=\dd \Lambda_1-\mu_2(B,\Lambda_0)+\mu_2(A,\Lambda_1)-\mu_1(\Lambda_2)-\chi_{\rm lp}(\Lambda_0,F)~,\\
 \delta C&=\dd \Lambda_2+\mu_2(A,\Lambda_2)+\mu_2(C,\Lambda_0)+\mu_2(B,\Lambda_1)+\mu_1(\Lambda_3)-\\
 &\phantom{{}={}}-\chi_{\rm lp}(F,\Lambda_1)+\chi_{\rm lp}(\Lambda_0,H)~,\\
 \delta D&=\dd \Lambda_3+\mu_2(A,\Lambda_3)-\mu_2(D,\Lambda_0)+\chi_{\rm lp}(F,\Lambda_2)-\chi_{\rm lp}(\Lambda_0,G)-\\
 &\phantom{{}={}}-\chi_{\rm lp}(F,\chi_{\rm lp}(A,\Lambda_1))-\chi_{\rm lp}(F,\chi_{\rm lp}(\Lambda_0,B))~,\\
\end{aligned}
\end{equation}
and 
\begin{equation}
 \begin{aligned}
  \delta F&=-\mu_2(F,\Lambda_0)-\mu_1(\chi_{\rm lp}(\Lambda_0,F))~,~~~&\delta H&=0~,\\
  \delta G&=0~,~~~&\delta I&=0
 \end{aligned}
\end{equation}
with gauge parameters (again, still carrying ghost degrees)
\begin{equation}
    \begin{aligned}
         \Lambda_0&\in \Omega^0(U)\otimes(P_0\frg\oplus \FR^*)~,~~~&\Lambda_1&\in\Omega^1(U)\otimes(L_0\frg\oplus  (L_0\frg)^*\oplus\FR\oplus \FR^*)[1]~,\\
         \Lambda_2&\in \Omega^2(U)\otimes((P_0\frg)^*\oplus\FR)[2]~,~~~&\Lambda_3&\in\Omega^3(U)\otimes\frg^*[3]~.
    \end{aligned}
\end{equation}
\end{subequations}

In both cases, the truncation from $\aghsk^\omega$- and $\aghl^\omega$-connection objects to actual differential string structures over $U$ is achieved by setting the $\FR_q$-components of $C$ and $G$ to zero. This is evidently consistent since $\langle q,\hat q\rangle$ is a differential ideal in both $\aghsk^\omega$ and $\aghl^\omega$.

We note that the kinematical data for ``ordinary'' higher gauge theory with gauge Lie 4-algebras $\aghsk^\omega$ and $\aghl^\omega$ are recovered by putting $F=H=G=0$ everywhere.

\section{Applications and outlook}\label{sec:outlook}

In this last section, we describe a number of immediate applications of our constructions as well as an outlook on future research directions.

\subsection{Self-dual 3-forms, self-dual strings and supersymmetry}\label{ssec:sd3forms}

Our original motivation for studying string structures is certainly the application in the description of non-abelian or rather interacting self-dual strings as discussed in~\cite{Saemann:2017rjm} and the interacting generalization of the free action for a single M5-brane as given in~\cite{Saemann:2017rjm,Saemann:2017zpd}. This action was based on the observation~\cite{Saemann:2017rjm} that metric string structures for the skeletal model are special cases of the gauge structure of the $\CN=(1,0)$ superconformal field theory constructed in~\cite{Samtleben:2011fj,Samtleben:2012mi,Samtleben:2012fb}, which was inspired by the observation made in~\cite{Palmer:2013pka} and later in~\cite{Lavau:2014iva} that this gauge structure was an $L_\infty$-algebra endowed with extra structure.

Ideally, the field theory presented in~\cite{Saemann:2017rjm,Saemann:2017zpd} should be formulated in a way that is agnostic about the model underlying the construction of the string structure, and this is certainly one of our future goals. A first step has been made with the full clarification of metric string structures for both the skeletal and the loop models in this paper.

One technical and more immediate goal was to clarify the precise form of the 4- and 5-form curvatures. Our results in~\cite{Saemann:2017rjm} deviated from those in~\cite{Samtleben:2011fj}, where, however, the forms were only specified up to terms in a particular subspace. We conclude that the curvatures found above in~\eqref{eq:higher_curvs_sk},
\begin{equation}
\begin{aligned}
 G&=\dd C+\mu_2(A,C)+\tfrac12\mu_3(A,A,B)+\mu_1(D)~,\\
 I&=\dd D+\mu_2(A,D)+\chi_{\rm sk}(F,C)+\chi_{\rm sk}(F,A,B)+\tfrac12 \chi_{\rm sk}(A,A,H)~,
\end{aligned}
\end{equation}
differ from the ones used in~\cite{Samtleben:2011fj},
\begin{equation}
\begin{aligned}
 G&=\dd C+\mu_2(A,C)+B_s(F,-)+\mu_1(D)+c_0\mu_4(A,A,A,A)~,\\
 I&=\dd D+\mu_2(A,D)+c_1~,
\end{aligned}
\end{equation}
where $c_0\in \FR$ and $c_1\in \Omega^4(U)\otimes (\frg^*[2]\oplus\FR_q)$ are not specified any further. We note that these are consistent curvatures arising from an adjusted Weil algebra, if $c_0=0$. However, this Weil algebra is {\em not} compatible with the cyclic structure induced by the symplectic form~\eqref{eq:symplectic_cyclic}. For the action given in~\cite{Samtleben:2011fj}, this compatibility is irrelevant, as no pairings of $\frg_{-2}$ and $\frg_0$ are present. In the PST-extended version, however, such pairings do exist~\cite{Saemann:2017zpd}. In general, the compatibility with the full symplectic form guarantees for mathematical consistency irrespective of the form of the action.

Recall that the curvatures of the model of~\cite{Saemann:2017rjm} appear in the equations of motion in a supercovariantized form. In particular, the self-dual 3-form $H$ for a skeletal string structure (without metric extension) is supercovariantized to
\begin{equation}
 H^-=\tfrac12 (H-*H)~~\longrightarrow~~ \CCH^-\coloneqq \tfrac12(H-*H)-(\lambdab,\gamma_{(3)}\lambda)~,
\end{equation}
where $\lambda$ is the spinor field of the vector multiplet and $\gamma_{(3)}\coloneqq \gamma_{\mu\nu\kappa}\dd x^\mu\wedge \dd x^\nu\wedge \dd x^\kappa$. This supercovariantized form is derived from the superspace version of the Bianchi identities for string structures on the $\CN=(n,0)$ superspace $\FR^{1,5|4n}$, cf.~\cite{Bergshoeff:1996qm}. The latter are readily derived from our perspective if we extend the image of our dga-morphism defining the string structure $\CA:\sW(\aghsk)\rightarrow \sW(\FR^{1,5})=\Omega^\bullet(\FR^{1,5})$ to the Weil algebra of the $\CN=(1,0)$-superspace $\FR^{1,5|8}$ consisting of the superforms on $\FR^{1,5|8}$. There are additional flatness conditions that need to be imposed on the supercurvatures, which is also the case when supersymmetric Yang--Mills theory is described as a partially flat dga-morphism~\cite{Ritter:2015ymv}.

\subsection{Relation to the tensor hierarchy in gauged supergravity}\label{ssec:tensor_hierarchies}

Particularly important Lie 2-algebras for the construction of six-dimensional superconformal field theories are those appearing in the tensor hierarchy of gauged supergravity. This was also the starting point for the original model of~\cite{Samtleben:2011fj}, which admitted the truncation to string structures presented in~\cite{Saemann:2017rjm,Saemann:2017zpd}.

Recall that gauged supergravities~(see~\cite{Trigiante:2016mnt} and references therein) are constructed by promoting a Lie subalgebra $\frh$ of the Lie algebra $\hat \frh$ of global symmetry (usually given by a split real form of the complex Lie algebra $\fre_{11-d}$) to a local symmetry. The Lie algebra $\frh$ is encoded in the image of a linear map $\Theta:V\rightarrow \hat \frh$ in $\hat \frh$, where $V$ is a $\hat \frh$-module and $\Theta$ is the {\em embedding tensor} satisfying 
\begin{equation}
 [\Theta(v_1),\Theta(v_2)]=\Theta(\Theta(v_1)\acton v_2)~.
\end{equation}
This relation guarantees that $\frh=\im(\Theta)$ is indeed a Lie algebra. It also implies that $V$ carries a Leibniz algebra structure defined by
\begin{equation}\label{eq:Leibniz-product}
 v_1\bullet v_2\coloneqq \Theta(v_1)\acton v_2~,
\end{equation}
as explained in detail in~\cite{Lavau:2017tvi,Hohm:2018ybo,Kotov:2018vcz}. The gauge potential 1-form $A$ now takes values in $V$. Because $V$ is not a Lie algebra, the curvature of $A$ does not transform covariantly, and 2-form potentials are introduced to compensate for this. These, again, may have curvatures that do not transform covariantly, leading to even higher forms and ultimately to what is known as the tensor hierarchy~\cite{deWit:2008ta}.

An appropriate description of this structure is given in terms of $EL_\infty$-algebras~\cite{Roytenberg:0712.3461}. Recall that the general categorification of a Lie algebra to a weak Lie 2-algebra involves a lift of the Jacobi identity by a natural transformation called the {\em Jacobiator} and a lift of the antisymmetry property by a natural transformation called the {\em Alternator}. If we take the Moore complex of such a categorified weak Lie 2-algebra, we arrive at a 2-term $EL_\infty$-algebra.
\begin{definition}[\cite{Roytenberg:0712.3461}]
 A \underline{2-term $EL_\infty$-algebra} is a 2-term chain complex $\fre:\fre_{-1}\xrightarrow{~\eps_1~}\fre_0$ equipped with a chain map $\eps_2:\fre\otimes \fre\rightarrow \fre$ as well as chain homotopies\footnote{Here, $\pi_{12}$ denotes the obvious permutation.}
 \begin{equation}
 \begin{aligned}
  \mathsf{alt}&:\eps_2(-,-)+\eps_2(-,-)\circ \pi_{12} \rightarrow 0~,\\
  \mathsf{jac}&:\eps_2(-,\eps_2(-,-))-\eps_2(\eps_2(-,-),-)-\eps_2(-,\eps_2(-,-))\circ \pi_{12}\rightarrow 0~,
 \end{aligned}
 \end{equation}
 which are explicitly given by maps
 \begin{equation}
  \mathsf{alt}:\fre_0\otimes \fre_0\rightarrow \fre_{-1}\eand
  \mathsf{jac}:\fre_0\otimes \fre_0\otimes \fre_0\rightarrow \fre_{-1}~,
 \end{equation}
 satisfying
 \begin{equation}
  \begin{gathered}
 \eps_2(x_1,x_2)+\eps_2(x_2,x_1)=\eps_1(\mathsf{alt}(x_1,x_2))~,~~~\eps_2(x_1,y)+\eps_2(y,x_1)=\mathsf{alt}(x_1,\eps_1(y))\\
 \eps_2(x_1,\eps_2(x_2,x_3))-\eps_2(\eps_2(x_1,x_2),x_3)-\eps_2(x_2,\eps_2(x_1,x_3))=\eps_1(\mathsf{jac}(x_1,x_2,x_3))~,\\
 \eps_2(x_1,\eps_2(x_2,y))-\eps_2(\eps_2(x_1,x_2),y)-\eps_2(x_2,\eps_2(x_1,y))=\mathsf{jac}(x_1,x_2,\eps_1(y))~,\\
 \end{gathered}
 \end{equation}
(because they are chain homotopies) as well as 
 \begin{equation}
  \begin{aligned}
 &\eps_2(x_1,\mathsf{jac}(x_2,x_3,x_4))+\mathsf{jac}(x_1,\eps_2(x_2,x_3),x_4)+\mathsf{jac}(x_1,x_3,\eps_2(x_2,x_4))+\\
 &\hspace{1cm}+\eps_2(\mathsf{jac}(x_1,x_2,x_3),x_4)+\eps_2(x_3,\mathsf{jac}(x_1,x_2,x_4))\\
 &\hspace{0.2cm}=\mathsf{jac}(x_1,x_2,\eps_2(x_3,x_4))+\mathsf{jac}(\eps_2(x_1,x_2),x_3,x_4)+\eps_2(x_2,\mathsf{jac}(x_1,x_3,x_4))\\
 &\hspace{1cm}+\mathsf{jac}(x_2,\eps_2(x_1,x_3),x_4)+\mathsf{jac}(x_2,x_3,\eps_2(x_1,x_4))~,\\
 &\mathsf{jac}(x_1,x_2,x_3)+\mathsf{jac}(x_2,x_1,x_3)=-\eps_2(\mathsf{alt}(x_1,x_2),x_3)~,\\
 &\mathsf{jac}(x_1,x_2,x_3)+\mathsf{jac}(x_1,x_3,x_2)=\eps_2(x_1,\mathsf{alt}(x_2,x_3))-\mathsf{alt}(\eps_2(x_1,x_2),x_3)\\
 &\hspace{8cm}-\mathsf{alt}(x_2,\eps_2(x_1,x_3))~,\\
 &\mathsf{alt}(x_1,\eps_2(x_2,x_3))=\mathsf{alt}(\eps_2(x_2,x_3),x_1)
  \end{aligned}
 \end{equation}
 for all $x_i\in \fre_0$ and $y\in \fre_{-1}$.
\end{definition}
\noindent Note that 2-term $EL_\infty$-algebras with trivial alternator $\mathsf{alt}$ are 2-term $L_\infty$-algebras and the Lie 2-algebras they describe are sometimes called {\em semistrict Lie 2-algebras}. If the Jacobiator $\mathsf{jac}$ is trivial, one speaks of {\em hemistrict Lie 2-algebras}. There is now a projection from general 2-term $EL_\infty$-algebras to 2-term $L_\infty$-algebras~\cite[Theorem 3.2]{Roytenberg:0712.3461}.

It is not hard to see that the failure of the Leibniz product~\eqref{eq:Leibniz-product} to be antisymmetric is in the kernel of $\Theta$, thus giving rise to a hemistrict $EL_\infty$-algebra structure on 
\begin{equation}
 \fre=(\,\fre_{-1}\,\rightarrow\,\fre_0\,)=(\,\ker(\Theta)\,\hookrightarrow\,V\,)~.
\end{equation}
The projection from a general 2-term $EL_\infty$-algebra to a 2-term $L_\infty$-algebra then results in an $L_\infty$-algebras structure via antisymmetrization.

The fact that the Leibniz algebra structure arising from the embedding tensor leads to a canonical 2-term $L_\infty$-algebra was explained also in~\cite{Sheng:2015:1-5} and~\cite{Kotov:2018vcz} independently of the results in~\cite{Roytenberg:0712.3461}. In~\cite{Kotov:2018vcz}, the authors also provided the differential graded associative algebra picture.

There is now indeed an example of a hemistrict Lie 2-algebra, which gives rise to the string Lie 2-algebra relevant to our model. Consider the chain complex
\begin{equation}
 \fre=(\,\fre_{-1}\rightarrow \fre_0\,)=(\,\FR\xrightarrow{~0~} \frg\,)
\end{equation}
for $\frg$ some finite-dimensional Lie algebra and $\eps_2:\fre_0\otimes \fre_0\rightarrow \fre_0$ given by the commutator and trivial on other arguments. The Killing form $(-,-)$ then yields an alternator $\mathsf{alt}(x_1,x_2)=(x_1,x_2)$. The projection on the corresponding 2-term $L_\infty$-algebra yields the Jacobiator
\begin{equation}
 \mathsf{jac}(x_1,x_2,x_3)=(x_1,[x_2,x_3])~,
\end{equation}
and we recover the string Lie 2-algebra.

It thus seems that the alternator of a hemistrict Lie 2-algebra provides the additional structure that is needed to construct an adjustment of the Weil algebra of an $L_\infty$-algebra. This point should be further explored, in particular in the context of the tensor hierarchy. Also the connection to the description of the tensor hierarchy in terms of Borcherds--Kac--Moody superalgebras as discussed in~\cite{Palmkvist:2011vz,Palmkvist:2013vya,Greitz:2013pua} should be studied in detail. Finally, let us point out that our perspective should also be relevant in the context of exceptional field theory, where similar structures arise~\cite{Hohm:2019wql}.

\subsection{Beyond local string structures}

Recall that minimal models of Lie 2-algebras can be classified~\cite[Theorem 57]{Baez:2003aa} in terms of a Lie algebra $\frg_0$, a vector space $\frg_{-1}$ carrying a representation $\rho$ of $\frg$ together with a cocycle $k(-,-,-)\in H^3(\frg_0,\frg_{-1})$, leading to the Lie 2-algebra
\begin{equation}
\begin{gathered}
 \frg_{-1}\xrightarrow{~0~}\frg_0~,\\
 \mu_2:\frg_0\wedge \frg_0\to\frg_0~,~~~\mu_2(a_1,a_2)=[a_1,a_2]~,\\
 \mu_2:\frg_0\wedge \frg_{-1}\to \frg_{-1}~,~~~\mu_2(a_1,b)=\rho_{a_1}(b)~,\\
 \mu_3:\frg_0\wedge \frg_0\wedge \frg_0\to \frg_{-1}~,~~~\mu_3(a_1,a_2,a_3)=k(a_1,a_2,a_3)
\end{gathered}
\end{equation}
for $a_{1,2,3}\in \frg_0$ and $b\in \frg_{-1}$. We should, in fact, extend our considerations to such more general Lie 2-algebras. This is particularly important for reproducing the correct branching of Lie 2-algebras which is expected from splitting a stack of $N=N_1+N_2$ M5-branes into two well-separated stacks of $N_1$ and $N_2$ M5-branes, cf.~\cite{Saemann:2019leg}. Also the above discussion of the Leibniz algebras arising from the tensor hierarchy suggests to look beyond string Lie 2-algebras.

An obvious problem to attack is thus a classification of $L_\infty$-algebras which admit an adjusted Weil algebra. This is particularly important since an extended parallel transport that does allow for corresponding interacting field theories is most likely to be only possible for such $L_\infty$-algebras. The study of the consistency of a parallel transport based on string structures as well as of higher generalizations will also be part of our future studies.

Clearly, we do not merely want to discuss metric string structures locally, but we wish to formulate a consistent global picture, ideally in terms of a general differential cocycle. Such a cocycle would consist of a \v Cech cocycle defining the principal 4-bundle underlying the metric string structure as well as the local $\astring(\frg)$-connection object and further differential forms taking values in subspaces of the Lie 4-algebra $\aghsk^\omega$ or $\aghl^\omega$ gluing all these together. We hope to report on progress in these directions soon.

\section*{Acknowledgements}

We would like to thank Domenico Fiorenza, Jo{\~ a}o Faria Martins, Alexander Schenkel, Urs Schreiber and Martin Wolf for useful discussions on various aspects of this paper. The work of CS is partially supported by the Leverhulme Research Project Grant RPG-2018-329 ``The Mathematics of M5-Branes.'' The work of LS was supported by an STFC PhD studentship and by MOE Tier~2 grant R-144-000-396-112.

\appendices

\subsection{Compositions of 2-morphisms and quasi-isomorphisms}\label{app:A:concatenation}

Let us briefly explore how 2-morphisms between morphisms of differential graded algebras and dual quasi-isomorphisms between these compose.

We start with vertical composition of 2-morphisms. Consider 2-morphisms $(\Phi_1,\Phi_2,\eta_{12})$ and $(\Phi_2,\Phi_3,\eta_{23})$ with $\Phi_i:\sCE(\frg)\rightarrow \sCE(\frh)$ for some $L_\infty$-algebras $\frg$ and $\frh$ with
\begin{equation}
\begin{aligned}
 Q_{\sCE(\frh)}\eta_{12}+\eta_{12}Q_{\sF(\frg)}&=(\Phi_1-\Phi_2)\circ i^*\circ \Upsilon~,\\
 Q_{\sCE(\frh)}\eta_{23}+\eta_{23}Q_{\sF(\frg)}&=(\Phi_2-\Phi_3)\circ i^*\circ \Upsilon~.
\end{aligned}
\end{equation}
Diagrammatically, we have
\begin{equation}
\begin{tikzcd}
&\sCE(\frg)\arrow[dl,bend right=30,"\Phi_1",swap]&&\frg[2]^*\\
\sCE(\frh) &\sCE(\frg)\arrow[l,"\Phi_2"{name=D1},swap] &\sW(\frg)\arrow[ur,hookleftarrow]\arrow[ul,"i^*"{name=U1,above}]\arrow[l,"i^*"{name=U2,above}]\arrow[dl,"i^*"]&\sF(\frg)\arrow[l,swap,"\Upsilon"]\\
&\sCE(\frg)\arrow[ul,bend left=30,"\Phi_3"{name=D2}]&& \arrow[Rightarrow,"\eta_{12}",from=U1,to=D1,start anchor={[xshift=-3ex,yshift=1ex]},end anchor={[yshift=1ex,xshift=-3ex]}]
\arrow[Rightarrow,"\eta_{23}",from=U2,to=D2,start anchor={[xshift=1ex,yshift=-2ex]},end anchor={[yshift=1ex,xshift=1ex]}]
\end{tikzcd}
\end{equation}
The composed 2-morphism is then
\begin{equation}
 (\Phi_2,\Phi_3,\eta_{23})\circ(\Phi_1,\Phi_2,\eta_{12})=(\Phi_1,\Phi_3,\eta_{12}+\eta_{23})
\end{equation}
with
\begin{equation}
\begin{aligned}
Q_{\sCE(\frh)}(\eta_{12}+\eta_{23})+(\eta_{12}+\eta_{23})Q_{\sF(\frg)}&=(\Phi_1-\Phi_2+\Phi_2-\Phi_3)\circ i^*\circ \Upsilon\\
&=(\Phi_1-\Phi_3)\circ i^*\circ \Upsilon~.
\end{aligned}
\end{equation}

Horizontal composition of 2-morphisms is slightly more cumbersome. Let $\frg_{1,2,3}$ be some $L_\infty$-algebras and $(\Phi_{12},\Psi_{12},\eta_{12})$ and $(\Phi_{23},\Psi_{23},\eta_{23})$ be 2-morphisms between them. Now we have
\begin{equation}
\begin{aligned}
 Q_{\sCE(\frg_1)}\eta_{12}+\eta_{12}Q_{\sF(\frg_2)}&=(\Phi_{12}-\Psi_{12})\circ i^*_2\circ \Upsilon_2~,\\
 Q_{\sCE(\frg_2)}\eta_{23}+\eta_{23}Q_{\sF(\frg_3)}&=(\Phi_{23}-\Psi_{23})\circ i^*_3\circ \Upsilon_3
\end{aligned} 
\end{equation}
and the diagram
\begin{equation}
\begin{tikzcd}[column sep=3ex]
&\sCE(\frg_2)\arrow[dl,bend right=30,"\Phi_{12}",swap]&&\frg_2[2]^*&&\sCE(\frg_3)\arrow[dl,bend right=30,"\Phi_{23}",swap]&&\frg_3[2]^*\\
\sCE(\frg_1) &&\sW(\frg_2)\arrow[ur,hookleftarrow]\arrow[ul,"i^*_2"{name=U,above}]\arrow[dl,"i^*_2"]&\sF(\frg_2)\arrow[l,swap,"\Upsilon_2"]&\sCE(\frg_2) &&\sW(\frg_3)\arrow[ur,hookleftarrow]\arrow[ul,"i^*_3"{name=U2,above}]\arrow[dl,"i^*_3"]&\sF(\frg_3)
\arrow[llll,"\hat\sF(\Phi_{23})", controls={+(1,-2) and +(0,-2)},start anchor={[xshift=0ex]},end anchor={[yshift=-4ex,xshift=2ex]}]\arrow[l,swap,"\Upsilon_3"]\\
&\sCE(\frg_2)\arrow[ul,bend left=30,"\Psi_{12}"{name=D}]&& \arrow[Rightarrow,"\eta_{12}",from=U,to=D,start anchor={[xshift=-1ex]},end anchor={[yshift=1ex,xshift=1ex]}]&&\sCE(\frg_3)
\arrow[ul,bend left=30,"\Psi_{23}"{name=D2}]&& \arrow[Rightarrow,"\eta_{23}",from=U2,to=D2,start anchor={[xshift=-1ex]},end anchor={[yshift=1ex,xshift=1ex]}]
\end{tikzcd}
\end{equation}
Here, 
\begin{equation}
 \hat\sF(\Phi_{23})\coloneqq \Upsilon^{-1}_2\circ \hat \Phi_{23}\circ \Upsilon_3~,
\end{equation}
where $\hat \Phi_{23}:\sW(\frg_3)\rightarrow \sW(\frg_2)$ is the lift of the map $\Phi_{23}$ to the Weil algebras, cf.~\eqref{eq:lift_CE_map_to_W}. The horizontal composition of the 2-morphisms now reads as
\begin{equation}
 (\Phi_{12},\Psi_{12},\eta_{12})\otimes(\Phi_{23},\Psi_{23},\eta_{23})=\big(\Phi_{12}\circ \Phi_{23},\Psi_{12}\circ \Psi_{23},\eta_{12}\circ \hat \sF(\Phi_{23})+\Psi_{12}\circ \eta_{23}\big)~.
\end{equation}
Indeed, we have
\begin{equation}
 \begin{aligned}
 Q_{\sCE(\frg_1)}&(\eta_{12}\circ \hat \sF(\Phi_{23})+\Psi_{12}\circ \eta_{23})+(\eta_{12}\circ \hat \sF(\Phi_{23})+\Psi_{12}\circ \eta_{23})Q_{\sF(\frg_3)}\\
 &=\big((\Phi_{12}-\Psi_{12})\circ i_2^*\circ \Upsilon_2-\eta_{12}\circ Q_{\sF(\frg_2)}\big)\circ\hat \sF(\Phi_{23})+\Psi_{12}\circ Q_{\sCE(\frg_2)}\circ \eta_{23}+ \\
 &\hspace{1cm}\eta_{12}\circ Q_{\sF(\frg_2)}\circ \hat \sF(\Phi_{23})+\Psi_{12}\circ\big((\Phi_{23}-\Psi_{23})\circ i_3^*\circ \Upsilon_3-Q_{\sCE(\frg_2)}\circ\eta_{23}\big) \\
  &=(\Phi_{12}-\Psi_{12})\circ i_2^*\circ \Upsilon_2\circ\hat \sF(\Phi_{23})+\Psi_{12}\circ(\Phi_{23}-\Psi_{23})\circ i_3^*\circ \Upsilon_3 \\
  &=\big((\Phi_{12}-\Psi_{12})\circ \Phi_{23}+\Psi_{12}\circ (\Phi_{23}-\Psi_{23})\big)\circ i_3^*\circ \Upsilon_3\\
  &=(\Phi_{12}\circ \Phi_{23}-\Psi_{12}\circ \Psi_{23})\circ i_3^*\circ \Upsilon_3~,
 \end{aligned}
\end{equation}
where we have used the fact that 
\begin{equation}
i_2^*\circ \Upsilon_2\circ\hat \sF(\Phi_{23})=\Phi_{23}\circ i_3^*\circ \Upsilon_3~.
\end{equation}

By composition of quasi-isomorphisms, we mean a chain of (dual) quasi-isomorphism
\begin{equation}
 \begin{tikzcd}
    \sCE(\frg_1) \arrow[r,bend left=30]{}{\Phi_{21}} & \sCE(\frg_2) \arrow[l,bend left=30]{}{\Psi_{12}} \arrow[r,bend left=30]{}{\Phi_{32}} & \sCE(\frg_3) \arrow[l,bend left=30]{}{\Psi_{23}}
 \end{tikzcd}
\end{equation}
with
\begin{equation}
\begin{aligned}
 {}[Q,\eta_{121}]&=\big(\Psi_{12}\circ \Phi_{21}-\id_{\sCE(\frg_1)}\big)\circ i^*_1\circ \Upsilon_1~,\\{}[Q,\eta_{212}]&=\big(\Phi_{21}\circ \Psi_{12}-\id_{\sCE(\frg_2)}\big)\circ i^*_2\circ \Upsilon_2~,\\
 {}[Q,\eta_{232}]&=\big(\Psi_{23}\circ \Phi_{32}-\id_{\sCE(\frg_2)}\big)\circ i^*_2\circ \Upsilon_2~,\\
 {}[Q,\eta_{323}]&=\big(\Phi_{32}\circ \Psi_{23}-\id_{\sCE(\frg_3)}\big)\circ i^*_3\circ \Upsilon_3~.
\end{aligned} 
\end{equation}
The composition of the two quasi-isomorphisms is the quasi-isomorphism $(\Phi_{31},\Psi_{13},\eta_{131},$ $\eta_{313})$ with
\begin{equation}
\begin{gathered}
\Phi_{31}=\Phi_{32}\circ \Phi_{21}~,~~~\Psi_{13}=\Psi_{12}\circ \Psi_{23}~,\\
\eta_{131}=\Psi_{12}\circ \eta_{232}\circ \hat \sF(\Phi_{21})+\eta_{121}~,~~~\eta_{313}=\Phi_{31}\circ \eta_{212}\circ \hat \sF(\Psi_{23})+\eta_{323}~,
\end{gathered}
\end{equation}
as one readily verifies: for example, we have
\begin{equation}
\begin{aligned}
 Q_{\sCE(\frg_1)}&\circ \eta_{131}+\eta_{131}\circ Q_{\sW(\frg_1)}\\
 &=\Psi_{12}\circ Q_{\sCE(\frg_2)}\circ \eta_{232}\circ \hat \sF(\Phi_{21})+\Psi_{12}\circ \eta_{232}\circ Q_{\sF(\frg_2)}\circ \hat \sF(\Phi_{21})+[Q,\eta_{121}]\\
 &=\Psi_{12}\circ(\Psi_{23}\circ \Phi_{32}-\id_{\sCE(\frg_2)})\circ i^\ast_2\circ\Upsilon_2\circ \hat \sF(\Phi_{21})\\
&\phantom{{}={}}+\big(\Psi_{12}\circ \Phi_{21}-\id_{\sCE(\frg_1)}\big)\circ i^*_1\circ \Upsilon_1\\
 &=\big(\Psi_{12}\circ\Psi_{23}\circ \Phi_{32}\circ\Phi_{21}-\id_{\sCE(\frg_1)}\big)\circ i_1^*\circ \Upsilon_1~.
\end{aligned}
\end{equation}

\subsection{Symplectic completion}\label{app:B:symplectic_extension}

In this appendix, we briefly summarize the computations behind the symplectic completions leading to metric string structures. Our conventions for differential forms and the Cartan calculus on graded manifolds will be those of~\cite{Cattaneo:2010re}. That is, each coordinate function $z^A$ of homogeneous degree $|z^A|$ yields a 1-form $\dd z^A$ of homogeneous degree $|z^A|+1$ (and the wedge product is graded commutative with respect to this degree). The contraction with the vector field $\der{z^A}$ satisfies $\iota_{\der{z^A}}\dd z^B=\delta^B_A$ and $\iota_{\der{z^A}}$ is a graded derivation of the algebra of differential forms of degree~$-|z^A|-1$.

In general, given a graded manifold $\frg$ with homogeneously graded coordinates $z^A$, we can symplectically complete it to a grade-shifted tangent bundle $T^*[k]\frg$ with additional homogeneously graded fiber coordinates\footnote{In the BV formalism, which corresponds to the special case $k=-1$ of this construction, the $z_A^\dagger$ are the antifields.} $z^\dagger_A$ are the  $z^\dagger_A$ of degree $|z^\dagger_A|=k-|z^A|$. On $T^*[k]\frg$, we have the canonical symplectic form
\begin{equation}
 \omega=\dd z^A\wedge \dd z_A^\dagger
\end{equation}
of degree~$2+k$ (or, if one wishes, bidegree~$(2,k)$). Starting from a homological vector field $Q=Q^A\der{z^A}$ on $\frg$, we construct the Hamiltonian 
\begin{equation}
 \hat \CQ=(-1)^{|z^A|+1} Q^A z_A^\dagger~.
\end{equation}
A quick computation checks that the Hamiltonian vector field of $\hat \CQ$, 
\begin{equation}
\hat Q=\hat Q^A\der{z^A}+\hat Q_A\der{z^\dagger_A}\ewith\iota_{\hat Q}\omega=\dd \hat \CQ~,  
\end{equation}
satisfies $\hat Q^A=Q^A$ and $\hat Q_A\sim z^\dagger_A$. Thus, upon factoring by the ideal generated by the $z_A^\dagger$, we recover the dg-manifold $(\frg,Q)$. It remains to ensure that
\begin{equation}
 \hat Q^2=0\Leftrightarrow \iota_{\hat Q}\dd\hat \CQ=\hat Q\hat \CQ=0~.
\end{equation}
We compute
\begin{equation}
 \iota_{\hat Q}\dd \hat \CQ=\iota_{\hat Q}\iota_{\hat Q}\omega=\hat Q^A\hat Q_A~,
\end{equation}
and thus we have to ensure $\hat Q^A\hat Q_A=0$, potentially by introducing quadratic and higher corrections in $z^\dagger_A$ to $\hat \CQ$.

In the case of the symplectic graded manifold  $T^*[4]\aghsk[1]$ defined in~\eqref{eq:graded_skeletal} we have the symplectic form
\begin{equation}
 \omega=\dd t^\alpha\wedge \dd u_\alpha +\dd q\wedge \dd p+\dd r\wedge \dd s~.
\end{equation}
The Hamiltonian is 
\begin{equation}
 \hat \CQ=-\tfrac12 f^\alpha_{\beta\gamma} t^\beta t^\gamma u_\alpha+\tfrac1{3!}f_{\alpha\beta\gamma}t^\alpha t^\beta t^\gamma s+qs~,
\end{equation}
which induces the homological vector field
\begin{equation}
 \hat Q=-\tfrac12 f^\alpha_{\beta\gamma}t^\beta t^\gamma\der{t^\alpha}+\left(\tfrac{1}{3!}f_{\alpha\beta\gamma}t^\alpha t^\beta t^\gamma +q\right)\der{r}-s\der{p}+\left(-f^\beta_{\alpha\gamma}t^\gamma u_\beta-\tfrac12 f_{\alpha\beta\gamma}t^\beta t^\gamma s\right)\der{u_\alpha}~.
\end{equation}
We have
\begin{equation}
 \hat Q^A\hat Q_A=\hat Q^\alpha \hat Q_\alpha=0~,
\end{equation}
because $f^\alpha_{\beta\gamma}$ satisfies the Jacobi identity.

In the case of the loop model $\aghl^\omega$, we have the symplectic form
\begin{equation}
 \omega=\int_0^1 \dd \tau~\left(\delta t^{\alpha \tau}\wedge \delta u_{\alpha \tau}+\delta r^{\alpha \tau}\wedge \delta s_{\alpha \tau}\right)+\delta q\wedge \delta p+\delta r_0\wedge \delta s_0~,
\end{equation}
and the minimal Hamiltonian reads as
\begin{equation}
\begin{aligned}
 \hat \CQ&=\int_0^1\dd \tau\,\left( (-\tfrac12 f^\alpha_{\beta\gamma}t^{\beta \tau}t^{\gamma \tau}-r^{\alpha \tau})u_{\alpha \tau}-f^\alpha_{\beta\gamma}t^{\beta \tau}r^{\gamma \tau}s_{\alpha \tau}\right)+\\
 &\phantom{=}+\left(2\int_0^1\dd \tau\, \kappa_{\alpha \beta}t^{\alpha \tau} \dot r^{\beta \tau} +q\right)s_0
\end{aligned}
\end{equation}
with Hamiltonian vector field
\begin{equation}
\begin{aligned}
 \hat Q&=\int_0^1\dd \tau\, \left(\big(-\tfrac12 f^\alpha_{\beta\gamma}t^{\beta\tau}t^{\gamma\tau}-r^{\alpha\tau}\big)\der{t^{\alpha\tau}}-f^\alpha_{\beta\gamma}t^{\beta\tau}r^{\gamma\tau}\der{r^{\alpha\tau}}+\right.\\
 &\hspace{2cm}+\left(-f^\gamma_{\alpha\beta}t^{\beta\tau}u_{\gamma\tau}-f^\gamma_{\alpha\beta}r^{\beta\tau}s_{\gamma\tau}+2\kappa_{\alpha\beta}\dot r^{\beta\tau}s_0\right)\der{u_{\alpha\tau}}+\\
 &\hspace{2cm}\left.+\left(-f^\gamma_{\alpha\beta}t^{\beta\tau}s_{\gamma\tau}+2\kappa_{\alpha\beta}\dot t^{\beta\tau}s_0+u_{\alpha\tau}\right)\der{s_{\alpha\tau}}\right)+\\
 &\phantom{=}+\left(2\int_0^1\dd \tau\, \kappa_{\alpha \beta}t^{\alpha \tau} \dot r^{\beta \tau}+q\right)\der{r_0}-s_0\der{p}~,
\end{aligned} 
\end{equation}
where we used 
\begin{equation}
 \int_0^1\dd \tau\, \kappa_{\alpha \beta}t^{\alpha \tau} \dot r^{\beta \tau}=\kappa_{\alpha\beta}\left(t^{\alpha1}r^{\beta 1}-t^{\alpha 0}r^{\beta 0}\right)-\int_0^1\dd \tau\, \kappa_{\alpha \beta}\dot t^{\alpha \tau} r^{\beta \tau}=-\int_0^1\dd \tau\, \kappa_{\alpha \beta}\dot t^{\alpha \tau} r^{\beta \tau}
\end{equation}
since $r^{\beta1}=r^{\beta 0}=0$.

\subsection{Kernel extension of an \texorpdfstring{$L_\infty$}{L-infinity}-algebra}\label{app:C}

Let us briefly discuss the extension of a $k$-term $L_\infty$-algebra 
\begin{equation}
\frg~=~(~\frg_{-k+1}\xrightarrow{~\mu^{-k+1}_1~}\dots \xrightarrow{~\mu^{-1}_1~}\frg_0~) 
\end{equation}
by the kernel of its left-most differential, 
\begin{equation}\label{eq:kernel_extension}
\hat \frg~=~(~\ker(\mu^{-k+1}_1)\xhookrightarrow{~~e~~}\frg_{-k+1}\xrightarrow{~\mu^{-k+1}_1~}\dots \xrightarrow{~\mu^{-1}_1~}\frg_0~)~.
\end{equation}
Note that such a $\hat \frg$ is necessarily quasi-isomorphic to a $k-1$-term $L_\infty$-algebra. In this paper, we considered such extensions of the string Lie 2-algebra to a Lie 3-algebra and of its symplectic completion to a Lie 4-algebra.

For simplicity, we shall discuss the extension in the higher product formulation. The higher products on $\frg$ induce higher products on $\hat \frg$:
\begin{equation}\label{eq:mu_i}
 \hat \mu_i|_{\wedge^i\frg}\coloneqq \mu_i~,
\end{equation}
and we define 
\begin{equation}\label{eq:mu_1_extension}
 \hat \mu_1(b)\coloneqq e(b)~,~~~b\in \hat \frg_{-k}=\ker(\mu_1^{-k+1})~.
\end{equation}
To fulfill the homotopy Jacobi identities for $\hat \frg$, we first note that clearly $\hat \mu_1\circ \hat \mu_1=0$. The only other homotopy Jacobi identities in $\hat \frg$ that now differ from those in $\frg$ due to definition~\eqref{eq:mu_1_extension} are the ones which contain terms of the form
\begin{equation}
 \hat \mu_{j+1}(\hat \mu_1(b),a_1,\dots,a_j)~,~~~b\in \hat \frg_{-k}~.
\end{equation}
Because of their degrees, the only higher products taking arguments in $\frg_{-k+1}$ are $\mu_1$ and $\mu_2$, and therefore the only affected homotopy Jacobi identity is
\begin{equation}
 \hat \mu_1(\hat \mu_2(a,b))=\hat \mu_2(a,\hat \mu_1(b))~,~~~a\in \frg_0~,~b\in \hat \frg_{-k}~.
\end{equation}
Since $\hat \mu_1$ is injective and $\mu_2(a,\hat \mu_1(b))\in \ker(\mu_1^{-k+1})$ due to the Jacobi identity in $\frg$, we can define
\begin{equation}\label{eq:mu_2_extension}
 \hat \mu_2(a,b)\coloneqq e^{-1}(\mu_2(a,e(b)))~,~~~a\in \frg_0~,~b\in \hat \frg_{-k}~,
\end{equation}
where $e^{-1}$ is the inverse of $e$ on $\ker(\mu_1^{-k+1})$. For $a_1,a_2\in \frg_0$ and $a_3\in \ker(\mu_1^{-k+1})\subset \frg_{-k}$, we have the homotopy Jacobi identity
\begin{equation}
\begin{aligned}
 0&=\mu_2(\mu_2(a_1,a_2),a_3)\pm\mu_2(\mu_2(a_1,a_3),a_2)\pm\mu_2(\mu_2(a_2,a_3),a_1)\\
 &~~~\pm\mu_1(\underbrace{\mu_3(a_1,a_2,a_3)}_{=0})\pm\mu_3(\underbrace{\mu_1(a_3)}_{=0},a_1,a_2)~,
\end{aligned}
\end{equation}
which translates to the new homotopy Jacobi identity
\begin{equation}
 0=\hat \mu_2(\hat \mu_2(a_1,a_2),b)\pm\hat \mu_2(\hat \mu_2(a_1,b),a_2)\pm\hat \mu_2(\hat \mu_2(a_2,b),a_1)
\end{equation}
for $b\in \hat \frg_{-k}$. Thus, the higher products $\hat \mu_i$ on $\hat \frg$ satisfy the homotopy Jacobi identities.

\begin{proposition}
 Any $k$-term $L_\infty$-algebra $\frg$ possesses a kernel extension to a $k+1$-term $L_\infty$-algebra $\hat \frg$ as in~\eqref{eq:kernel_extension}. The higher products $\hat \mu_i$ on $\hat \frg$ are given by equations~\eqref{eq:mu_i},~\eqref{eq:mu_1_extension}, and~\eqref{eq:mu_2_extension}.
\end{proposition}

\bibliography{bigone}

\bibliographystyle{latexeu}

\end{document}